\documentclass[prd,aps,showpacs,nofootinbib,twocolumn]{revtex4-2}%

\usepackage[justification=justified]{caption}
\usepackage{graphicx}
\usepackage{amssymb}
\usepackage{amsmath}
\usepackage{color}
\usepackage{float}
\usepackage{graphicx}
\usepackage{graphicx}
\usepackage[dvipsnames]{xcolor}
\usepackage{amsfonts}
\usepackage[colorlinks=true,pdfstartview=FitV,linkcolor=blue,citecolor=blue,urlcolor=blue,breaklinks=true]{hyperref}
\usepackage{subfig}%
\usepackage{cancel}
\usepackage{dsfont}
\providecommand{\U}[1]{\protect\rule{.1in}{.1in}}

\newcommand{\orcid}[1]{\href{https://orcid.org/#1}{\includegraphics[width=10pt]{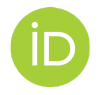}}}

\begin{document}

\title{Nonminimal planar electrodynamics modified by higher-derivative terms}
\author{Let\'{\i}cia Lisboa-Santos\orcid{0000-0003-4939-3856}$^a$}
\email{leticia.lisboa@discente.ufma.br}
\author{Jo\~{a}o A.A.S. Reis\orcid{0000-0002-2831-5317}$^b$}
\email{joao.reis@uesb.edu.br}
\author{Marco Schreck\orcid{0000-0001-6585-4144}$^{a,c}$}
\email{marco.schreck@ufma.br}
\author{Manoel M. Ferreira, Jr.\orcid{0000-0002-4691-8090}$^{a,c}$}
\email{manojr.ufma@gmail.com, manoel.messias@ufma.br}
\affiliation{$^a$Programa de P\'{o}s-gradua\c{c}\~{a}o em F\'{i}sica, Universidade Federal do Maranh\~{a}o, Campus
	Universit\'{a}rio do Bacanga, S\~{a}o Lu\'is (MA), 65080-805, Brazil}
\affiliation{$^b$Departamento de Cie\^ncias Exatas e Naturais, Universidade Estadual do Sudoeste da Bahia, Campus Juvino Oliveira, Itapetinga (BA), 45700-00, Brazil,}
\affiliation{$^c$Departamento de F\'{\i}sica, Universidade Federal do Maranh\~{a}o, Campus
	Universit\'{a}rio do Bacanga, S\~{a}o Lu\'{\i}s (MA), 65080-805, Brazil}

\begin{abstract}
We consider a $(2+1)$-dimensional modified electrodynamics endowed with terms that are either Lorentz-invariant or Lorentz-violating and involve an ever increasing number of derivatives. Our construction relies on $\mathit{U}(1)$ gauge invariance and the Abelian Chern-Simons term poses the starting point. The structure of the nonminimal Standard-Model Extension (SME) in $(3+1)$ spacetime dimensions serves as an inspiration for our pursuit. For elaborate studies and applications we particularly focus on the second term of the operator series in the general framework, which is the first contribution with additional derivatives. The latter forms the essential ingredient for several models of modified planar electrodynamics to be examined. The propagators of the models constitute the foundation for us deriving the physical propagating modes as well as for drawing conclusions on unitarity in the quantum regime. We are also interested in identifying parameter regions of sub- and superluminal mode propagation and determine classical solutions of the field equations for the planar models introduced. Moreover, a duality between an extended Chern-Simons theory and a subset of the fermion sector coefficients in the nonminimal SME is pointed out, as well. Finally, the integer quantum Hall effect is chosen as a testbed to demonstrate the applicability of our findings to real physical systems. Predictions on momentum- and direction-dependent corrections of the Hall resistivity are made at the level of effective field theory, which could be tested in experiments. Thus, the $(2+1)$-dimensional models proposed are potentially applicable to model electromagnetic phenomena in certain planar condensed-matter systems.
\end{abstract}

\maketitle

\section{Introduction}

Pioneering developments on higher-derivative electrodynamics due to Bopp \cite{Bopp} in 1940 and Podolsky~\cite{Podolsky1,Podolsky2} in 1942 opened a rich line of research that has remained fruitful until today. Podolsky's second-order derivative term, $\theta^{2}\partial_{\alpha}F^{\alpha\beta} \partial_{\sigma}F^{\sigma}_{\phantom{\sigma}\beta}$, with the electromagnetic field strength tensor $F_{\mu\nu}=\partial_{\mu}A_{\nu}-\partial_{\nu}A_{\mu}$ of the \textit{U}(1) vector field $A_{\mu}$ and the dimensionful parameter~$\theta$, when included into the Maxwell Lagrangian, provides two distinct dispersion relations for electromagnetic waves. After quantization, one corresponds to an Abelian, massless vectorboson, which can be interpreted as the usual Maxwell photon, and the other describes a massive vectorboson whose mass is proportional to~$\theta^{-1}$.
The Maxwell-Podolsky model has been addressed in several respects, including its constraint structure \cite{Galvao}, its quantization based on different procedures \cite{Barcelos}, the problems of self-force and self-interaction \cite{Gratus, Zayats}, its Green's functions and classical solutions \cite{Lazar}, the multipole expansion for fields in the static regime~\cite{Bonin}, the conservation of the energy-momentum tensor \cite{Fan}, its consistency based on the BRST approach \cite{Dai}, its quantum field theoretic properties \cite{Pimentel,Zambrano}, and other aspects \cite{Granado}.

Lee-Wick electrodynamics, which is another higher-derivative model, is characterized by the dimension-6 term $F_{\mu\nu
}\partial_{\alpha}\partial^{\alpha}F^{\mu\nu}$ \cite{Lee-Wick-1}. Incorporating this modification into the Maxwell Lagrangian implies energy instabilities at the classical level and negative-norm states in the Hilbert space at the quantum level. Therefore, this kind of term requires a decoupling mechanism that separates the negative-norm states from the physical Hilbert space to assure unitarity. The latter is reestablished in the context of the Lee-Wick Standard Model \cite{LW1,LW2}, which has always been of vast interest to the scientific community~\cite{Turcati,Accioly2,Barone1,Borges,Anselmi,Manavella}.

Field theories with higher-derivative terms play a relevant role in quantization attempts of gravity, too, where their inclusion assures renormalizability \cite{Stelle}. However, the drawback is then that ghosts and an indefinite Hilbert space metric emerge \cite{Asorey}.

Higher-derivative models have also been investigated systematically in the context of Lorentz-violating theories in the absence of gravity. First of all, the minimal nongravitational Standard-Model Extension (SME) as a comprehensive field theory framework for parameterizing Lorentz violation was constructed in Ref.~\cite{Colladay}. The latter is endowed with gauge invariance, translation invariance, and power-counting renormalizability, i.e., it only involves field operators of mass dimensions 3 and~4. Dynamical Standard-Model (SM) fields are coupled to nondynamical, tensor-valued background fields, which guarantees coordinate invariance of the Lagrange density. The background fields are tensors under coordinate transformations, whose components are frequently denoted as controlling coefficients in the SME literature.

The SME involves all particle sectors of the Standard Model. Its Dirac fermion sector has a spin-degenerate part with a single dispersion relation for both spin-up and spin-down fermions. The complementary spin-nondegenerate piece is equipped with two distinct dispersion relations dependent on the spin projection~\cite{Kostelecky:2000mm}. The electromagnetic sector of the SME exhibits a \textit{CPT}-odd piece, represented by the Carroll-Field-Jackiw (CFJ) term, which has been studied extensively due to its variety of peculiar properties~\cite{CFJ,CFJ2,CFJ3,CFJ4,CFJ5,CFJ6,CFJ7}. One of the most intriguing features of this setting is vacuum birefringence, which makes the rotation plane of linearly polarized light rotate \textit{in vacuo}. There are also \textit{CPT}-even contributions~\cite{KM,Escobar,Belich,Brett1} that decompose into a birefringent sector and a nonbirefringent one at leading order in the coefficients. The latter have been subject to broad experimental searches implying strict constraints~\cite{Klinkhamer:nonbirefringent}. Several aspects of Lorentz-violating versions of quantum electrodynamics (QED) have been studied, as well; see, e.g.,~Ref.~\cite{Sobreiro}. Finally, we would like to highlight projects pursued on Lorentz-violating modifications of scalar field theories, which are based on the SME Higgs sector, such as those on Bose-Einstein condensation~\cite{Casana} and several other aspects~\cite{Maluf}.

Myers and Pospelov were the first to introduce a dimension-5 higher-derivative model \cite{Myers,Marat1,Marat2,Marat5} describing Lorentz violation for scalars, photons, and Dirac fermions. In the aftermath, higher-derivative extensions of the minimal SME including terms of mass dimensions greater than four were constructed generically \cite{Kostelecky,KosteleckyDing,Mewes,Schreck}, which gave rise to the nonminimal SME in Minkowski spacetime. The latter comprises an infinite number of such contributions. Over the recent years, a subset of the lowest-dimensional terms has been investigated thoroughly. In particular, the focus was on classical aspects of the modes, e.g., causality and stability as well as certain indispensable properties of the corresponding quantized theories such as unitarity.

For example, Ref.~\cite{Leticia2} rests upon a \textit{CTP}-even dimension-6 term of the electromagnetic sector of the SME. The latter analysis is complemented by Ref.~\cite{Leticia1}, which is on a \textit{CPT}-odd dimension-5 CFJ-type structure. The dimension-5 term can be constrained by radiative corrections arising in other sectors \cite{Brett}. It has also been used to evaluate the interaction energy between electromagnetic sources \cite{Borges2022} and to examine optical effects in a continuous dielectric medium \cite{Pedro2021}. From a phenomenological viewpoint, higher-derivative theories have been tightly constrained recently via optical polarization data from active galactic nuclei~\cite{Kislat}. There is further literature on classical solutions \cite{Borges-Ferrari}, the thermodynamic properties of electrodynamic systems~\cite{Filho-Maluf}, radiative corrections~\cite{Ferrari}, and developments in QED~\cite{Thiago}.

Our particular interest in the current paper is on modified electrodynamics in $(2+1)$ spacetime dimensions. In this realm one highly fascinating class of theories beyond conventional Maxwell electrodynamics emerges that are related to a concept known as the Chern-Simons (CS) form. The origin of the latter is found in the study of topological invariants of manifolds such as the Euler characteristic, which are of great significance in pure and applied mathematics. Chern-Simons forms originally arose in an attempt made by Simons to obtain a combinatorial formula for another topological invariant for $4n$-dimensional manifolds ($n\in\mathbb{N}$), which is called the signature. The starting point of this endeavor was a four-dimensional manifold and --- to Simons's surprise --- it turned out to be futile due to the emergence of a three-dimensional boundary term that was intractable. Simons discovered that the latter had intriguing properties on its own right, whereupon Chern and Simons generalized these results to manifolds of any odd number of dimensions~\cite{Chern:1974}. In the early 1980s physicists adopted these findings~\cite{Deser,Hagen} and since then they have found hitherto unexpected applications in a large number of different arenas such quantum field theory \cite{Witten:1988hf}, string theory \cite{Witten:1992fb}, and condensed-matter physics \cite{Fradkin:1997ge,Qi:2008ew,Tong:2016kpv}, to just mention a few.

By considering a $\mathit{U}(1)$ bundle over $(2+1)$-dimensional Minkowski spacetime, the celebrated CS action reads
	\begin{equation}
		S_{\mathrm{CS}}=\frac{k}{2}\int A\wedge \mathrm{d}A=\frac{k}{2}\int \mathrm{d}^{3}x\,\varepsilon^{\mu\nu\rho}A_{\mu}%
		\partial_{\nu}A_{\rho}\,,
        \label{eq:CST-1}%
	\end{equation}
with the vector potential 1-form $A$, its exterior derivative $\mathrm{d}A$, and the wedge product $\wedge$. Thus, the integrand of the latter corresponds to an Abelian version of the CS form. In the physics literature, this CS action is often written in components with the completely antisymmetric Levi-Civita symbol $\varepsilon^{\mu\nu\rho}$ in $(2+1)$ spacetime dimensions. The parameter $k$ is dimensionful and remains unspecified at this moment, although it will later be subject to a set of conditions when gauge invariance is imposed. As the vector potential has mass dimension 1/2 in $(2+1)$ spacetime dimensions, a contribution of the form $A_{\mu}\partial_{\nu}A_{\rho}$ has mass dimension 2. Therefore, we conclude that $[k]=1$ in Eq.~\eqref{eq:CST-1}.

The particular term of Eq.~\eqref{eq:CST-1} yields a planar electrodynamics. Since it also has topological characteristics, the associated electrodynamics is remarkably different from planar Maxwell theory. When coupled to an external current density, the CS term connects the charge density $\rho$ to the magnetic field $B$ via the relationship $\rho=k B$~\cite{CS1}. The latter is crucial for magnetic-flux quantization, which provides an appealing description of the Aharonov-Bohm effect. Due to this property, Eq.~\eqref{eq:CST-1} also plays a significant role as an effective theory for the integer quantum Hall effect~\cite{Tong:2016kpv} and is even capable of effectively describing quasi-particles that obey a fractional statistics and are known as anyons~\cite{Wilczek:1982wy,Wilczek:1990}. This makes CS theory applicable to the fractional quantum Hall effect, too.

Chern-Simons electrodynamics has been extensively studied in both the classical and the quantum regime~\cite{Dunne,JNair,CSNR,Loop}. The construction of topological defects in the context of CS-Higgs electrodynamics was also examined with great interest \cite{CSvortex,VLcs2}. Another interesting feature is that the CS term is of first order in spacetime derivatives, rendering its canonical structure significantly different from that of Maxwell theory.

At a first glance, it is not obvious that the CS term is gauge-invariant, since it explicitly depends on $A_{\mu}$. In fact, by applying a gauge transformation $A\mapsto A+\mathrm{d}\omega$, or in component form, $A_{\mu}\mapsto A_{\mu}+\partial_{\mu}\omega$, to Eq.~\eqref{eq:CST-1}, the action changes by a total derivative:
\begin{subequations}
\begin{align}
S_{\mathrm{CS}}&\mapsto S_{\mathrm{CS}}+\delta S_{\mathrm{CS}}\,, \displaybreak[0]\\[2ex]
\delta S_{\mathrm{CS}}&=\frac{k}{2}\int \mathrm{d}\omega\wedge\mathrm{d}A=\frac{k}{2}\int \mathrm{d}(\omega\wedge\mathrm{d}A) \notag \\
&=\frac{k}{2}\int \mathrm{d}^{3}x\,\partial_{\mu}(\omega\varepsilon^{\mu\nu\rho}\partial_{\nu}A_{\rho})\,.
\end{align}
\end{subequations}
The latter does not contribute as long as the fields vanish sufficiently fast far from the origin, whereupon gauge invariance is established. However, if the gauge group has nontrivial topology, i.e., if there are noncontractible loops, $\omega$ can be multi-valued~\cite{Tong:2016kpv,CS1}. This does not pose a problem \textit{per se}, as $\omega$ is a gauge degree of freedom and it is wavefunctions and fields that should be single-valued. In our conventions, this requirement then leads to the essential finding that $2\pi k\in\mathbb{Z}$, i.e., $k$ must be quantized in units of $1/(2\pi)$.

In 1999 Deser and Jackiw (DJ) proposed an extension of CS theory with higher-order derivatives~\cite{CS2} by including a term into the action that contains the d'Alembertian $\square\equiv \partial_{\mu}\partial^{\mu}$:
\begin{equation}
S_{\mathrm{DJ}}=\frac{\vartheta}{2}\int A\wedge \mathrm{d}(\square A)=\frac{\vartheta}{2}\int\mathrm{d}^3x\,\varepsilon_{\mu\nu\rho}A^{\mu}\partial^{\nu}\square A{}^{\rho}\,,
\label{DJT1}
\end{equation}
which is $\mathit{SO}(2,1)$-invariant and has mass dimension 4, so that $[\vartheta]=-1$.
Adding the DJ term to the Maxwell term in $(2+1)$ spacetime dimensions results in a modification of Maxwell theory that could be coined Maxwell-Deser-Jackiw (MDJ) theory. This altered electrodynamics exhibits two modes. One of the modes is massless and can be associated with the standard photon after quantization. The other is massive, originates from the presence of the higher-derivative term, and corresponds to a ghost.

An additional important aspect to be highlighted is on the nature of the integrand in Eq.~\eqref{DJT1}. While the CS term is of topological nature, the DJ term involves the metric tensor \cite{CS2}. This nonstandard contribution can be generated in the context of a noncommutative massive planar QED by integrating out the fermionic fields in the effective action~\cite{Sanchez,Anacleto,Bufalo}.

Since DJ electrodynamics was first proposed, it has been addressed from several perspectives, including its Lagrangian and Hamiltonian formulations \cite{Kumar}, dualities between different higher-derivative theories \cite{Banerjee,Bazeia}, conservation laws and stability \cite{Kaparulin,Abakumova} as well as classical stationary solutions in the	presence of sources describing pointlike charges and Dirac points \cite{Borges1}. Having in mind that MDJ electrodynamics exhibits a ghost mode that implies an indefinite metric in the Hilbert space, the canonical quantization and Hamiltonian structure of this theory had to be examined in detail, which was accomplished in Ref.~\cite{Avila}. The latter article also provides additional analyses of microcausality and unitarity.

In the recent paper \cite{Joca}, a nonminimal Lorentz-violating planar electrodynamics containing field operators of arbitrary mass dimensions was constructed by applying a procedure known as dimensional reduction to the nonminimal electromagnetic sector of the SME~\cite{Kostelecky:2009zp}. The corresponding Lagrange density is composed of a {\em CPT}-even sector parameterized by the tensor-valued operator $(\hat{k}_{F})^{\kappa\lambda\mu\nu}$. Moreover, it contains a {\em CPT}-odd piece, which is a generalization of the CFJ term and depends on the vector-valued background field operator $(\hat{k}_{AF})^{\kappa}$. Note that the latter as well as its $(3+1)$-dimensional counterpart are sometimes also called Maxwell-Chern-Simons terms. It should be emphasized, though, that they are not genuine CS terms, as they require the presence of a vector-valued background field. Such theories are substantially different from the modified electrodynamics that we will be referring to as Maxwell-Chern-Simons~(MCS) theories, which do not incorporate any violation of Lorentz invariance:
\begin{align}
S_{\mathrm{MCS}}&=\int \bigg(-\frac{1}{4}F\wedge (*F)+\frac{k}{2}A\wedge \mathrm{d}A\bigg) \notag \\
&=\int\mathrm{d}^3x\,\left(-\frac{1}{4}F_{\mu\nu}F^{\mu\nu}{}+\frac{k}{2}\varepsilon_{\mu\nu\rho}A^{\mu}\partial^{\nu}A{}^{\rho}\right)\,,
\end{align}
where $*$ is the Hodge dual.

The framework presented in Ref.~\cite{Joca} involves a planar electromagnetic field $A_{\mu}$ and a scalar field $\phi$, which originates from dimensional reduction. Both fields couple to each other. The procedure applied also gives rise to five classes of distinct Lorentz-violating operators that can be understood as infinite sums over suitable contractions of controlling coefficients and additional derivatives, whose number successively increases by 2. The general structure of this theory and some its properties were scrutinized in the latter paper.

In the present work we propose a nonminimal planar electrodynamics by including higher derivatives into the CS term of Eq.~\eqref{eq:CST-1}. To do so, the third-rank tensor $k\varepsilon^{\mu\nu\varrho}$ is generalized to a tensor-valued background operator, say $\hat{\mathcal{Q}}^{\mu\nu\rho}$, which implies the violation of $\mathit{SO}(2,1)$ invariance. The latter is suitably contracted with an increasing number of additional derivatives. In the following, after examining basic aspects of the higher-derivative structures, we propose a general extended Lagrange density, which encompasses several new possibilities of planar models and contains the usual Maxwell and CS terms as special cases. From the gauge field propagator, we obtain the dispersion relations that allow us to analyze whether or not the velocity of signal propagation exceeds the speed of light. One interesting observation is that the higher-derivative CS term provides dynamics to the planar theory even in absence of the kinetic Maxwell term. Our ultimate goal is to level the ground for studies of electromagnetic aspects in $(2+1)$-dimensional condensed-matter systems. First applications are to be found in the quantum Hall effect.

This paper is organized as follows. In Sec.~\ref{sec:nonminimal-CS}, the basic aspects of the higher-derivative operator constructed from the CS structure are presented, including its symmetries. In Sec.~\ref{sec:modified-planar-electrodynamics-with-LV}, we intend to review important properties of several versions of planar electrodynamics modified by higher-derivative terms, but with Lorentz invariance intact. Section~\ref{sec:MCS-electrodynamics-higher-derivatives} is dedicated to extended models of planar electrodynamics that do not only involve higher derivatives, but also Lorentz-violating contributions. Section~\ref{sec:duality} presents an intriguing duality between a higher-derivative Lorentz-violating electrodynamics and a modified Dirac theory in $(2+1)$ spacetime dimensions. These rather technical findings culminate in an application to the quantum Hall effect in Sec.~\ref{sec:application-quantum-Hall-effect}. Finally, Sec.~\ref{sec:conclusions} provides a conclusion of all the results obtained. Worthwhile calculational details are relegated to App.~\ref{eq:green-function-spacelike-configuration-space-details}. Natural units with $\hbar=c=1$ are employed unless otherwise stated. The Minkowski metric $\eta_{\mu\nu}$ has signature $(+,-,-,-)$ in our conventions.

\section{Extended higher-derivative Chern-Simons term}
\label{sec:nonminimal-CS}

In this section, we introduce a higher-derivative extension of the $(2+1)$-dimensional CS term by following the systematic procedure employed for constructing nonminimal Lorentz-violating theories \cite{Kostelecky,KosteleckyDing,Mewes,Schreck,Joca}. Such nonminimal terms arise in replacing $k\varepsilon^{\mu\nu\varrho}$ in Eq.~\eqref{eq:CST-1} by a tensor-valued operator in $(2+1)$ spacetime dimensions, which contains arbitrary powers of spacetime derivatives. This operator will be denoted as $\hat{\mathcal{Q}}^{\mu\nu\rho}$ and will be implemented into the bilinear CS action as follows. So we obtain an extended, nonminimal CS action that we indicate by a prime:
\begin{equation}
S_{\mathrm{CS}}'=\frac{1}{2}\int \mathrm{d}^{3}x\,A_{\mu}\hat{\mathcal{Q}}^{\mu\nu\rho}\partial_{\nu}A_{\rho}\,,
\label{APCS}%
\end{equation}
where the operator $\hat{\mathcal{Q}}^{\mu\nu\rho}$ is defined as a sum of terms with a number of derivatives successively increasing by~\textcolor{violet}{1}:
\begin{equation}
	\hat{\mathcal{Q}}^{\mu\nu\rho}=\sum_{d\geq 2}^{\infty}\mathcal{Q}^{(d)\mu\nu\rho\alpha_{1}\dots\alpha_{(d-2)}}%
	\partial_{\alpha_{1}}\dots\partial_{\alpha_{(d-2)}}\,,
\label{Q}%
\end{equation}
in accordance with the pattern of similar operators that occur in nonminimal Lorentz-violating theories, such as $(\hat{k}_{AF})^{\kappa}$ in the electromagnetic sector of the nonminimal SME \cite{Kostelecky}. Here, the $\mathcal{Q}^{(d)\mu\nu\rho\alpha_{1}\dots\alpha_{(d-2)}}$ are frequently referred to as controlling coefficients. Note that in Ref.~\cite{Joca} we took over the notation from the electromagnetic sector of the SME in $(3+1)$ spacetime dimensions that served as the very base of the paper. Therefore, the label $d$ in the latter article does not relate to the mass dimension of field operators, but is simply inherited from the SME. However, since the current construction in $(2+1)$ spacetime dimensions is independent of the SME, $d$ now stands, in fact, for the mass dimension of the field operator in $(2+1)$ spacetime dimensions, which includes all derivatives and is contracted with a particular coefficient.

By counting the mass dimensions properly, the operator $\hat{\mathcal{Q}}^{\mu\nu\rho}$ is deduced to have mass dimension 1, for consistency: $[\hat{\mathcal{Q}}^{\mu\nu\rho}]=1$. In the sum of Eq.~\eqref{Q}, the tensor-valued coefficients of $d=2$ do not come with additional derivatives. Hence, this sector represents the conventional CS term of Eq.~\eqref{eq:CST-1}, i.e., $\mathcal{Q}^{(2)\mu\nu\rho}=k\varepsilon^{\mu\nu\rho}$. Now, the operator of Eq.~\eqref{Q} can be further decomposed as follows:
\begin{subequations}
\begin{align}
\label{QkappaT1}
\hat{\mathcal{Q}}^{\mu\nu\rho}&=\hat{\mathcal{K}}^{\mu\nu\rho}%
+\hat{\mathcal{T}}^{\mu\nu\rho}\,, \\[2ex]
\hat{\mathcal{K}}^{\mu\nu\rho}&=\sum_{\substack{d\geq
2\\\text{\textrm{even}}}}^{\infty}\mathcal{K}^{(d)  \mu
\nu\rho\alpha_{1}\dots\alpha_{(d-2)}}\partial_{\alpha_{1}%
}\dots\partial_{\alpha_{(d-2)}}\,, \\
\hat{\mathcal{T}}^{\mu\nu\rho}&=\sum_{\substack{d\geq
3\\\mathrm{odd}}}^{\infty}\mathcal{T}^{(d)\mu\nu\rho\alpha
_{1}\dots\alpha_{(d-2)}}\partial_{\alpha_{1}}\dots
\partial_{\alpha_{(d-2)}}\,, \label{K}%
\end{align}
\end{subequations}
where the first operator is endowed with an even number of derivatives and the second with an odd number. The latter are contracted with an ever rising number of additional Lorentz indices of the controlling coefficients with mass dimensions
\begin{equation}
\left[  \mathcal{K}^{(d)\mu\nu\rho\alpha_{1}\dots
\alpha_{(d-2)}}\right]=\left[\mathcal{T}^{(d)\mu\nu\rho\alpha_{1}\dots
\alpha_{(d-2)}}\right]=3-d\,.
\end{equation}
So to maintain $[\hat{\mathcal{Q}}^{\mu\nu\rho}]=1$, the mass dimensions of the coefficients decrease when additional derivatives are contracted with the latter.
In principle, we do not impose any symmetry on the tensor operator $\hat{\mathcal{Q}}$. However, we need to require that the action be invariant under the gauge transformation $A_{\mu}\mapsto A_{\mu}+\partial_{\mu}\omega$, whose application to Eq.~$\eqref{APCS}$ provides
\begin{align}
S_{\mathrm{CS}}'&\mapsto \frac{1}{2}\int \mathrm{d}^{3}x\,\Big( A_{\mu}\hat{\mathcal{Q}}^{\mu\nu\rho}\partial_{\nu}A_{\rho}+\partial_{\mu}\omega\hat{\mathcal{Q}}^{\mu\nu\rho}\partial_{\nu}A_{\rho} \notag \\
&\phantom{{}={}}\hspace{1.6cm}+A_{\mu}\hat{\mathcal{Q}}^{\mu\nu\rho}\partial_{\nu}\partial_{\rho}\omega
+\partial_{\mu}\omega\hat{\mathcal{Q}}^{\mu\nu\rho}\partial_{\nu}%
\partial_{\rho}\omega\Big)\,.
\end{align}
Imposing that $\hat{\mathcal{Q}}$ be antisymmetric in the last two indices, we obtain
\begin{align}
S_{\mathrm{CS}}'&\mapsto\frac{1}{2}\int \mathrm{d}^{3}x\,\Big(A_{\mu}\hat{\mathcal{Q}}^{\mu\nu\rho}\partial_{\nu}A_{\rho}+\partial_{\mu}\omega\hat{\mathcal{Q}}^{\mu\nu\rho}\partial_{\nu}A_{\rho}\Big)
\notag \\
&=S_{\mathrm{CS}}'+\frac{1}{2}\int\mathrm{d}^{3}x\,\partial_{\mu}\omega\hat{\mathcal{Q}}^{\mu\nu\rho}\partial_{\nu}A_{\rho}\,.
\label{actionCSQ2}
\end{align}
Furthermore, it is beneficial to introduce a total derivative via
\begin{equation}
\partial_{\mu}\left(  \omega\hat{\mathcal{Q}}^{\mu\nu\rho}\partial_{\nu
}A_{\rho}\right)  =\partial_{\mu}\omega\hat{\mathcal{Q}}^{\mu\nu\rho
}\partial_{\nu}A_{\rho}+\omega\hat{\mathcal{Q}}^{\mu\nu\rho}\partial_{\mu
}\partial_{\nu}A_{\rho}\,,
\end{equation}%%
which allows us to rewrite Eq.~\eqref{actionCSQ2} as
\begin{align}
S_{\mathrm{CS}}'&\mapsto S_{\mathrm{CS}}'+\frac{1}{2}\int
\mathrm{d}^{3}x\,\Big[\partial_{\mu}\left(\omega\hat{\mathcal{Q}}^{\mu\nu\rho}\partial_{\nu}A_{\rho}\right) \notag \\
&\phantom{{}={}}\hspace{2.6cm}-\omega\hat{\mathcal{Q}}^{\mu\nu\rho}\partial_{\mu}\partial_{\nu}A_{\rho}\Big]\,.
\label{ActionCSHD}
\end{align}
Now we assume that $\hat{\mathcal{Q}}$ is antisymmetric also in the first two indices, which implies $\hat{\mathcal{Q}}^{\mu\nu\rho}\partial_{\mu}\partial_{\nu}A_{\rho}=0$. Then,
\begin{equation}
S_{\mathrm{CS}}'\mapsto S_{\mathrm{CS}}'+\frac{1}{2}\displaystyle\oint \mathrm{d}S_{\mu}\,\left(\omega\hat{\mathcal{Q}}^{\mu\nu\rho}\partial_{\nu}A_{\rho
}\right)\,,
\end{equation}
where the second term on the right-hand side of Eq.~\eqref{ActionCSHD} has been recast into a surface integral. This finding assures gauge invariance of the action of Eq.~\eqref{APCS} in analogy to how it is demonstrated in CS theory. Note that the antisymmetry in the first two and last two indices of $\hat{\mathcal{Q}}^{\mu\nu\rho}$, when considered simultaneously, guarantees that the operator is completely antisymmetric with respect to the interchange of any neighbouring indices. This can be inferred from the properties of the three-dimensional representation of the permutation group. So the symmetry of $\hat{\mathcal{Q}}^{\mu\nu\rho}$ is directly inherited from that of the $(2+1)$-dimensional Levi-Civita symbol, as it is the essential property that guarantees gauge invariance of the CS action of Eq.~\eqref{eq:CST-1} --- at least for gauge transformations that do not wind around the gauge group.

Using the parameterization of Eq.~\eqref{QkappaT1}, the higher-derivative Lagrange density of Eq.~\eqref{APCS} reads
\begin{equation}
	\mathcal{L}_{\mathrm{CS}}'=\frac{1}{2}A_{\mu}\hat{\mathcal{Q}}^{\mu\nu\rho
	}\partial_{\nu}A_{\rho}=\frac{1}{2}A_{\mu}\left(  \hat{\mathcal{K}%
	}^{\mu\nu\rho}+\hat{\mathcal{T}}^{\mu\nu\rho}\right)  \partial_{\nu
	}A_{\rho}\,.
\label{LagrangianCSQ}
\end{equation}
The equation of motion for this model is obtained through the general Euler-Lagrange equation for \textit{U}(1) gauge theories with higher-order derivatives described by the Lagrange density $\mathcal{L}$:
\begin{align}
0&=\frac{\partial\mathcal{L}}{\partial A_{\beta}}-\partial_{\sigma}\left(
\frac{\partial\mathcal{L}}{\partial\left(  \partial_{\sigma}A_{\beta}\right)
}\right)  +\partial_{\gamma}\partial_{\sigma}\left(  \frac{\partial
\mathcal{L}}{\partial\left(  \partial_{\gamma}\partial_{\sigma}A_{\beta
}\right)  }\right)  -\dots \notag \\
&\phantom{{}={}}+\left(  -1\right)  ^{n}\partial_{\mu_{1}}%
\dots\partial_{\mu_{n}}\left(  \frac{\partial\mathcal{L}}{\partial\left(
\partial_{\mu_{1}}\dots\partial_{\mu_{n}}A_{\beta}\right)  }\right)\,,
\label{E-L}%
\end{align}
which, applied to Eq.~\eqref{LagrangianCSQ}, provides
\begin{subequations}
\begin{equation}
\label{CSQME1}
0=\frac{1}{2}\left(  \hat{\mathcal{T}}^{\beta\nu\rho}+\hat
{\mathcal{K}}^{\beta\nu\rho}\right)  \partial_{\nu}A_{\rho}+\frac{1}{2}\left(\hat{\mathcal{T}}^{\rho\nu\beta}-\hat{\mathcal{K}}^{\rho\nu\beta}\right)\partial_{\nu}A_{\rho}\,,
\end{equation}
or equivalently,
\begin{equation}
\label{EM1B}
0=\frac{1}{2}\left(  \hat{\mathcal{T}}^{\beta\nu\rho}+\hat
{\mathcal{T}}^{\rho\nu\beta}\right)  \partial_{\nu}A_{\rho}+\frac{1}%
{2}\left(  \hat{\mathcal{K}}^{\beta\nu\rho}-\hat{\mathcal{K}}^{\rho
\nu\beta}\right)  \partial_{\nu}A_{\rho}\,.
\end{equation}
\end{subequations}
The reversed sign of the fourth term in Eq.~\eqref{CSQME1}, in comparison to the second, originates from $\hat{\mathcal{K}}^{\beta\nu\rho}$ and $\hat{\mathcal{T}}^{\beta\nu\rho}$ being associated with an even and odd number of derivatives, respectively. Since the operators $\hat{\mathcal{K}}^{\beta\nu\rho}$ and $\hat
{\mathcal{T}}^{\beta\nu\rho}$ are completely  antisymmetric, the outcome of Eq.~\eqref{EM1B} simplifies as
\begin{equation}
\hat{\mathcal{K}}^{\beta\nu\rho}F_{\nu\rho}=0\,,
\label{kappaME1}
\end{equation}
which is obviously gauge-invariant, as expected.
We also notice that the operator $\hat{\mathcal{T}}^{\beta\nu\rho}$ does not
contribute to the equations of motion, suggesting that it can be rewritten as a total derivative in the action. As a consequence, the higher-derivative operator in Eq.~\eqref{Q} can now be recast into
\begin{equation}
	\hat{\mathcal{Q}}^{\mu\nu\rho}=\hat{\mathcal{K}}^{\mu\nu\rho}=\sum_{\substack{d\geq
			2\\\text{\textrm{even}}}}^{\infty}\mathcal{K}^{(d)\mu\nu\rho\alpha_{1}\dots\alpha_{(d-2)}}\partial_{\alpha_{1}%
	}\dots\partial_{\alpha_{(d-2)}}\,. \label{K2}%
\end{equation}
Considering that $\hat{\mathcal{K}}^{\mu\nu\rho}$ is totally antisymmetric, its simplest form corresponds to the term with the label $d=2$ in the sum on the right-hand side, that is, the product of the $(2+1)$-dimensional Levi-Civita symbol and a scalar operator $\hat{\mathcal{K}}$:
\begin{subequations}
\begin{equation}
\hat{\mathcal{K}}^{\beta\nu\rho}=\varepsilon^{\beta\nu\rho
}\hat{\mathcal{K}}\,,
\label{kappa5}
\end{equation}
such that
\begin{equation}
\hat{\mathcal{K}}=\frac{1}{3!}\varepsilon_{\alpha\beta\gamma}%
\hat{\mathcal{K}}^{\alpha\beta\gamma}\,,
\label{eq:K-equation}%
\end{equation}
\end{subequations}
contains an even number of additional derivatives as does $\hat{\mathcal{K}}^{\beta\nu\rho}$.
Inserting Eq.~\eqref{kappa5}, the field equations stated in Eq.~\eqref{kappaME1} are rewritten as
\begin{equation}
\varepsilon^{\beta\nu\rho}\hat{\mathcal{K}}F_{\nu\rho}=0\,,
\end{equation}
which have a form analogous to the field equations of CS theory, as expected.
For $d=4$, the operator $\hat{\mathcal{K}}^{\mu\nu\varrho}$ involves two additional derivatives and represents a DJ-like term to be examined in the forthcoming sections. It is worthwhile to mention that the operator of Eq.~\eqref{K2} is more general than that of Eq.~\eqref{kappa5}, as it exhibits an additional index structure starting from $d=4$.

Finally, we point out that the operator $\hat{\mathcal{K}}$ of Eq.~\eqref{kappa5} provides a contribution to the Lagrange density of the type $A_{\mu}\hat{\mathcal{K}}^{\mu\nu\rho}\partial_{\nu}A_{\rho}=A_{\mu}\varepsilon^{\mu\nu\rho
}\hat{\mathcal{K}}\partial_{\nu}A_{\rho}$. The latter exhibits the same structure as the term $A_{\rho}\varepsilon^{\rho\mu\nu}(\hat{k}_{AF})F_{\mu\nu}$ with the operator
\begin{equation}
\hat{k}_{AF}=\sum_{\substack{d\geq 3 \\ \text{odd}}}(k_{AF}^{(d)})^{\alpha_{1}\dots\alpha_{(d-3)}}\partial_{\alpha_{1}}\dots \partial_{\alpha_{(d-3)}}\,,
\label{KAF}
\end{equation}
which is part of the planar modified electrodynamics obtained in Ref.~\cite{Joca} via dimensional reduction from the electromagnetic sector of the nonminimal SME in $(3+1)$ spacetime dimensions. Note that $d$ in Eq.~\eqref{KAF} is, indeed, the mass dimension inherited from the field operators of the SME. So our present construction is contained in the generic $(2+1)$-dimensional framework of Ref.~\cite{Joca}, as expected.

\subsection{Higher-derivative Deser-Jackiw-like term}

The higher-derivative operator of lowest order identified within Eq.~\eqref{LagrangianCSQ} is dominant at low energies. Therefore, with our attention restricted to the latter, we propose the following gauge-invariant higher-derivative (hd) CS-like structure, which exhibits $\mathit{SO}(2,1)$ violation:
\begin{subequations}
\begin{align}
S_{\mathrm{hdCS}}&=\frac{\vartheta_1}{2}\int A\wedge \mathrm{d}(\hat{\mathcal{K}}A)=\int\mathrm{d}^3x\,\mathcal{L}_{\mathrm{hdCS}}\,, \\[2ex]
\mathcal{L}_{\mathrm{hdCS}}&=\frac{\vartheta_{1}}{2}\varepsilon_{\mu\nu\rho}A^{\mu}\partial^{\nu}\hat{\mathcal{K}}A^{\rho}\,,
\end{align}
with
\begin{equation}
	\hat{\mathcal{K}}=\sum_{\substack{d\geq 2 \\ \text{even}}} \hat{\mathcal{K}}^{(d)}\,,\quad  \hat{\mathcal{K}}^{(d)}=K^{(d)\alpha_{1}\dots\alpha_{(d-2)}}\partial_{\alpha_{1}}\dots\partial_{\alpha_{(d-2)}}\,.
\label{KHO1}%
\end{equation}
\end{subequations}
The latter is expressed in terms of tensor-valued controlling coefficients $K^{(d)\alpha_{1}\dots\alpha_{(d-2)}}$ and a dimensionful parameter $\vartheta_1$ whose mass dimension is chosen properly after restricting the action to a specific subset of operators. Being interested in the lowest-order operator and recalling that $\hat{\mathcal{K}}$ has an even number of additional derivatives suitably contracted with the indices of the controlling coefficients, the simplest operator contained in Eq.~\eqref{KHO1} carries the label $d=4$ and involves two derivatives:
\begin{equation}
	\hat{\mathcal{K}}^{(4)}=K^{(4)\alpha\beta}\partial_{\alpha}\partial_{\beta}\equiv K^{\alpha\beta}\partial_{\alpha}\partial_{\beta}\,,
\label{kappa2}
\end{equation}
where, for simplicity, we will drop the mass dimension label from the controlling coefficients, which are symmetric by construction. Thus, the lowest-order extension of the conventional CS term includes the operator in Eq.~(\ref{kappa2}) as follows:
\begin{equation} \mathcal{L}_{\mathrm{hdCS}}^{(4)}=\frac{\vartheta_{1}}{2}A_{\mu}\varepsilon^{\mu\nu\rho}K^{\alpha\beta}\partial_{\alpha}\partial_{\beta}\partial_{\nu}A_{\rho}\,,
	\label{LCSN}
\end{equation}
such that the field operator is of mass dimension 4. This Lagrange density represents a Lorentz-violating and gauge-invariant CS-like higher-derivative term. For consistency, $[\vartheta_{1}K^{\alpha\beta}]=-1$, which can be fulfilled by supposing that $[\vartheta_{1}]=-1$, while $[K^{\alpha\beta}]=0$.
With such a choice, the operator of Eq.~\eqref{kappa2} has mass dimension 2. It is worthwhile to identify the latter within the generic higher-derivative operator of Eq.~\eqref{Q}. By doing so,
\begin{equation} \hat{\mathcal{K}}^{(4)\mu\nu\rho}=\varepsilon^{\mu\nu\rho}K^{\alpha\beta}\partial_{\alpha}\partial_{\beta}=\varepsilon^{\mu\nu\rho}\hat{\mathcal{K}}^{(4) }\,.
\label{Khat5}
\end{equation}
Eventually, it is easy to note that the Lagrange density of Eq.~\eqref{LCSN} provides the usual DJ contribution, if the background is chosen to be equal to the metric tensor: $K^{\alpha\beta}=\eta^{\alpha\beta}$. In fact,
\begin{equation}
	\mathcal{L}_{\mathrm{hdCS}}^{(4)}|_{K=\eta}=\frac
	{\vartheta_{1}}{2}A_{\mu}\varepsilon^{\mu\nu\rho}\square\partial_{\nu}A_{\rho}\,.
	\label{LCSN2}
\end{equation}
In other words, this equivalence states that the Lagrange density of Eq.~\eqref{LCSN} is a Lorentz-violating extension of the DJ term, which may even be anisotropic.
This fact also motivates the convention $[\vartheta_1]=-1$, which corresponds to $[\vartheta]=-1$ for DJ theory in Eq.~\eqref{DJT1}.

For now, we will bring these more technical deliberations to a close and focus on phenomenological aspects of a modified electrodynamics in $(2+1)$ spacetime dimensions. Our ultimate goal will be to find suitable models having potential applicability in condensed-matter systems of two spatial dimensions.

\section{Modified $\boldsymbol{(2+1)}$-dimensional electrodynamics with Lorentz invariance}
\label{sec:modified-planar-electrodynamics-with-LV}

In this section we intend to review some aspects of a planar electrodynamics that involves the CS term and higher-derivative CS-type contributions, including the DJ term, where Lorentz invariance is to be maintained. In particular, we are going to state the corresponding gauge field propagators, whose poles provide the dispersion relations, which will allow for further investigations. Another goal is to examine static solutions of the magnetic field for a pointlike charge in pure CS theory endowed with an additional DJ term. Note that stationary classical solutions in the context of a higher-derivative extension of planar MCS theory have been in the spotlight, recently~\cite{Borges1}.

\subsection{Maxwell-Chern-Simons-Deser-Jackiw electrodynamics}
\label{sec:MCSDJ-electrodynamics}

The starting point is the MCS Lagrange density endowed with the DJ term, which we call Maxwell-Chern-Simons-Deser-Jackiw (MCSDJ) electrodynamics \cite{CS2}:
\begin{subequations}
\begin{align}
S_{\mathrm{MCSDJ}}&=\int \bigg(-\frac{1}{4}F\wedge (*F)+\frac{k}{2}A\wedge\mathrm{d}A+\frac{\vartheta}{2}A\wedge \mathrm{d}(\square A) \notag \\
&\phantom{{}={}}\hspace{0.7cm}+\frac{1}{2\xi}*\!(\mathrm{i}_{\partial}A)^2\bigg) \notag \\
&=\int\mathrm{d}^3x\,(\mathcal{L}_{\mathrm{MCSDJ}}+\mathcal{L}_{\mathrm{gf}})\,, \\[2ex]
\label{MCSDJ1}
\mathcal{L}_{\mathrm{MCSDJ}}&=-\frac{1}{4}F_{\mu\nu}F^{\mu\nu}+\frac{k}{2}\varepsilon_{\mu\nu\rho}A^{\mu}\partial^{\nu}A{}^{\rho} \notag \\
&\phantom{{}={}}+\frac{\vartheta}{2}\varepsilon_{\mu\nu\rho}A^{\mu}\partial^{\nu}\square A{}^{\rho}\,, \\[2ex]
\label{eq:gauge-fixing}
\mathcal{L}_{\mathrm{gf}}&=\frac{1}{2\xi}\left(\partial_{\mu}A^{\mu}\right)^{2}\,,
\end{align}
\end{subequations}
where $\mathrm{i}_V$ stands for the interior product with a vector field $V$. The last term, $\mathcal{L}_{\mathrm{gf}}$, is added to fix the gauge freedom, where $\xi$ is a real gauge fixing parameter taken as arbitrary. The Lagrange density of Eq.~\eqref{MCSDJ1} can be reformulated as
\begin{subequations}
	\begin{equation}
		\tilde{\mathcal{L}}_{\mathrm{MCSDJ}}=\frac{1}{2}A^{\mu}\Lambda_{\mu\nu}A^{\nu}\,,
	\end{equation}
	with the following tensor operator sandwiched between gauge fields,
	\begin{equation}
    \label{eq:sandwiched-operator-Lambda}
        \Lambda_{\mu\nu}=\square\Theta_{\mu\nu}+(k+\vartheta\square)L_{\mu\nu}-\frac{1}{\xi}\square\Omega_{\mu\nu}\,,
	\end{equation}
	containing the CS operator $L_{\mu\nu}$ as well as longitudinal and transverse projectors $\Theta_{\mu\nu}$ and $\Omega_{\mu\nu}$, respectively. In particular,
	\begin{align}
    \label{eq:CS-operator}
	L_{\mu\nu}&=\varepsilon_{\mu\rho\nu}\partial^{\rho}=-L_{\nu\mu}\,, \\[2ex]
    \Omega_{\mu\nu}&=\frac{\partial_{\mu}\partial_{\nu}}{\square}\,,\quad \Theta_{\mu\nu}=\eta_{\mu\nu}-\Omega_{\mu\nu}\,.
	\label{projectors}
	\end{align}
\end{subequations}
	These form the closed algebra of Tab.~\ref{tab:algebra-tensor-projectors}.
\begin{table}[t]
	\centering
	\begin{tabular}[c]{cccc}
    \toprule
		& $\Theta_{\phantom{\mu}\alpha}^{\mu}$ & $L_{\phantom{\mu}\alpha}^{\mu}$ & $\Omega_{\phantom{\mu}\alpha}^{\mu}$ \\
    \colrule
		$\Theta_{\nu\mu}$ & $\Theta_{\nu\alpha}$ & $L_{\nu\alpha}$ & 0 \\
		$L_{\nu\mu}$ & $L_{\nu\alpha}$ & $-\square\Theta_{\nu\alpha}$ & 0 \\
		$\Omega_{\nu\mu}$ & 0 & 0 & $\Omega_{\nu\alpha}$ \\
    \botrule
	\end{tabular}
	\caption{Algebra of tensor operators.}%
    \label{tab:algebra-tensor-projectors}
\end{table}%%
To obtain the propagator of Eq.~\eqref{MCSDJ1}, we need to invert $\Lambda_{\mu\nu}$ of Eq.~\eqref{eq:sandwiched-operator-Lambda}. The following \textit{ansatz} for the inverse, which is expressed in terms of the operators in Tab.~\ref{tab:algebra-tensor-projectors}, is valuable:%%
\begin{subequations}
	\begin{equation}
		\Delta_{\phantom{\mu}\alpha}^{\mu}=a\Theta_{\phantom{\mu}\alpha}^{\mu}+bL_{\phantom{\mu}\alpha}^{\mu}+c\Omega_{\phantom{\mu}\alpha}^{\mu}\,,
	\end{equation}
with parameters $a,b,c$ to be determined from the identity
	\begin{equation}
		\Lambda_{\mu\alpha}\Delta_{\phantom{\alpha}\nu}^{\alpha}=\eta_{\mu\nu}\,.
\end{equation}
\end{subequations}%%
Solving the resulting system of equations for the parameters results in the propagator of MCSDJ theory:
	\begin{equation}
		\Delta_{\mu\alpha}=\frac{\square\Theta_{\mu\alpha}-(k+\vartheta\square)
		L_{\mu\alpha}-\xi[\square+(k+\vartheta\square)^{2}]\Omega_{\mu\alpha}}{\square [\square+(k+\vartheta\square)^{2}]}\,,
	\end{equation}
which in momentum space with $\partial_{\mu}=-\mathrm{i}p_{\mu}$ reads
\begin{align}
	\Delta_{\mu\alpha}(p)&=-\frac{1}{p^{2}[p^{2}-(k-\vartheta p^{2})^2]} \notag \\
&\phantom{{}={}}\times \bigg\{p^2 \Theta_{\mu\alpha}(p)-\mathrm{i}(k-\vartheta p^{2})\varepsilon_{\mu\sigma\alpha}p^{\sigma} \notag \\
&\phantom{{}={}}\hspace{0.6cm}-\xi[p^{2}-(k-\vartheta p^{2})^2]\frac{p_{\mu}p_{\alpha}}{p^{2}}\bigg\}\,.
\label{PMCSDJ}
\end{align}%%
Note that this propagator is not symmetric in its indices, which is a property attributed to CS and CS-like theories such as those considered in Ref.~\cite{Leticia1}. As we commented in the latter paper, such propagators are symmetric under the combined operations of switching the indices and reversing the direction of the four-momentum, i.e., $p_{\mu}\mapsto -p_{\mu}$. Thus, $\Delta_{\mu\alpha}(p)=\Delta_{\alpha\mu}(-p)$.

The dispersion equations can be read off the poles of the propagator. Poles occurring in terms proportional to at least one uncontracted four-momentum are related to gauge degrees of freedom and are discarded in the study of physical-signal propagation. By taking this into account, we obtain the dispersion equations
\begin{equation}
	p^{2}=0\,, \quad p^{2}-(k-\vartheta p^{2})^2 =0\,,
\end{equation}%%
associated with two modes of the theory. The first dispersion relation represents a massless mode,
\begin{equation}
	p_{0}^{2}=\mathbf{p}^{2}\,,
	\label{masslessmode1}
\end{equation}
and it occurs twice, in fact, which is a point that we will come back to later. The second, rewritten as
\begin{equation}
{\vartheta}^2 p^{4}-(1+2 k\vartheta) p^{2}+k^2=0\,,
\end{equation}
represents two massive modes:
\begin{subequations}
\label{massmode1}
\begin{equation}
	p_{0}^{2}=\mathbf{p}^{2}+m_{\pm}^2(\vartheta,k)\,,
\end{equation}
with the squared masses
\begin{equation}
m_{\pm}^2(\vartheta,k)=\frac{1}{2{\vartheta}^2}\left(1+2k\vartheta \pm\sqrt{1+4k\vartheta}\right)\,.
\label{MassMCSDJ}
\end{equation}
\end{subequations}
These depend on the two parameters $k,\vartheta$. As long as $k\vartheta\geq -1/4$, which is what we will assume to be the case, $m_{\pm}^2$ are not only real, but also manifestly positive definite. The masses themselves can be expressed as
\begin{equation}
\label{eq:masses-MCSDJ}
m_{\pm}=\frac{1}{2|\vartheta|}\left|1\pm \sqrt{1+4k\vartheta}\right|\,.
\end{equation}
Expansions in the parameters $k$ and $\vartheta$ are worthwhile to consider. On the one hand, for $|k|\ll 1/|\vartheta|$ we have
\begin{subequations}
\begin{align}
m_+&=\frac{1}{|\vartheta|}+\mathrm{sgn}(\vartheta)k+\dots\,, \\[1ex]
m_-&=(1-\vartheta k)|k|+\dots\,,
\end{align}
\end{subequations}
with the sign function $\mathrm{sgn}(x)$, which shows that both modes are perturbative in $k$. On the other hand, for $\vartheta\ll 1/|k|$ with $\vartheta>0$ we obtain
\begin{subequations}
\begin{align}
m_+&=\frac{1}{\vartheta}+k(1-k\vartheta)+\dots\,, \\[1ex]
m_-&=|k|(1-k\vartheta)+\dots\,,
\end{align}
\end{subequations}
whereupon the first massive mode is nonperturbative in $\vartheta$. Hence, the latter strongly deviates from the CS modes when $\vartheta\mapsto 0$.

Now, an interesting discovery can be made based on Eq.~\eqref{eq:masses-MCSDJ}. Let MCS theory be coupled to a scalar Higgs field in $(2+1)$ spacetime dimensions, which gives rise to Maxwell-Chern-Simons-Higgs (MCSH) theory given by
\begin{align}
\label{sec:MCSH-theory}
\mathcal{L}_{\mathrm{MCSH}}&=-\frac{1}{4}F_{\mu\nu}F^{\mu\nu}+\frac{m_{\mathrm{CS}}}{2}\varepsilon^{\mu\nu\varrho}A_{\mu}\partial_{\nu}A_{\varrho} \notag \\
&\phantom{{}={}}+(D_{\mu}\phi)^{*}D^{\mu}\phi-V(|\phi|)\,,
\end{align}
with the CS mass $m_{\mathrm{CS}}$, the complex Higgs field $\phi$, the gauge-covariant derivative $D_{\mu}$, and the Higgs potential~$V$. Then, the gauge field becomes massive by spontaneous symmetry breaking. Two massive modes emerge~\cite{CS1} whose masses are given by
\begin{equation}
M_{\pm}=\frac{m_{\mathrm{CS}}}{2}\left(\sqrt{1+\frac{4m_H^2}{m_{\mathrm{CS}}^2}}\pm 1\right)\,,
\end{equation}
with the Higgs mass $m_H^2=2v^2$, where $v$ is the vacuum expectation value of the Higgs field. The similarities between the latter and Eq.~\eqref{eq:masses-MCSDJ} are evident. An identification between these and our parameters of the MCSDJ model provides $k=m_H^2/m_{\mathrm{CS}}$ and $\vartheta=1/m_{\mathrm{CS}}$. Therefore, there is an intriguing relationship between MCSH and MCSDJ theories. In fact, the possibility exists that both settings are dual to each other, which could be an interesting project to be studied elsewhere. Having pointed this out, we will dedicate ourselves to further relevant questions.

\subsubsection{Sub- and superluminal signal propagation}

The setting of a modified electrodynamics in $(2+1)$ spacetime dimensions is envisioned to find applications in two-dimensional condensed-matter systems. While, in principle, Lorentz invariance is explicitly broken in such systems due to the presence of the lattice background, there are materials such as graphene that exhibit what is called an emergent Lorentz invariance. For example, near the $\Gamma$ point, i.e., for vanishing momentum, the dispersion relation of electrons may have a cone-like structure, which makes these particles behave like massless Weyl or Dirac fermions. Thus, they can be described by a relativistic field theory and the corresponding fields transform under suitable representations of the Lorentz group. The defining velocity of the latter is not the speed of light $c$ \textit{in vacuo}, but the Fermi velocity $v_F\ll c$. Similarly, electromagnetic waves propagate through materials with the speed of light in the medium, $c_m<c$.

In general, Lorentz-violating electromagnetism can exhibit regions in parameter space with a propagation velocity $\lessgtr c_m$, which we will be referring to as subluminal and superluminal regimes, respectively. As the emergent Lorentz symmetry in condensed-matter systems is not fundamental, superluminal propagation is not expected to imply issues with classical causality. Nevertheless, it is worthwhile to identify and distinguish between sub- and superluminal regimes, as these are characterized by vastly different physical behaviors.

We will call a theory subluminal, when both the group and front velocity do not exceed the speed of light, and superluminal, if the opposite holds true. In this context we will be referring to the medium speed of light $c_m=1$ in natural units. Based on the dispersion relation $p_0=p_0(\mathbf{p})$ of a mode, these characteristic velocities are defined by \cite{Brillouin:1960}
\begin{equation}
	\mathbf{u}_{\mathrm{gr}}\equiv\frac{\partial p_{0}}{\partial\mathbf{p}}\,,\quad u_{\mathrm{fr}}\equiv\lim_{|\mathbf{p}|\mapsto\infty}\frac{p_{0}}{|\mathbf{p}|}\,,
	\label{uguf1}
\end{equation}
respectively. On the one hand, the massless dispersion relation of Eq.~\eqref{masslessmode1} obviously yields $u_{\mathrm{gr}}=1=u_{\mathrm{fr}}$, as expected. On the other hand, the group velocities of the massive modes read
\begin{subequations}
	\begin{equation}
		\mathbf{u}_{\mathrm{gr}}^{(\pm)}=\frac{\mathbf{p}}{\sqrt{\mathbf{p}^{2}+m_{\pm}^2(\vartheta,k)}}\,,
		\label{groupvelocityCSDJM1}
	\end{equation}
with magnitudes
\begin{equation}
u_{\mathrm{gr}}^{(\pm)}=\frac{|\mathbf{p}|}{\sqrt{\mathbf{p}^{2}+m_{\pm}^2(\vartheta,k)}}\,,
\end{equation}
\end{subequations}
which are typical of massive modes. Since the masses squared are always positive --- at least in the regime $k\vartheta<-1/4$, where the square root in Eq.~\eqref{MassMCSDJ} is real --- the norms of $u_{\mathrm{gr}}$ are always $<1$. As for the front velocity, we obtain $u_{\mathrm{fr}}=1$, independently of any parameters. Thus, we conclude that MCSDJ theory is subluminal.

\subsubsection{Unitarity of the quantum theory at tree-level}
\label{eq:unitarity-MCSDJ-theory}

Another point of interest is the presence of states of negative norm in the Hilbert space of the quantized theory, which usually appear in higher-derivative theories and spoil unitarity. For a tree-level examination, we can consider a scalar quantity composed of the propagator in Eq.~\eqref{PMCSDJ}, contracted with suitable external currents, which is often called the saturated gauge propagator in the literature:
\begin{equation}
	\mathit{SP}\equiv J^{\mu}\mathrm{i}\Delta_{\mu\nu}J^{*,\nu}\,. \label{Sat2}%
\end{equation}
Here, $J^{\mu}$ is an external conserved current, which is generically taken as complex, i.e., $J^{*,\mu}$ is the complex conjugate of the latter. This current satisfies the continuity equation $\partial_{\mu}J^{\mu}=0$ or $p_{\mu}J^{\mu}=0$ in momentum space. Now, Eq.~\eqref{Sat2} is one possibility of stating a generic forward-scattering amplitude at tree-level without resorting to a particular scattering process. The outgoing current is then directly related to the incoming one by a time reversal transformation, i.e., a complex conjugation in the context of quantum theory. Contracting the propagator with conserved currents gets rid of contributions that were proportional to uncontracted momenta initially. Thus, all gauge-dependent pieces are eliminated by doing so.

Now, unitarity is assured whenever the imaginary part of the residue of the saturation \textit{SP}, evaluated at the poles of the propagator, is nonnegative \cite{Veltman}. For the propagator of Eq.~(\ref{PMCSDJ}), the saturation reads
\begin{equation}
	\mathit{SP}=-\mathrm{i}\frac{p^2|J|^2+(k-\vartheta p^2)\Omega}{p^2[p^{2}-(k-\vartheta p^{2})^2]}\,,
	\label{eq:saturated-propagator-MCSDJA}
\end{equation}
with the quantities
\begin{subequations}
\begin{align}
\label{eq:norm-squared-current}
|J|^2&=J_{\mu}J^{*,\mu}=\eta_{\mu\nu}J^{\mu}J^{*\nu}\,, \\[2ex]
\label{eq:scalar-triple-product}
\Omega&=\varepsilon_{\mu\sigma\nu}p^{\sigma}\mathrm{Im}(J^{\mu}J^{*,\nu})\,,
\end{align}
\end{subequations}
which will occur in all analyses of this kind. Note that $|J|^2$ can be interpreted as the Lorentzian norm squared of a complex three-current and clearly $|J|^2 \neq |J^2|$. Furthermore, $\Omega$ is a kind of Lorentzian scalar triple product of the three-vectors $\{J^{\mu},J^{\mu,*},p^{\mu}\}$. Both $|J|^2$ and $\Omega$ are manifestly real quantities. For a dispersion relation that satisfies $p^0\geq |\mathbf{p}|$, current conservation implies that
\begin{equation}
|J|^2=\frac{|\mathbf{J}\cdot\mathbf{p}|^2}{p_0^2}-|\mathbf{J}|^2\leq \left(\frac{\mathbf{p}^2}{p_0^2}-1\right)|\mathbf{J}|^2\leq 0\,,
\end{equation}
i.e., $|J|^2<0$ when $\mathbf{J}$ and $\mathbf{p}$ are not (anti)parallel to each other. In contrast, the sign of $\Omega$ is unclear and depends on the orientation of the vectors with respect to each other.

Now, $\Omega$ can be evaluated explicitly as follows. First of all, let us define the following set of vectors living in a three-dimensional Euclidean vector space:
\begin{equation}
\mathbf{P}=\begin{pmatrix}
p^0 \\
p^1 \\
p^2 \\
\end{pmatrix}\,,\quad \mathbf{V}=\mathrm{Re}\begin{pmatrix}
J^0 \\
J^1 \\
J^2 \\
\end{pmatrix}\,,\quad \mathbf{W}=\mathrm{Im}\begin{pmatrix}
J^0 \\
J^1 \\
J^2 \\
\end{pmatrix}\,.
\end{equation}
Then, it is possible to recast Eq.~\eqref{eq:scalar-triple-product} into
\begin{equation}
\label{eq:Omega-rewritten}
\Omega=-2\mathbf{P}\cdot (\mathbf{V}\times\mathbf{W})\,.
\end{equation}
Note that the components of $\mathbf{P}$, $\mathbf{V}$, and $\mathbf{W}$ correspond to the components of the contravariant four-vectors $p^{\mu}$, $\mathrm{Re}(J^{\mu})$, and $\mathrm{Im}(J^{\mu})$, respectively. However, $\{\mathbf{P},\mathbf{V},\mathbf{W}\}$ is now treated as a set of Euclidean vectors, i.e., `$\cdot$' stands for the Euclidean scalar product and `$\times$' for the Euclidean vector product, respectively, in three dimensions. From this alternative of writing up $\Omega$ it is obvious that the latter does not involve the Minkowski metric, which corroborates the topological nature of $\Omega$. Clearly, $\Omega=0$ when the current is assumed to be real. Moreover, when $\mathbf{V}$ and $\mathbf{W}$ point along the same Euclidean direction or when $\mathbf{P}$ lies in the plane spanned by linearly independent $\mathbf{V}$ and $\mathbf{W}$, then $\Omega=0$, too. However, whether $\Omega>0$ or $\Omega<0$ cannot be deduced in a straightforward manner from a generic form of the external current, which will render the analysis of tree-level unitarity nontransparent compared to similar studies \cite{Leticia2,Leticia1} carried out in the past.

The result of Eq.~\eqref{eq:saturated-propagator-MCSDJA} can be suitably rewritten as
\begin{equation}
	\mathit{SP}=\mathrm{i}\frac{1}{\vartheta^2(p^{2}-m_{+}^2)(p^{2}-m_{-}^2)}\left(|J|^2+\frac{k-\vartheta p^2}{p^2}\Omega\right)\,,
\label{eq:saturated-propagator-MCSDJ}
\end{equation}
with $m_{\pm}^2$ given in Eq.~\eqref{MassMCSDJ}.
Observe that the pole $p^2=0$ cancels in the first contribution of Eq.~\eqref{eq:saturated-propagator-MCSDJA}, but it is still present in the second. Our interpretation of this observation is as follows. In the propagator of MCSDJ theory, Eq.~\eqref{PMCSDJ}, the massless dispersion equation $p^2$ occurs to the first power in the denominator of the CS contribution, but to the second power in the denominator of the gauge-dependent part. Thus, this theory is equipped with two massless modes: one is unphysical and the second is physical, but related to phenomena in the infrared. The first is eliminated in the saturated propagator, whereas the second remains in the term proportional to $\Omega$.

Now, the residues of Eq.~\eqref{eq:saturated-propagator-MCSDJ} at the poles $\{0,m_{\pm}^2\}$ are
\begin{subequations}
\begin{align}
	\mathrm{Res}(\mathit{SP})|_{p^{2}=0}&=\mathrm{i}\frac{k}{\vartheta^2m_+^2m_-^2}\Omega|_{p^2=0}\,, \\[1ex]
	\mathrm{Res}(\mathit{SP})|_{p^{2}=m_{+}^2}&=\frac{\mathrm{i}}{\vartheta^2 (m_{+}^2-m_{-}^2)} \notag \\
&\phantom{{}={}}\times \left(|J|^{2}-\mathrm{sgn}(\vartheta)\frac{\Omega}{m_+}\right)_{p^{2}=m_{+}^2}\,, \\[1ex]
	\mathrm{Res}(\mathit{SP})|_{p^{2}=m_{-}^2}&=\frac{\mathrm{i}}{\vartheta^2 (m_{-}^2-m_{+}^2)}\bigg[|J|^{2} \notag \\
&\phantom{{}={}}+\mathrm{sgn}\left(\frac{\vartheta}{\sqrt{1+4k\vartheta}-1}\right)\frac{\Omega}{m_-}\bigg]_{p^{2}=m_{-}^2}\,.
\end{align}
\end{subequations}
As mentioned above, for a real current, the CS term does not affect unitarity, at all, which renders the behavior of the above residues rather straightforward. Specifically,
\begin{equation}
\mathrm{Im}\bigg[\mathrm{Res}(\mathit{SP})\bigg|_{\substack{\Omega=0 \\ p^{2}=m_{\pm}^2}}\bigg]\lessgtr 0\,,
\end{equation}
in accordance with $m_{\pm}^2-m_{\mp}^2=\pm\sqrt{1+4k\vartheta}/\vartheta^2$.
For a complex current, though, it is the CS term that makes the behavior more involved. Depending on the sign and size of $\Omega$, the imaginary part of each of the three residues can be positive, zero or negative. The latter outcome suggests unitarity violation for the corresponding excitations. However, note that excitations with such properties may be split from the physical modes of the theory by using a field redefinition as usually done in the context of the Lee-Wick theories \cite{Lee-Wick-1}. For a detailed investigation of such and related questions in DJ electrodynamics at tree- and one-loop level, the reader is suggested to consult Ref.~\cite{Avila}.

Let us also emphasize that $|J|^{2}$ of Eq.~\eqref{eq:norm-squared-current} is a geometric quantity, as it involves the metric. Unlike the latter, $\Omega$ of Eq.~\eqref{eq:scalar-triple-product} does not depend on the metric and can be interpreted as a topological quantity, in particular, a generalized ``volume form'' for a set of three partially complex three-vectors. So the CS term governs unitarity via a kind of topological property, which does not come as a surprise.

\subsection{Maxwell-Deser-Jackiw electrodynamics}

In the absence of the CS term, the action of Eq.~\eqref{MCSDJ1} recovers the Maxwell-Deser-Jackiw (MDJ) model,
\begin{subequations}
\begin{align}
S_{\mathrm{MDJ}}&=\int \left(-\frac{1}{4}F\wedge (*F)+\frac{\vartheta}{2}A\wedge \mathrm{d}(\square A)+\frac{1}{2\xi}*\!(\mathrm{i}_{\partial}A)^2\right) \notag \\
&=\int\mathrm{d}^3x\,(\mathcal{L}_{\mathrm{MDJ}}+\mathcal{L}_{\mathrm{gf}})\,, \displaybreak[0]\\[1ex]
\mathcal{L}_{\mathrm{MDJ}}&=-\frac{1}{4}F_{\mu\nu}F^{\mu\nu}+\frac{\vartheta}{2}\varepsilon_{\mu\nu\rho}A^{\mu}\partial^{\nu}\square A^{\rho}\,,
\label{MDJ1}
\end{align}
\end{subequations}
with the gauge fixing term of Eq.~\eqref{eq:gauge-fixing}. This theory has the following propagator in momentum space, which results from Eq.~\eqref{PMCSDJ} by setting $k=0$:
\begin{align}
	\Delta_{\mu\alpha}(p)&=-\frac{1}{p^{2}(1-\vartheta^2 p^{2})} \notag \\
&\phantom{{}={}}\times \left[\Theta_{\mu\alpha}(p)+\mathrm{i}\vartheta\varepsilon_{\mu\sigma\alpha}p^{\sigma}-\xi(1-\vartheta^2 p^{2}) \frac{p_{\mu}p_{\alpha}}{p^{2}}\right]\,.
\label{eq:propagador-MDJ-momentum-space}
\end{align}
The physical dispersion equations are
	\begin{equation}
		p^{2}=0\,, \quad   p^{2}=\frac{1}{\vartheta^2}\,,
\end{equation}
as these occur in the gauge-independent part of Eq.~\eqref{eq:propagador-MDJ-momentum-space}.
In contrast to MCSDJ theory there is only a single massive mode, whereas before we had 2; cf.~Eq.~\eqref{massmode1}. We thus notice that the presence or absence of the conventional CS term in the MDJ Lagrange density changes the mode as well as the pole structure. A further crucial difference to Eq.~\eqref{PMCSDJ} is that one of the massless modes is now also contained in the gauge-independent term proportional to $\Theta_{\mu\alpha}(p)$. Thus, this mode is not necessarily related to physics in the infrared, anymore.

So the corresponding positive-energy dispersion relations read
\begin{equation}
p_0=|\mathbf{p}|\,,\quad p_0=\sqrt{\mathbf{p}^2+\frac{1}{\vartheta^2}}\,.
\label{eq:dispersion-relations-MDJ-theory}
\end{equation}
The latter are relatively simple and immediately imply that there is no superluminal signal propagation. On the contrary, the first mode propagates with the speed of light and the second with velocities lower than the speed of light. Moreover, the analysis of unitarity can be taken over directly from Sec.~\ref{eq:unitarity-MCSDJ-theory} of MCSDJ theory where $k=0$ has to be inserted. These results are expressed in terms of the masses $m_{\pm}$ of Eq.~\eqref{eq:masses-MCSDJ}. Now, we observe that $m_+=1/|\vartheta|$ and $m_-=0$, as expected from Eq.~\eqref{eq:dispersion-relations-MDJ-theory}. The propagator saturated with external conserved currents reads
\begin{equation}
\mathit{SP}=\mathrm{i}\frac{|J|^2-\vartheta\Omega}{\vartheta^2p^2(p^2-m_+^2)}\,,
\end{equation}
which corresponds to that of Eq.~\eqref{eq:saturated-propagator-MCSDJ} for $k=0$. So,
\begin{subequations}
\label{eq:residues-MDJ}
\begin{align}
\mathrm{Res}(\mathit{SP})|_{p^2=m_+^2}=\mathrm{i}(|J|^2-\vartheta\Omega)_{p^2=m_+^2}\,, \\[2ex]
\mathrm{Res}(\mathit{SP})|_{p^2=0}=-\mathrm{i}(|J|^2-\vartheta\Omega)_{p^2=0}\,.
\end{align}
\end{subequations}
For a real current, $\Omega=0$, it is straightforward to deduce that
\begin{equation}
\mathrm{Im}\bigg[\mathrm{Res}(\textit{SP})\bigg|_{\substack{\Omega=0 \\ p^2=\{m_+^2,0\}}}\bigg]\lessgtr 0\,.
\end{equation}
In analogy to what we found in Sec.~\ref{eq:unitarity-MCSDJ-theory}, issues with unitarity in MDJ theory are expected to occur for the massive mode associated with $m_+$. In contrast, the massless mode is well-behaved. However, when $\Omega \neq 0$, the presence of the CS term again complicates the behavior of the saturation. Nevertheless we can directly deduce from Eq.~\eqref{eq:residues-MDJ} that both residues have opposite signs --- independently of the sign of $\Omega$. Thus, if the imaginary part of the first is positive, the imaginary part of the second is automatically negative and vice versa. So there are indications for unitarity violation at tree-level for either $p^2=m_+^2$ or $p^2=0$.

\subsection{Chern-Simons-Deser-Jackiw electrodynamics}
\label{sec:CSDJ-theory}

An alternative model, whose study could be worthwhile, emerges from the Lagrange density of Eq.~\eqref{MCSDJ1} in the absence of the Maxwell term. We will refer to the latter as Chern-Simons-Deser-Jackiw (CSDJ) theory:
\begin{subequations}
\begin{align}
S_{\mathrm{CSDJ}}&=\int \left(\frac{k}{2}A\wedge \mathrm{d}A+\frac{\vartheta}{2}A\wedge \mathrm{d}(\square A)+\frac{1}{2\xi}*\!(\mathrm{i}_{\partial}A)^2\right) \notag \\
&=\int\mathrm{d}^3x\,(\mathcal{L}_{\mathrm{CSDJ}}+\mathcal{L}_{\mathrm{gf}})\,, \\[2ex]
\label{eq:CSDJ-theory}
\mathcal{L}_{\mathrm{CSDJ}}&=\frac{k}{2}\varepsilon_{\mu\nu\rho}A^{\mu}\partial^{\nu}A{}^{\rho}+\frac{\vartheta}{2}\varepsilon_{\mu\nu\rho}A^{\mu}\partial^{\nu}\square A{}^{\rho}\,,
\end{align}
\end{subequations}
with $\mathcal{L}_{\mathrm{gf}}$ of Eq.~\eqref{eq:gauge-fixing}. Its propagator in momentum space is readily obtained as follows:
\begin{equation}
\Delta_{\mu\alpha}(p) =-\frac{1}{p^{2}\left(k-\vartheta p^{2}\right)  }\left[  \mathrm{i}\varepsilon_{\mu\sigma\alpha}p^{\sigma}-\xi(k-\vartheta p^{2})  \frac{p_{\mu}p_{\alpha}}{p^{2}}\right]\,.
\label{eq:propagator-CSDJ}
\end{equation}
The physical dispersion relations are
\begin{equation}
	p_{0}=|\mathbf{p}|\,,\quad p_{0}=\sqrt{\mathbf{p}^{2}+\frac{k}{\vartheta}}\,,
	\label{massmode2}
\end{equation}
representing a massless, Abelian vector boson and a massive particle, respectively, with mass squared $k/\vartheta$. The absence of the Maxwell term does not imply a pole structure different from that of MDJ theory, but it is still composed of a massless and a massive mode, respectively. Furthermore, we note that the DJ term generates dynamics for the CS Lagrangian, assuring the existence of propagating modes by itself. Signal propagation in this model is subluminal, since the group velocity is given by Eq.~\eqref{groupvelocityCSDJM1} with $m^2(\vartheta,k)=k/\vartheta$ and the front velocity is equal to 1.

Contracting the propagator of Eq.~\eqref{eq:propagator-CSDJ} with two external conserved currents eliminates the second term, which is gauge-dependent and unphysical. Thus, the saturation is now completely governed by the CS term:
\begin{equation}
\mathit{SP}=\frac{\mathrm{i}\Omega}{p^2(k-\vartheta p^2)}=-\frac{\mathrm{i}\Omega}{\vartheta p^2(p^2-m^2)}\,.
\end{equation}
So the following statements are made on the residues at the poles:
\begin{subequations}
\label{eq:residues-CSDJ}
\begin{align}
\mathrm{Res}(\mathit{SP})|_{p^2=0}&=\frac{\mathrm{i}\Omega|_{p^2=0}}{k}\,, \\[2ex]
\mathrm{Res}(\mathit{SP})|_{p^2=m^2}&=-\frac{\mathrm{i}\Omega|_{p^2=m^2}}{k}\,.
\end{align}
\end{subequations}
Depending on the signs of $\Omega$ and $k$, it is either the massless or the massive mode that may cause unitarity issues. This behavior is reminiscent of that for the residues of MDJ theory, cf.~Eq.~\eqref{eq:residues-MDJ}. Since the parameter $\vartheta$ is absent in Eq.~\eqref{eq:residues-CSDJ}, the DJ term does not have any direct impact on the properties of the residues.

\subsubsection{Classical field equations and solutions}
\label{sec:field-equations-solutions-CSDJ}

Classical solutions of the field equations coupled to external physical sources provide the bread and butter to any more sophisticated study of a field theory, e.g., at the quantum level. Therefore, our interest in the forthcoming paragraphs is to solve the field equations of CSDJ theory coupled to the external conserved current $J^{\mu}$, i.e., we consider
\begin{equation}
\mathcal{L}=\mathcal{L}_{\mathrm{CSDJ}}-J_{\mu}A^{\mu}\,,
\end{equation}
with $\mathcal{L}_{\mathrm{CSDJ}}$ taken from Eq.~\eqref{eq:CSDJ-theory}.
The Euler-Lagrange equations of Eq.~\eqref{E-L} for $n=3$ imply the field equations:
\begin{equation}
	k\varepsilon _{\mu \nu \rho }\partial ^{\nu }A{}^{\rho }+\vartheta
	\varepsilon _{\mu \nu \rho }\partial ^{\nu }\square A{}^{\rho }=J_{\mu}\,.
\end{equation}
The component $\mu=0$ corresponds to an equation for the magnetic field $B=\varepsilon^{ij}\partial ^{i}A{}^{j}$ with the Levi-Civita symbol $\varepsilon^{ij}$ in two spatial dimensions:
\begin{equation}
kB+\vartheta \square B=\rho\,,
\end{equation}
where $\rho=J^0$ is the charge density.
Interestingly, the DJ term is able to turn the magnetic field dynamical. Indeed, the latter now fulfills a wave equation:
\begin{equation}
\left( \square +\frac{k}{\vartheta }\right) B=\frac{\rho}{\vartheta}\,,
\end{equation}%
which for a static configuration reads
\begin{equation}
	\left( \nabla ^{2}-\frac{k}{\vartheta }\right) B=-\frac{\rho}{\vartheta}\,.
	\label{MagneticCS1}
\end{equation}
It is worthwhile to point out that the relationship $B=\rho/k$, which is characteristic of CS theory, is no longer valid, which is due to the magnetic field becoming dynamical. An equation like the latter can be solved by the Green's function method. Let $G(\mathbf{R})$ be a Green's function such that
\begin{equation}
	\left( \nabla^{2}-\frac{k}{\vartheta}\right) G(\mathbf{R})=\delta^{(2)}(\mathbf{R})\,,
    \label{eq:Green-function-equation}
\end{equation}
with the Dirac function $\delta^{(2)}(\mathbf{R})$ in two spatial dimensions and $\mathbf{R}=\mathbf{r}-\mathbf{r}^{\prime}$. The magnetic field is then given by
\begin{equation}
\mathbf{B}(\mathbf{r})=-\frac{1}{\vartheta}\int\mathrm{d}^{2}\mathbf{r}^{\prime}\,G(\mathbf{r}-\mathbf{r}^{\prime})\rho(\mathbf{r}^{\prime})\,.
\end{equation}
To determine the Green's function, we start from its usual Fourier expansion in two dimensions:
\begin{align}
	G(\mathbf{R}) & =\frac{1}{(2\pi)^{2}}\int\mathrm{d}^{2}p\,G(\mathbf{p})\exp(-\mathrm{i}\mathbf{R}\cdot\mathbf{p}) \notag \\
	              & =\frac{1}{(2\pi)^{2}}\int\mathrm{d}^{2}p\,G(\mathbf{p})\exp(-\mathrm{i}Rp\cos\phi)\,,
	\label{GreenF1a}
\end{align}
where we choose a polar-coordinate system with the horizontal positive axis pointing along $\mathbf{R}$ such that $\mathrm{d}^{2}p=\mathrm{d}\phi\mathrm{d}p\,p$, with $\phi$ being the angle between $\mathbf{R}$ and $\mathbf{p}$, as well as $|\mathbf{R}|=R$ and $|\mathbf{p}|=p$. Inserting Eq.~\eqref{GreenF1a} into Eq.~\eqref{eq:Green-function-equation}, one obtains the Green's function in momentum space:
\begin{subequations}
\label{eq:green-function-momentum-space}
\begin{equation}
G(\mathbf{p})=-\frac{1}{p^{2}+m^{2}}\,,
\label{eq:green-function-momentum-space1}
\end{equation}
with the effective mass squared
\begin{equation}
	m^{2}=\frac{k}{\vartheta}\,.
\end{equation}
\end{subequations}
To transform the momentum space Green's function of Eq.~\eqref{eq:green-function-momentum-space1} to configuration space, we must evaluate the two-dimensional integral of Eq.~\eqref{GreenF1a}. To do so, we employ%%
\begin{subequations}
\begin{align}
	\int_{0}^{2\pi}\mathrm{d}\phi\,\exp(-\mathrm{i}Rp\cos\phi)&=2\pi J_{0}(R)\,, \\[2ex]
	\int_0^{\infty} \mathrm{d}p\,p\frac{J_{0}(pR)}{p^{2}+m^{2}}&=K_{0}(mR)\,,
\end{align}
\end{subequations}
cf.~Eqs.~(3.915.2) and (6.532.4), respectively, in Ref.~\cite{Gradshteyn:2007}. Here, $J_0(x)$ is the zeroth-order Bessel function of the first kind and $K_0(x)$ the zeroth-order modified Bessel function of the second kind. The Green's function in configuration space can then be cast into the form
\begin{equation}
\label{eq:green-function-configuration-space-first-model}
G(\mathbf{R})=-\frac{1}{2\pi}K_{0}(mR)\,.
\end{equation}
Now, the static magnetic field is written as
\begin{equation}
B(\mathbf{r})=\frac{1}{2\pi\vartheta}\int\mathrm{d}^{2}r^{\prime}\,K_{0}(m|\mathbf{r}-\mathbf{r}^{\prime}|)\rho(\mathbf{r}^{\prime})\,,
\end{equation}
which for a pointlike charge $q$ sitting at $\mathbf{r}^{\prime}=\mathbf{0}$ reads
\begin{align}
\label{eq:magnetic-field-SCDJ-theory}
B(\mathbf{r})&=\frac{1}{2\pi\vartheta}\int\mathrm{d}^{2}r^{\prime}\,K_{0}(m|\mathbf{r}-\mathbf{r}^{\prime}|)q\delta^{(2)}(\mathbf{r}^{\prime}), \notag \\
&=\frac{q}{2\pi\vartheta}K_{0}(mr)\,,
\end{align}
describing the radial behavior of an isotropic magnetic field.
Thus, one notices that this DJ solution for the magnetic field captures the effect of the Laplacian in Eq.~\eqref{MagneticCS1}, which stems from the Maxwell term. This modifies the usual behavior of the magnetic field of a point-like charge in CS theory, which is described by a Dirac function because of the direct proportionality between the magnetic field and the charge density:
\begin{equation}
B_{\mathrm{CS}}(\mathbf{r})=\frac{q}{k}\delta^{(2)}(\mathbf{r})\,.
\label{eq:magnetic-field-CS-theory}
\end{equation}
Now, integrating the magnetic field of Eq.~\eqref{eq:magnetic-field-CS-theory} over the entire plane provides the total magnetic flux permeating the plane, which amounts to the ratio of the total charge and the CS parameter $k$, that is,
\begin{align}
	\label{eq:integration-magnetic-field-CS}
	\int\mathrm{d}^2r\,B_{\mathrm{CS}}(\mathbf{r})&=\frac{q}{k}\,.
\end{align}
The latter property is still valid for Eq.~\eqref{eq:magnetic-field-SCDJ-theory} in CSDJ theory. Indeed, as it holds
\begin{equation}
	\label{eq:auxiliary-integral}
	\int_0^{\infty} \mathrm{d}x\,xK_0(ax)=\frac{1}{a^2}\,,\quad a>0\,,
\end{equation}
one obtains
\begin{align}
	\label{eq:integration-magnetic-field-CS2}
	\int\mathrm{d}^2r\,B(\mathbf{r})&=\frac{q}{\vartheta}\int_0^{\infty} \mathrm{d}r\,r K_{0}(mr)=\frac{q}{\vartheta}\left(\frac{k}{\vartheta}\right)^{-1}=\frac{q}{k}\,,
\end{align}
where the factor of $1/(2\pi)$ in Eq.~\eqref{eq:magnetic-field-SCDJ-theory} cancels due to the angular integration. The final result coincides with Eq.~\eqref{eq:integration-magnetic-field-CS}.
\begin{figure}
\includegraphics[scale=0.5]{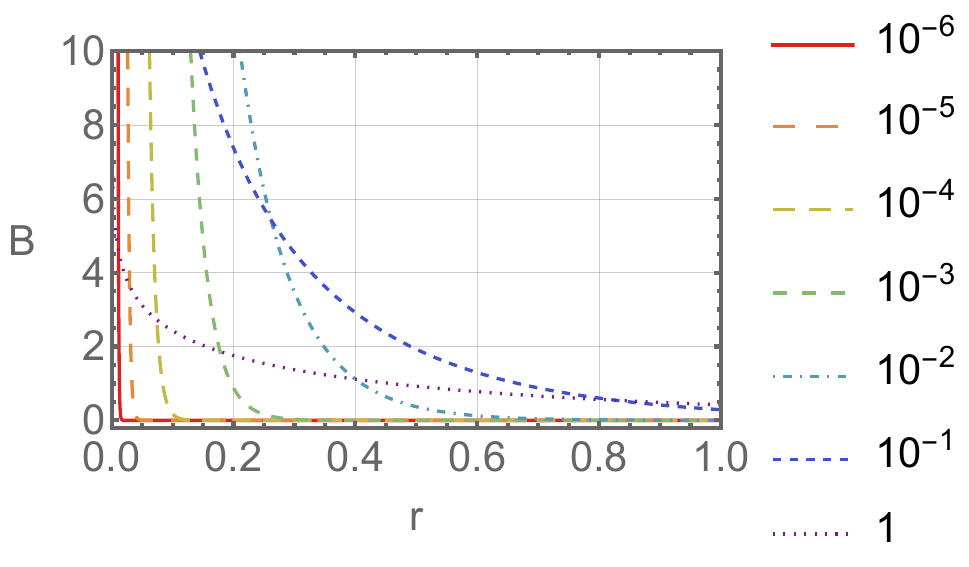}
\caption{Behavior of Eq.~\eqref{eq:magnetic-field-SCDJ-theory} for decreasing values of $\vartheta$ chosen in terms of rising inverse powers of 1/10. We observe that the magnetic field becomes more and more localized at $r=0$ when $\vartheta\mapsto 0$, which is the limit recovering a $\delta$-function behavior, as expected.}
\label{fig:magnetic-field}
\end{figure}
So supplementing CS theory by a DJ term regularizes the $\delta$-function type behavior of a point charge turning it into a spatially extended Bessel function, without altering the total magnetic flux. Note that the properties characteristic of a $\delta$-function are reproduced from Eq.~\eqref{eq:magnetic-field-SCDJ-theory} in the limit $\vartheta\mapsto 0$. The behavior of this function for decreasing values of $\vartheta$ is illustrated in Fig.~\ref{fig:magnetic-field}.

A similar static magnetic-field solution for a pointlike charge, which is characterized by the function $K_{0}(mr)$, occurs in the planar MCS theory, in the absence of spatial currents. The field equations in this case read
\begin{equation}
	(\nabla ^{2}-k^2)B= - k {\rho}\,,
	\label{MagneticMCS1}
\end{equation}
where the Laplacian originates from the Maxwell term. This finding confirms that the DJ term can effectively mimic the effects of a Maxwell term in the field equations.
%Do we have to provide a reference here?
In the next section we intend to examine a planar electrodynamics in the presence of higher-order derivatives that are contracted with nondynamical background fields. The latter give rise to preferred spacetime directions, which lead to a breakdown of $\mathit{SO}(2,1)$ symmetry, in general, and to spatial anisotropies in the plane, in particular.

\section{Modified $\boldsymbol{(2+1)}$-dimensional electrodynamics with Lorentz violation}
\label{sec:MCS-electrodynamics-higher-derivatives}

In general, the higher-derivative term in the Lagrange density of Eq.~\eqref{LCSN}, whose structure is $\varepsilon_{\mu\nu\varrho}A^{\mu}\partial^{\nu}(K^{\lambda\beta}\partial_{\lambda}\partial_{\beta})A^{\rho}$, corresponds to an extension of the DJ term $\varepsilon_{\mu\nu\varrho}A^{\mu}\partial^{\nu}\square A^{\rho}$ of mass dimension 4, which incorporates anisotropic contributions.
For the purpose of generality, we propose an extended higher-derivative planar electrodynamics in the presence of the standard terms, either endowed with the Maxwell term or not:
\begin{subequations}
\label{LMCSD5}
\begin{align}
S_{\mathrm{ext}}&=\int \bigg(-\frac{\varkappa}{4}F\wedge (*F)+\frac{k}{2}A\wedge \mathrm{d}A+\frac{\vartheta}{2}A\wedge \mathrm{d}(\square A) \notag \\
&\phantom{{}={}}\hspace{0.7cm}+\frac{\vartheta_1}{2}A\wedge \mathrm{d}(\hat{\mathcal{K}}^{(4)}A)+\frac{1}{2\xi}*\!(\mathrm{i}_{\partial}A)^2\bigg) \notag \\
&=\int\mathrm{d}^3x\,(\mathcal{L}_{\mathrm{ext}}+\mathcal{L}_{\mathrm{gf}})\,,
\end{align}
with the Lagrange density
\begin{align}
\mathcal{L}_{\mathrm{ext}}&=-\frac{\varkappa}{4}F_{\mu\nu}F^{\mu\nu}{}+\frac
{k}{2}\varepsilon_{\mu\nu\rho}A^{\mu}\partial^{\nu}A{}^{\rho}+\frac{\vartheta}{2}\varepsilon_{\mu\nu\rho}A^{\mu}\partial^{\nu}\square A{}^{\rho} \notag \\
&\phantom{{}={}}+\frac{\vartheta_{1}}{2}\varepsilon_{\mu\nu\rho}A^{\mu}\partial^{\nu}\hat{\mathcal{K}}^{(4)}A^{\rho}\,,
\end{align}
\end{subequations}
and the gauge fixing term of Eq.~\eqref{eq:gauge-fixing}. Here, we employ the operator $\hat{\mathcal{K}}^{(4)}$ of Eq.~\eqref{kappa2}, for brevity. Moreover, the parameter $\varkappa$ was introduced to permit the description of a nontrivial dielectric constant, i.e., such a setting can be neatly adopted to describing certain effects in a material medium, as we did in Refs.~\cite{Silva:2020dli,Pedro2021}. The choice $\varkappa=1$ reproduces the $(2+1)$-dimensional version of the usual Maxwell term \textit{in vacuo}.

To further analyze the features of the electrodynamics represented by the action of Eq.~\eqref{LMCSD5}, we calculate the propagator of the model whose physical poles provide the dispersion relations of the theory as well as information about the associated modes. As before, we cast the Lagrange density into bilinear form,
\begin{subequations}
\begin{equation}
\tilde{\mathcal{L}}_{\mathrm{ext}}=\frac{1}{2}A^{\nu}\Xi_{\nu\mu}A^{\mu}\,, \label{LMCSD5B2}
\end{equation}
with the tensor operator
\begin{align}
\label{eq:operator-Xi}
\Xi_{\nu\mu}&=\varkappa\square\Theta_{\nu\mu}{}+k\varepsilon_{\nu\rho\mu}\partial^{\rho}+\vartheta\varepsilon_{\nu\rho\mu}\partial^{\rho}\square \notag \\
&\phantom{{}={}}+\vartheta_{1}\varepsilon_{\nu\rho\mu}\hat{\mathcal{K}}^{(4)}\partial^{\rho}-\frac{1}{\xi}\square\Omega_{\nu\mu}\,,
\end{align}
which can also be reformulated as
\begin{equation}
\Xi_{\nu\mu}=\varkappa\square\Theta_{\nu\mu}+(k+\vartheta\square+\vartheta_{1}\hat{\mathcal{K}}^{(4)})  L_{\nu\mu}-\frac{1}{\xi}\square
\Omega_{\nu\mu}\,,
\end{equation}
\end{subequations}
by using the CS operator of Eq.~\eqref{eq:CS-operator} and the convenient projectors of Eq.~\eqref{projectors}. The propagator is obtained as before, but now its computation poses a greater technical challenge. We propose the \textit{ansatz}
\begin{equation}
\Delta_{\phantom{\mu}\alpha}^{\mu}=a\Theta_{\phantom{\mu}\alpha}^{\mu}+bL_{\phantom{\mu}\alpha}^{\mu}+c\Omega_{\phantom{\mu}\alpha}^{\mu}\,,
\end{equation}
with unknown parameters $a,b,c$. The latter must satisfy%%
\begin{equation}
\Xi_{\nu\mu}^{{}}\Delta_{\phantom{\mu}\alpha}^{\mu}=\eta_{\nu\alpha}\,.
\label{identity2}
\end{equation}%%
The tensor algebra of Tab.~\ref{tab:algebra-tensor-projectors} allows us to rewrite Eq.~\eqref{identity2} as
\begin{align}
\Theta_{\nu\alpha}+\Omega_{\nu\alpha}&=\varkappa a\square\Theta_{\nu\alpha}+b\varkappa\square L_{\nu\alpha}-\frac{1}{\xi}c\square\Omega_{\nu}{}_{\alpha} \notag \\
&\phantom{{}={}}+\left(k+\vartheta\square+\vartheta_{1}\hat{\mathcal{K}}^{(4)}\right)(aL_{\nu\alpha}-b\square\Theta_{\nu\alpha})\,,
\end{align}
which provides the following solutions for the parameters $a,b,c$ of the \textit{ansatz}:
\begin{subequations}
\begin{equation}
a=\frac{\varkappa}{\boxminus}\,,\quad b=-\frac{k+\vartheta \square+\vartheta_{1}\hat{\mathcal{K}}^{(4)}}{\square\boxminus}\,,\quad c=-\frac{\xi}{\square}\,,
\end{equation}
with
\begin{equation}
\boxminus=\varkappa^{2}\square+\left(  k+\vartheta\square
+\vartheta_{1}\hat{\mathcal{K}}^{(4)}\right)  ^{2}\,.
\end{equation}
\end{subequations}
Then, the inverse of Eq.~\eqref{eq:operator-Xi} is
\begin{align}
\Delta_{\mu\alpha}&=\frac{1}{\square\boxminus}\bigg[\varkappa\square
\Theta_{\mu\alpha}-\xi\boxminus\Omega_{\mu\alpha} \notag \\
&\phantom{{}={}}\hspace{0.8cm}-\left(k+\vartheta\square+\vartheta_{1}\hat{\mathcal{K}}^{(4)}\right)L_{\mu\alpha}\bigg]\,,
\end{align}
which is written in momentum space as
\begin{subequations}
\label{PropP}
\begin{align}
\Delta_{\mu\alpha}(p)&=-\frac{1}{p^{2}F(p)}\Big[  \varkappa p^{2}\Theta_{\mu\alpha}(p)-\xi F\left(  p\right)\frac{p_{\mu}p_{\alpha}}{p^{2}} \notag \\
&\phantom{{}={}}\hspace{1.5cm}+\left(k-\vartheta p^{2}+\vartheta_{1}\mathcal{K}^{(4)}(p)\right)L_{\mu\alpha}(p)\Big]\,,
\label{PMCSDJA}
\end{align}
where
\begin{equation}
\label{eq:denominator-propagator}
F(p)=\varkappa^{2}p^{2}-\left(k-\vartheta p^{2}+\vartheta_{1}\mathcal{K}^{(4)}(p)\right)^{2}\,,
\end{equation}
and
\begin{equation}
L_{\nu\alpha}(p)=-\mathrm{i}\varepsilon_{\nu\sigma\alpha}p^{\sigma}\,,\quad \mathcal{K}^{(4)}(p)=-K^{\lambda
\beta}p_{\lambda}p_{\beta}\,.
\end{equation}
\end{subequations}
The dispersion relations of the theory represented by the Lagrangian of Eq.~\eqref{LMCSD5} can be extracted from the poles of the propagator of Eq.~\eqref{PropP}, namely,
\begin{equation}
F(p)=0\,,
\label{prd}
\end{equation}
with $F(p)$ of Eq.~\eqref{eq:denominator-propagator}. To gain information about the physical modes, it is necessary to analyze the dispersion relations from Eq.~\eqref{prd} for some configurations of $K^{\alpha\beta}$. Before we dedicate ourselves to this endeavor, let us explore the properties of Eq.~\eqref{PropP} saturated by external conserved currents:
\begin{align}
\mathit{SP}&=-\mathrm{i}\frac{1}{\varkappa^{2}p^{2}-(k-\vartheta p^{2}+\vartheta_{1}\mathcal{K}^{(4)}(p))^{2}} \notag \\
&\phantom{{}={}}\times\left[\varkappa |J|^2+(k-\vartheta p^2+\vartheta_1\mathcal{K}^{(4)}(p))\frac{\Omega}{p^2}\right]\,,
	\label{SPMCSDJA}
\end{align}
where the pole $p^2$ cancels in the first contribution as in Eq.~\eqref{eq:saturated-propagator-MCSDJ} previously considered for MCSDJ theory. The imaginary part of the residues of Eq.~\eqref{SPMCSDJA} provide both negative and positive results, as will be seen ahead when certain parameter subsets of Eq.~\eqref{LMCSD5} will be studied.

It is also worthwhile to note that the general Lagrange density of Eq.~\eqref{LMCSD5} provides a smorgasbord of the usual terms of standard planar theories that can be examined more carefully. Thus, Eq.~\eqref{LMCSD5} encompasses several possibilities of planar models. The different models obtained from suitable choices of parameters are stated in Tab.~\ref{tab:planar-theories} and the corresponding Lagrange densities read as follows:%%
\begin{table}
\begin{tabular}{cccccc}
\toprule
$\varkappa$ & $k$ & $\vartheta$ & $\vartheta_1$ & Model          & Lagrange density \\
\colrule
1           &     &             &               & Extended MCSDJ & Eq.~\eqref{LMCSD5B} \\
1           &     & 0           &               & Extended MCS   & Eq.~\eqref{LMCSD5D} \\
1           & 0   &             &               & Extended MDJ   & Eq.~\eqref{LMCSD5C} \\
0           &     &             &               & Extended CSDJ  & Eq.~\eqref{LMCSD5A} \\
\botrule
\end{tabular}
\caption{Variety of different higher-derivative Lorentz-violating planar electrodynamics emerging from the generic proposal of Eq.~\eqref{LMCSD5} for appropriate choices of parameters. Parameters that remain arbitrary are not given explicitly.}
\label{tab:planar-theories}
\end{table}%%
\begin{subequations}
		\begin{align}
        \label{LMCSD5B}%
		\mathcal{L}'_{\mathrm{MCSDJ}}&=-\frac{1}{4}F_{\mu\nu}F^{\mu\nu}{}+\frac
		{k}{2}\varepsilon_{\mu\nu\rho}A^{\mu}\partial^{\nu}A{}^{\rho} \notag \\
&\phantom{{}={}}+\frac{\vartheta}{2}\varepsilon_{\mu\nu\rho}A^{\mu}\partial^{\nu}\square A{}^{\rho} \notag \\
&\phantom{{}={}}+\frac{\vartheta_{1}}{2}\varepsilon_{\mu\nu\rho}A^{\mu}\partial^{\nu}\hat{\mathcal{K}}^{(4)}
		A{}^{\rho}\,, \displaybreak[0]\\[2ex]
        \label{LMCSD5D}%
        \mathcal{L}'_{\mathrm{MCS}}&=-\frac{1}{4}F_{\mu\nu}F^{\mu\nu}{}+\frac
		{k}{2}\varepsilon_{\mu\nu\rho}A^{\mu}\partial^{\nu}A{}^{\rho} \notag \\
&\phantom{{}={}}+\frac{\vartheta_{1}}{2}\varepsilon_{\mu\nu\rho}A^{\mu}\partial^{\nu}\hat{\mathcal{K}}^{(4)}
		A{}^{\rho}\,, \displaybreak[0]\\[2ex]
		\label{LMCSD5C}%
		\mathcal{L}'_{\mathrm{MDJ}}&=-\frac{1}{4}F_{\mu\nu}F^{\mu\nu}+\frac{\vartheta
		}{2}\varepsilon_{\mu\nu\rho}A^{\mu}\partial^{\nu}\square A{}^{\rho} \notag \\
&\phantom{{}={}}+\frac{\vartheta_{1}}{2}\varepsilon_{\mu\nu\rho}A^{\mu}\partial^{\nu}\hat{\mathcal{K}}^{(4)}A{}^{\rho}\,, \displaybreak[0]\\[2ex]
        \label{LMCSD5A}
		\mathcal{L}'_{\mathrm{CSDJ}}&=\frac{k}{2}\varepsilon_{\mu\nu\rho}A^{\mu}\partial^{\nu}A{}^{\rho}+\frac{\vartheta
		}{2}\varepsilon_{\mu\nu\rho}A^{\mu}\partial^{\nu}\square A{}^{\rho} \notag \\
&\phantom{{}={}}+\frac{\vartheta_{1}}{2}\varepsilon_{\mu\nu\rho}A^{\mu}\partial^{\nu}\hat{\mathcal{K}}^{(4)}A{}^{\rho}\,.
	\end{align}
\end{subequations}
The latter are indicated by primes to distinguish them from their counterparts in Eqs.~\eqref{MCSDJ1}, \eqref{MDJ1}, \eqref{eq:CSDJ-theory} that do not involve Lorentz-violating contributions. In principle, we could take $\hat{\mathcal{K}}^{(4)}$ as traceless, as the contribution proportional to the trace corresponds to the DJ term, anyhow. However, to be as flexible as possible, we will not necessarily assume that $\hat{\mathcal{K}}^{(4)}$ is traceless. Note that MCS and MDJ theories are, in principle, special cases of MCSDJ theory that arise for suitable choices of the parameters.

\subsection{Extended MCS(DJ) theory}
\label{sec:extended-MCSDJ}

For the Lagrange density stated in Eq.~\eqref{LMCSD5B}, we extract the physical dispersion relations from the poles of the generic saturated propagator given by Eq.~\eqref{SPMCSDJA} where $\varkappa=1$. In the general case, they read
	\begin{equation}
		\label{DRMCSDJA}
	p^{2}-\left[k-\vartheta p^{2}+\vartheta_{1}\mathcal{K}^{(4)}(p)\right]^{2}=0\,.
	\end{equation}
As the structure $\hat{\mathcal{K}}^{(4)}$ reduces to the d'Alembertian when $K^{\lambda\beta}=\eta^{\lambda\beta}$, the higher-derivative term in the Lagrange density of Eq.~\eqref{LCSN} contains the DJ operator. Therefore, without loss of generality, we can consider Eq.~\eqref{LMCSD5B} with $\vartheta=0$ by writing
\begin{equation}
\mathcal{L}=\mathcal{L}'_{\mathrm{MCSDJ}}|_{\vartheta=0}+\mathcal{L}_{\mathrm{gf}}=\mathcal{L}'_{\mathrm{MCS}}+\mathcal{L}_{\mathrm{gf}}\,,
\label{LMCSD5B-theta-0}
\end{equation}
whose propagator is given by Eq.~\eqref{PropP} with $\vartheta=0$. The corresponding dispersion relation reads
\begin{equation}
	p^{2}-(k-\vartheta_{1}K^{\lambda\beta}p_{\lambda}p_{\beta})^{2}=0\,. \label{prd4}
\end{equation}
A dispersion relation of this form may be very intricate, since it involves higher powers of $p_0$. Specifically,
\begin{equation}
	p^{2}-\Big[k-\vartheta_{1}(K^{00}p_{0}^{2}-K^{0i}p_{0}p^{i}+K^{ij}p^{i}p^{j})\Big]^{2}=0\,.
\label{prd4B}
\end{equation}
To get rid of at least some of the terms depending on $p_0$, we discard the mixed coefficients, i.e., $K^{0i}=0$. Then, Eq.~\eqref{prd4B} yields
\begin{subequations}
\label{prd4D}
\begin{equation}
	p_{0}^{2}=\frac{1}{2\vartheta_{1}^2 (K^{00})^2}\left[1+2\vartheta_{1}K^{00}(k-\vartheta_{1}K^{ij}p^{i} p^{j}) \pm \sqrt{\Theta} \right]\,, \label{prd4C}
\end{equation}
with
\begin{align}
\Theta&=\Big[1+4\vartheta_{1}K^{00}(k-\vartheta_{1}K^{ij}p^{i} p^{j})-4\vartheta_{1}^2 (K^{00})^2 \mathbf{p}^2 \Big]\,.
\end{align}
\end{subequations}
The latter may still take an even more tractable form for specific choices of the spacelike tensor components $K^{ij}$.

\subsubsection{Isotropic timelike configuration}

For $K^{ij}=0$, for instance, Eq.~\eqref{prd4D} implies two positive-energy dispersion relations in an isotropic setting:
\begin{subequations}
	\label{prd4E1}
\begin{align}
(\omega^{(\pm)})^2&=\frac{1}{2\vartheta_{1}^2 (K^{00})^2}\left(1+2k\vartheta_{1}K^{00}\pm \Xi\right)\,, \\[2ex]
\Xi&=\sqrt{1+4\vartheta_1 K^{00}(k-\vartheta_1K^{00}\mathbf{p}^2)}\,.
\label{prd4E}
\end{align}
\end{subequations}
To assure real energies, the radicand in $\Xi$ must be nonnegative. Thus, the norm of the momentum should not exceed a particular value:
\begin{equation}
	|\mathbf{p}|\leq p_{\mathrm{max}}\,,\quad p_{\mathrm{max}}=\frac{\sqrt{1+4k\vartheta_{1} K^{00}}}{2|\vartheta_{1}K^{00}|}\,.
	 \label{prd4F}
\end{equation}
Furthermore, an expansion of Eq.~\eqref{prd4E} in $K^{00}$ reads
\begin{subequations}
\begin{align}
\omega^{(+)}&=\frac{1}{|\vartheta_1K^{00}|}+\mathrm{sgn}(\vartheta_1K^{00})k \notag \\
&\phantom{{}={}}-\frac{\mathrm{sgn}(\vartheta_1K^{00})}{2}(2k^2+\mathbf{p}^2)K^{00}+\dots\,, \\[2ex]
\omega^{(-)}&=\sqrt{\mathbf{p}^2+k^2}(1-k\vartheta_1K^{00})+\dots\,.
\end{align}
\end{subequations}
So the $(-)$ mode is a perturbation of the CS dispersion relation, whereas the $(+)$ mode is nonpertubative, i.e., it strongly deviates from the standard dispersion relation for $K^{00}$ small. Therefore, the occurrence of two modes in Eq.~\eqref{prd4E} cannot be ascribed to the emergence of birefringence in the usual sense. Note that an electrodynamics in $(2+1)$ spacetime dimensions does exhibit birefringence, which is a property that we were also able to observe in Ref.~\cite{Joca}. Thus, the nonperturbative $(+)$ mode is a consequence of the presence of higher-order time derivatives in the sector considered.

Next, let us determine the group velocity of each mode. By resorting to Eq.~\eqref{prd4E} their norms are written in succinct form as
\begin{equation}
u_{\mathrm{gr}}^{(\pm)}=\frac{1}{\Xi}\frac{|\mathbf{p}|}{\omega^{(\pm)}}\,.
\end{equation}
The latter posses sub- and superluminal regimes. Each branch of the group velocity exceeds 1 for momenta larger than certain limits given by%%
\begin{subequations}
\begin{align}
\label{eq:momentum-limit-1}
p_{\mathrm{lim}}^{(+)}&=\sqrt{\frac{k}{\vartheta_1K^{00}}}\,, \\[2ex]
\label{eq:momentum-limit-2}
p_{\mathrm{lim}}^{(-)}&=\frac{1}{4\sqrt{2}|\vartheta_1K^{00}|}\bigg((1+4k\vartheta_1K^{00})\Big[3-4k\vartheta_1K^{00} \notag \\
&\phantom{{}={}}\hspace{1.4cm}+\sqrt{1+4k\vartheta_1K^{00}}\sqrt{9+4k\vartheta_1K^{00}}\Big]\bigg)^{1/2}\,,
\end{align}
\end{subequations}
respectively. Note that the dispersion relations become complex when the momentum lies beyond the value of Eq.~\eqref{prd4F}. So for $|\mathbf{p}|\geq p_{\mathrm{max}}$, the dispersion relations are more appropriately recast into
\begin{align}
\label{eq:dispersion-relations-complex-form}
\omega^{(\pm)}&=\frac{1}{2|\vartheta_1K^{00}|}\bigg(\sqrt{2|\vartheta_1K^{00}|\sqrt{\mathbf{p}^2+k^2}+|1+2k\vartheta_1K^{00}|} \notag \\
&\phantom{{}={}}\hspace{0.65cm}\pm\mathrm{i}\sqrt{2|\vartheta_1K^{00}|\sqrt{\mathbf{p}^2+k^2}-|1+2k\vartheta_1K^{00}|}\,\bigg)\,.
\end{align}
The imaginary parts of the latter vanish for $p=p_{\mathrm{max}}$, as expected. Furthermore, the norms of the group velocities for both modes approach infinity for that value, as $\Xi=0$.

To evaluate the front velocity, we must consider the limit of infinite momenta where the dispersion relations take the complex form of Eq.~\eqref{eq:dispersion-relations-complex-form}. Since the imaginary part is commonly associated with attenuation, an adaptation of the definition for the front velocity restricted to the real part is reasonable:
\begin{equation}
u_{\mathrm{fr}}^{(\pm)}\equiv \lim_{|\mathbf{p}|\mapsto\infty} \frac{\mathrm{Re}(\omega^{(\pm)})}{|\mathbf{p}|}=0\,.
\end{equation}
So by looking at Eq.~\eqref{eq:dispersion-relations-complex-form}, it is not difficult to acknowledge that $u_{\mathrm{fr}}^{(\pm)}=0$. Thus, the front velocity does not exhibit a superluminal regime. After all, infinite-momentum excitations do not even propagate. The analysis of this sector of Eq.~\eqref{LMCSD5D} highlights the peculiar properties of dispersion relations for choices of controlling coefficients that are contracted with additional time derivatives.

The latter property of the presently studied dispersion relation also makes it more challenging to evaluate the saturated propagator. For $\varkappa=1$ and $\vartheta=0$, the saturation of Eq.~\eqref{SPMCSDJA} reads
	\begin{align}
		\mathit{SP}&=-\mathrm{i}\frac{|J|^2+(k+\vartheta_1\mathcal{K}^{(4)}(p)) \Omega/p^2}{p^{2}-(k-\vartheta_{1}K^{00}p_{0}^2)^{2}}\,,
		\label{SPMCSDJA2}
	\end{align}
which can also be rewritten as
	\begin{align}
	\mathit{SP}&=\mathrm{i}\frac{|J|^2+(k+\vartheta_1\mathcal{K}^{(4)}(p)) \Omega/p^2}{ \vartheta_{1}^2 (K^{00})^2 [p_{0}^{2}-(\omega^{(+)})^2][p_{0}^{2}-(\omega^{(-)})^2]}\,,
	\label{SPMCSDJA3}
\end{align}
where $\omega^{(\pm)}$ are the dispersion relations stated in Eq.~\eqref{prd4E1}. In contrast to what we did before, we will now compute the residues at $p_0=\omega^{(\pm)}$ directly. This leads to
\begin{align}
\mathrm{Res}(\textit{SP})|_{p_0=\omega^{(\pm)}}&=\pm\frac{\mathrm{i}}{2\omega^{(\pm)}\Xi}\bigg[|J|^2 \notag \\
&\phantom{{}={}}\hspace{0.8cm}-\frac{1+\Xi}{2(k-\vartheta_1K^{00}\mathbf{p}^2)}\Omega\bigg]_{p_0=\omega^{(\pm)}}\,.
\end{align}
Now, $\omega^{(\pm)}\geq |\mathbf{p}|$ where the equality sign holds for $|\mathbf{p}|=p_{\mathrm{lim}}^{(+)}$ of Eq.~\eqref{eq:momentum-limit-1}. Hence, for $\Omega=0$ we can immediately conclude that
\begin{equation}
\mathrm{Im}\bigg[\mathrm{Res}(\mathit{SP})\bigg|_{\substack{\Omega=0 \\ p_0=\omega^{(\pm)}}}\bigg]\lessgtr 0\,.
\end{equation}
So the nonperturbative mode $(+)$ is likely to imply unitarity problems. The perturbative mode $(-)$ is in accordance with unitarity, as expected. For $\Omega\neq 0$, though, it is unclear which one of the modes is problematic for unitarity.

\subsubsection{Anisotropic purely spacelike configuration}

An alternative sector of Eq.~\eqref{LMCSD5D} that may be worthwhile to study further is the purely spacelike one, i.e., we choose $K^{0i}=0$ and $K^{00}=0$ simultaneously. Going back to Eq.~\eqref{prd4B}, this yields the single dispersion relation
\begin{align}
	\omega&=\sqrt{\textbf{{p}}^2+(k-\vartheta_{1}K^{ij}p^{i} p^{j})^2} \notag \\
&=\sqrt{(\delta^{ij}-2k\vartheta_1K^{ij})p^ip^j+k^2+\vartheta_1^2(K^{ij}p^ip^j)^2}\,,
\label{prd4G}
\end{align}
which is relatively simple, as there are no additional time derivatives in this sector. This dispersion relation demonstrates the possibility of including anisotropies at both the second and fourth order in the momentum. The associated group velocity reads
\begin{equation}
\mathbf{u}_{\mathrm{gr}}=\frac{1}{\omega}\left[\mathbf{p}-2\vartheta_1(k-\vartheta_{1}K_{pp})\mathbf{K}_p\right]\,,
\end{equation}
with the scalar $K_{pp}\equiv K^{lm}p^{l}p^{m}$ and the spatial vector $\mathbf{K}_p$ having components $(K_p)^i\equiv K^{ij}p^j$. The norm of the group velocity is
\begin{align}
u_{\mathrm{gr}}&=\frac{1}{\omega}\Big(\mathbf{p}^2-4\vartheta_1(k-\vartheta_{1}K_{pp})K_{pp} \notag \\
&\phantom{{}={}}\hspace{0.5cm}+4\vartheta_1^2(k-\vartheta_{1}K_{pp})^2\mathbf{K}_p^2\Big)^{1/2}\,,
\label{eq:group-velocity-MCS-spacelike}
\end{align}
which can exceed 1 for sufficiently large momentum components. Also, for the front velocity we have that $u_{\mathrm{fr}}=\infty$. The fourth-order term in the momentum, which becomes dominant for large momenta, is responsible for both behaviors.

The saturated propagator of Eq.~\eqref{SPMCSDJA} simplifies as
\begin{align}
\mathit{SP}&=-\mathrm{i}\frac{1}{p^{2}-(k-\vartheta_{1}K^{ij}p^{i}p^{j})^2} \notag \\
&\phantom{{}={}}\times\left[|J|^2+(k-\vartheta_1K^{ij}p^{i}p^{j})\frac{\Omega}{p^2}\right]\,,
	\label{SPMCSDJA2}
\end{align}
where the residue evaluated at the pole given by Eq.~\eqref{prd4G}, which is equivalent to $p^{2}=(k-\vartheta_{1}K^{ij}p^{i}p^{j})^2$, reads
\begin{align}
	\mathrm{Res}&(\mathit{SP})|_{p^{2}=\omega^2-\mathbf{p}^2} \notag \\
&=-\mathrm{i}\bigg(|J|^{2}+\frac{\Omega}{k-\vartheta_{1} K^{ij}p^{i}p^{j}}\bigg)_{p^{2}=\omega^2-\mathbf{p}^2}\,,
\label{eq:saturation-residue-MCSDJ-theory}
\end{align}
with $\omega$ of Eq.~\eqref{prd4G}.
Note that a plethora of investigations of higher-derivative theories, see, e.g., Refs.~\cite{Schreck:2014qka,Leticia2,Leticia1}, indicate that additional time derivatives are likely to spoil unitary time evolution, which is a behavior that shows up in the saturated propagator evaluated at the corresponding poles. Since we implemented $K^{0i}=0$ and $K^{00}=0$, all additional time derivatives are eliminated. Therefore, based on the results on higher-derivative theories, which are available in the contemporary literature, our expectation would have been that unitarity is guaranteed for the configuration under study. However, the presence of of the second term in Eq.~\eqref{eq:saturation-residue-MCSDJ-theory} again obscures unitarity. The latter can be violated for suitable choices of the parameters.

\subsection{Extended MDJ theory}

In principle, the extended MDJ theory of Eq.~\eqref{LMCSD5C} is a special case of the extended MCSDJ theory of Eq.~\eqref{LMCSD5B} whose study we concluded in Sec.~\ref{sec:extended-MCSDJ}. At a technical level, $k=0$ must be inserted into the previous findings, which is not expected to be challenging and will probably lead to a vastly repetitive analysis. However, the physical interpretation of the results may differ from that of Sec.~\ref{sec:extended-MCSDJ}. Thus, we leave a more thorough examination of the properties of Eq.~\eqref{LMCSD5C} for a future occasion and will dedicate ourselves to the extended CSDJ theory of Eq.~\eqref{LMCSD5A} whose structure is fundamentally different from that of the extended MCSDJ and MDJ theories.

\subsection{Extended CSDJ theory}

Let us now explore the Lagrange density of Eq.~\eqref{LMCSD5A}.
The dispersion relations different from $p^0=|\mathbf{p}|$ follow from Eq.~\eqref{SPMCSDJA} by inserting $\varkappa=0$, which gets rid of the Maxwell term. Then,
\begin{equation}
\vartheta p^{2}-k-\vartheta_{1}\mathcal{K}^{(4)}(p)=0\,.
\label{RDCSD5A}
\end{equation}
By inserting $k=\vartheta_{1}=0$, the dispersion equation reads $p^2=0$, which corresponds to that of DJ theory; cf.~Sec.~\ref{sec:CSDJ-theory} for $k=0$.

Decomposing $K^{\alpha\beta}$ into the purely timelike, mixed and spacelike pieces $K^{00}$, $K^{0i}$, and $K^{ij}$, respectively, the dispersion equation turns into
\begin{align}
0&=(\vartheta+\vartheta_{1}K^{00})p_{0}^{2}-\vartheta_{1}K^{0i}p^{i}p_{0} \notag \\
&\phantom{{}={}}+\vartheta_{1}K^{ij}p^{i}p^{j}-\vartheta\mathbf{p}^{2}-k\,.
\end{align}
Its positive-energy solution is
\begin{subequations}
\label{eq:dispersion-relation-CSDJ}
\begin{equation}
\omega=\frac{1}{2(\vartheta+\vartheta_{1}K^{00})}\left(\sqrt{\Delta}+\vartheta_{1}K^{0i}p^{i}\right)\,,
\end{equation}
where
\begin{align}
\Delta&=4(\vartheta+\vartheta_{1}K^{00})[\vartheta\mathbf{p}^{2}+k-\vartheta_{1}K^{ij}p^{i}p^{j}] \notag \\
&\phantom{{}={}}+(\vartheta_{1}K^{0i}p^{i})^{2}\,.
\end{align}
\end{subequations}
An analysis of the properties of the latter is complicated by the presence of the background field coefficients. Thus, as we already did for the extended MCSDJ theory in Sec.~\ref{sec:extended-MCSDJ}, we will restrict ourselves to specific configurations of the background field.

Moreover, note that the propagator saturated with external currents, Eq.~\eqref{SPMCSDJA}, is only governed by the topological quantity $\Omega$ as well as the modified dispersion equations:
\begin{equation}
\mathit{SP}=\frac{\mathrm{i}\Omega}{p^2(k-\vartheta p^{2}+\vartheta_{1}\mathcal{K}^{(4)}(p))}\,.
	\label{eq:saturated-propagator-estended-CSDJ}
\end{equation}
We already observed a similar behavior for CSDJ theory in Sec.~\ref{sec:CSDJ-theory}, which is reproduced for a vanishing background field, $K^{\mu\nu}=0$. Nevertheless, the generic behavior is challenging to understand from the latter result, which provides further motivation for us considering special subsets of nonzero coefficients at a time.

\subsubsection{Isotropic timelike configuration}

We set $K^{0i}=K^{ij}=0$ and keep $K^{00}\neq0$ only, which eliminates all potential anisotropies. In this case, the dispersion relation of Eq.~\eqref{eq:dispersion-relation-CSDJ} drastically simplifies to
\begin{equation}
\omega=\sqrt{\frac{\vartheta\mathbf{p}^{2}+k}{\vartheta+\vartheta_{1}K^{00}}}\,.
\label{RDCSD5A1}
\end{equation}
To be able to make a statement on the velocity of signal propagation, for instance, we calculate the group and wavefront velocities in accordance with the definitions of Eq.~\eqref{uguf1}. For the group velocity, we obtain
\begin{equation}
\mathbf{u}_{\mathrm{gr}}=\frac{\vartheta\mathbf{p}}{\sqrt{(\vartheta+\vartheta_{1}K^{00})(\vartheta\mathbf{p}^{2}+k)}}\,,
\end{equation}
whose norm is bounded by 1 from above for nonnegative choices of the parameters $\vartheta_1$, $k$, and $K^{00}$. The wavefront velocity reads
\begin{equation}
u_{\mathrm{fr}}=\sqrt{\frac{\vartheta}{\vartheta+\vartheta_{1}K^{00}}}\,,
\end{equation}
where $u_{\mathrm{fr}}<1$ again for nonnegative choices of the parameters. Subluminal signal propagation is definitely assured in this case. However, there are also regions of the parameter space, e.g., for $\vartheta_1K^{00}<0$ where the front velocity exceeds 1, which implies superluminal regimes.

The saturated propagator of Eq.~\eqref{eq:saturated-propagator-estended-CSDJ} is reformulated as
\begin{equation}
\mathit{SP}=-\frac{\mathrm{i}\Omega}{(\vartheta+\vartheta_1K^{00})(p_0^2-\mathbf{p}^2)(p_0^2-\omega^2)}\,,
\end{equation}
where $\omega$ is given by Eq. \eqref{RDCSD5A1}. The residues evaluated at the poles read
\begin{subequations}
\label{eq:residues-isotropic-timelike-configuration}
\begin{align}
\mathrm{Res}(\mathit{SP})|_{p_0=|\mathbf{p}|}&=\frac{\mathrm{i}\Omega|_{p_0=|\mathbf{p}|}}{2|\mathbf{p}|(k-\vartheta_1K^{00}\mathbf{p}^2)}\,, \\[2ex]
\mathrm{Res}(\mathit{SP})|_{p_0=\omega}&=-\frac{\mathrm{i}\Omega|_{p_0=\omega}}{2\omega(k-\vartheta_1K^{00}\mathbf{p}^2)}\,.
\end{align}
\end{subequations}
Both residues involve the same factor in their denominators, but they have opposite signs. Thus, taking $\omega>0$ and $|\mathbf{p}|>0$, if the imaginary part of the first residue is positive, the latter is negative for the second residue. Unitarity issues at tree-level are then expected to arise for any choice of parameters.

\subsubsection{Anisotropic mixed configuration}
\label{sec:anisotropic-mixed-configuration}

\begin{figure}[t]
	\centering
    \includegraphics[scale=0.7]{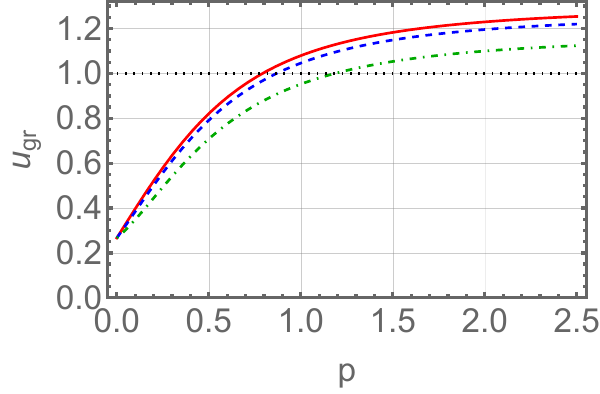}
	\caption{Norm of the group velocity stated in Eq.~\eqref{eq:group-velocity-anisotropic-mixed-case} for the anisotropic configuration of Sec.~\ref{sec:anisotropic-mixed-configuration}. We employed the values $|\mathbf{C}|=0.2$, $k=0.2$, $\vartheta=0.3$, and $\vartheta_1=0.8$ (expressed in suitable units). Furthermore, $\phi=\{0,\pi/6,\pi/3\}$ for the red (plain), blue (dashed), and green (dashed-dotted) curves. The black (dotted) line indicates the speed of light $c_m=1$.}
	\label{fig:group-velocity-spacelike}%
\end{figure}

Here we discard both $K^{00}$ and $K^{ij}$, but we keep $K^{0i}\neq 0$. The dispersion relation is then readily obtained from Eq.~\eqref{eq:dispersion-relation-CSDJ}:
\begin{equation}
\omega=\frac{1}{2\vartheta}\left(\sqrt{(\vartheta_{1}K^{0i}p^{i})^{2}+4\vartheta(\vartheta\mathbf{p}^{2}+k)}+\vartheta_{1}K^{0i}p^{i}\right)\,.
\label{RDCSD5A2}
\end{equation}
For brevity, it is reasonable to define the vector-valued quantity $\mathbf{C}$ with components $C^{i}\equiv K^{0i}$. Then, the group velocity reads
\begin{equation}
	\mathbf{u}_{\mathrm{gr}}=\frac{1}{2\vartheta }\left(\frac{\vartheta _{1}^{2}( \mathbf{C}\cdot \mathbf{p}) \mathbf{C}+4\vartheta ^{2}\mathbf{p}}{\sqrt{\vartheta _{1}^{2}(
			\mathbf{C}\cdot \mathbf{p})^{2}+4\vartheta (\vartheta \mathbf{p}^{2}+k)}}+\vartheta _{1}\mathbf{C}\right)\,,
\end{equation}
whose magnitude corresponds to a lengthy expression. Let $\phi$ be the angle between $\mathbf{C}$ and $\mathbf{p}$ such that
\begin{equation}
\mathbf{C}\cdot\mathbf{p}=\vert \mathbf{C}\vert\vert \mathbf{p}\vert \cos\phi\,,
\end{equation}
which allows us to express the norm of the group velocity as
\begin{align}
\label{eq:group-velocity-anisotropic-mixed-case}
u_{\mathrm{gr}}&=\frac{1}{2\vartheta}\bigg(\frac{\vartheta _{1}^{2}( 8\vartheta ^{2}+\vartheta _{1}^{2}\mathbf{C}%
			^{2}) \mathbf{C}^{2}\mathbf{p}^{2}\cos ^{2}\phi +16\vartheta
			^{4}\vert \mathbf{p}\vert ^{2}}{\vartheta _{1}^{2}\mathbf{C}^{2}%
			\mathbf{p}^{2}\cos ^{2}\phi +4\vartheta(\vartheta \mathbf{p}^{2}+k)} \notag \\
&\phantom{{}={}}+\vartheta _{1}^{2}\mathbf{C}^{2}+\frac{2\vartheta _{1}( \vartheta _{1}^{2}\mathbf{C}%
			^{2}+4\vartheta ^{2}) \vert \mathbf{C}\vert \vert\mathbf{p}\vert \cos \phi }{\sqrt{\vartheta_{1}^{2}\mathbf{C}^{2}%
				\mathbf{p}^{2}\cos^{2}\phi +4\vartheta(\vartheta \mathbf{p}^{2}+k)}}\bigg)^{1/2}\,.
\end{align}
The wavefront velocity is comparably simple:
\begin{equation}
u_{\mathrm{fr}}=\frac{1}{2\vartheta}\left[\sqrt{\vartheta_{1}^{2}\vert\mathbf{C}\vert^{2}\cos^{2}\phi+4\vartheta^{2}}+\vartheta_{1}\vert \mathbf{C}\vert \cos\phi\right]\,.
\end{equation}
The behavior for the absolute value of the group velocity as a function of the momentum is presented in Fig.~\ref{fig:group-velocity-spacelike} for various angles $\phi$, which shows that its norm can exceed~1. The wavefront velocity can also be greater than 1 when the angle $\phi$ is chosen appropriately. These characteristics indicate superluminal signal propagation.
%%

Here, the propagator of Eq.~\eqref{eq:saturated-propagator-estended-CSDJ}, which has been contracted with conserved currents, takes the form
\begin{equation}
\mathit{SP}=\frac{\mathrm{i}\Omega}{p^2(k-\vartheta p^2+2\vartheta_1p_0\mathbf{C}\cdot\mathbf{p})}\,.
\end{equation}
The residues at the poles read
\begin{subequations}
\begin{align}
\mathrm{Res}(\mathit{SP})|_{p_0=|\mathbf{p}|}&=\frac{\mathrm{i}\Omega|_{p_0=|\mathbf{p}|}}{2|\mathbf{p}|(k+2\vartheta_1\mathbf{C}\cdot\mathbf{p}|\mathbf{p}|)}\,, \\[2ex]
\mathrm{Res}(\mathit{SP})|_{p_0=\omega}&=-\frac{\mathrm{i}\Omega|_{p_0=\omega}}{2[\vartheta_1\mathbf{C}\cdot\mathbf{p}(\omega^2+\mathbf{p}^2)+k\omega]}\,,
\end{align}
\end{subequations}
with $\omega$ given in Eq. \eqref{RDCSD5A2}. Now, this case is more challenging to analyse compared to the previous ones. However, it is quite palpable that the denominators can take positive as well as negative values for specific choices of the parameters and the momentum. Since $\Omega$ is not positive definite, too, unitarity violations are also inherent to this particular sector.

\subsubsection{Anisotropic purely spacelike configuration}
\label{sec:anisotropic-purely-spacelike-configuration}

We now discard the isotropic coefficient $K^{00}$ as well as the mixed ones $K^{0i}$ and only keep $K^{ij}\neq0$. The dispersion relation results from Eq.~\eqref{eq:dispersion-relation-CSDJ}:
\begin{equation}
\omega=\sqrt{\left(\delta^{ij}-\frac{\vartheta_1}{\vartheta}K^{ij}\right)p^ip^j+\frac{k}{\vartheta}}\,,
\label{RDCSD5A3}
\end{equation}
and is clearly anisotropic. Note the similarity to Eq.~\eqref{prd4G}, but the absence of fourth-order terms in the momentum. In the following, it will be beneficial to employ the quantities defined under Eq.~\eqref{eq:group-velocity-MCS-spacelike}. Now, the associated group velocity reads
\begin{equation}
\mathbf{u}_{\mathrm{gr}}=\frac{1}{\omega}\left(\mathbf{p}-\frac{\vartheta_{1}}{\vartheta}\mathbf{K}_p\right)\,,
\end{equation}
and its norm is given by
\begin{equation}
\label{eq:purely-spacelike-group-velocity}
u_{\mathrm{gr}}=\frac{1}{\omega}\sqrt{\mathbf{p}^{2}
-2\frac{\vartheta_{1}}{\vartheta}K_{pp}+\left(\frac{\vartheta_{1}}{\vartheta}\right)^{2}\mathbf{K}_p^2}\,.
\end{equation}
For the wavefront velocity, we obtain
\begin{equation}
\label{eq:purely-spacelike-front-velocity}
u_{\mathrm{fr}}=\sqrt{1-\frac{\vartheta_{1}}{\vartheta}K^{ij}\hat{p}^{i}\hat{p}^{j}}\,,
\end{equation}
where $\hat{p}^i$ denotes the $i$-th component of the unit vector $\hat{\mathbf{p}}\equiv \mathbf{p}/p$ pointing along the momentum. The latter is an anisotropic nondispersive deviation from the conventional case $u_{\mathrm{fr}}=1$. Since the expression of Eq.~\eqref{eq:purely-spacelike-group-velocity}, in particular, is nontransparent, a reasonable possibility is to choose a specific parameterization of the background field. We express the latter in terms of the components of two spatial vectors $\mathbf{D}$ and~$\mathbf{F}$:
\begin{equation}
K^{ij}=\frac{1}{2}(D^{i}F^{j}+D^{j}F^{i})\,.
\label{eq:parameterization-spatial-K}
\end{equation}
This implies
\begin{equation}
\label{eq:parameterization-spatial-K-front-velocity}
u_{\mathrm{fr}}=\sqrt{1-\frac{\vartheta_{1}}{\vartheta}(\mathbf{D}\cdot\hat{\mathbf{p}})(\mathbf{F}\cdot\hat{\mathbf{p}})}\,.
\end{equation}
In the plane, let $\Psi$ be the angle between the vectors $\mathbf{D}$ and $\mathbf{F}$: $\mathbf{D}\cdot\mathbf{F}=|\mathbf{D}||\mathbf{F}|\cos\Psi$.
Furthermore, we introduce angles $\alpha$ and $\beta$ as follows:
\begin{equation}
\mathbf{D}\cdot\mathbf{p}=\vert \mathbf{D}\vert\vert \mathbf{p}\vert \cos\alpha\,,\quad \mathbf{F}\cdot\mathbf{p}=\vert \mathbf{F}\vert\vert \mathbf{p}\vert \cos\beta\,.
\end{equation}
\begin{figure}
\subfloat[]{\label{fig:spacelike-vectors-configuration-1}\includegraphics{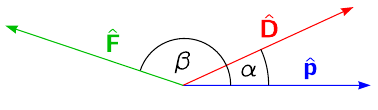}} \\
\subfloat[]{\label{fig:spacelike-vectors-configuration-2}\includegraphics{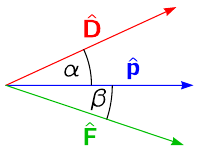}}
\caption{Configurations of unit vectors $\hat{\mathbf{p}}$, $\hat{\mathbf{D}}$, and $\hat{\mathbf{F}}$ that must be distinguished from each other. \protect\subref{fig:spacelike-vectors-configuration-1} Momentum vector outside of sector formed from $\hat{\mathbf{D}}$ and $\hat{\mathbf{F}}$. \protect\subref{fig:spacelike-vectors-configuration-2} Momentum vector enclosed by $\hat{\mathbf{D}}$ and $\hat{\mathbf{F}}$.}
\label{fig:spacelike-vectors-configurations}
\end{figure}%%
To express the angle between $\mathbf{D}$ and $\mathbf{F}$ in terms of $\alpha$ and $\beta$, we must distinguish between the two configurations illustrated in Fig.~\ref{fig:spacelike-vectors-configurations}. This is possible by defining the quantity
\begin{equation}
\Xi\equiv \frac{p^1D^2-p^2D^1}{p^1F^2-p^2F^1}\,.
\end{equation}
The latter is, in principle, the ratio of the third components of cross products between $\{\mathbf{p},\mathbf{D}\}$ and $\{\mathbf{p},\mathbf{F}\}$ when these vectors are extended to an auxiliary third spatial dimension. Due to the arrangement of the vectors with respect to each other, these cross products are of equal signs for the configuration of Fig.~\ref{fig:spacelike-vectors-configuration-1}, where their signs disagree for the configuration of Fig.~\ref{fig:spacelike-vectors-configuration-2}. Therefore, $\Xi$ provides a means to distinguish between these two arrangements of $\{\mathbf{p},\mathbf{D},\mathbf{F}\}$. Then,
\begin{equation}
\Psi=\Psi(\alpha,\beta)=\left\{\begin{array}{ccl}
|\alpha-\beta| & \text{for} & \Xi\geq 0 \\
\alpha+\beta & \text{for} & \Xi<0,\alpha+\beta<\pi \\
2\pi-(\alpha+\beta) & \text{for} & \Xi<0,\alpha+\beta>\pi\,. \\
\end{array}\right.
\end{equation}
Now, in terms of the angles $\alpha$ and $\beta$, the wavefront velocity of Eq.~\eqref{eq:parameterization-spatial-K-front-velocity} can be cast into
\begin{equation}
u_{\mathrm{fr}}=\sqrt{1-\frac{\vartheta_{1}}{\vartheta}\vert\mathbf{D}\vert\vert \mathbf{F}\vert \cos\alpha\cos\beta}\,.
\end{equation}
Let us assume that $\vartheta_1>0$ and $\vartheta>0$. If either $\alpha$ or $\beta$ lies within $(\pi/2,\pi)$, the front velocity exceeds 1. Otherwise, it is less than 1. The situation changes, of course, when the signs of $\vartheta_1$ and $\vartheta$ are different.

\begin{figure}
	\centering
    \includegraphics[scale=0.7]{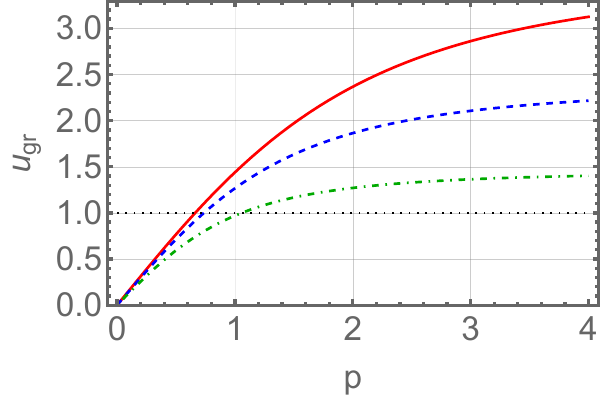}
	\caption{Norm of the group velocity given in Eq.~\eqref{eq:group-velocity-anisotropic-spacelike-case} for the anisotropic configuration of Sec.~\ref{sec:anisotropic-purely-spacelike-configuration}. Here, we used $\vartheta=\vartheta_1=0.5$ and $|\mathbf{D}|=|\mathbf{F}|=0.9$ as well as $k=1.2$ (expressed in suitable units). Moreover, $\alpha\in \{0,\pi/6,\pi/3\}$ and $\beta=\pi$ for the red (plain), blue (dashed), and green (dashed-dotted) curves. The black (dotted) line represents the speed of light $c_m=1$.}
	\label{fig:group-velocity-spacelike-2}
\end{figure}

Now, based on the parameterization of Eq.~\eqref{eq:parameterization-spatial-K}, the norm of the group velocity of Eq.~\eqref{eq:purely-spacelike-group-velocity} reads
\begin{align}
\label{eq:group-velocity-anisotropic-spacelike-case}
u_{\mathrm{gr}}&=\left(\vert \mathbf{p}\vert ^{2}+(k/\vartheta)-(\vartheta_{1}/\vartheta)\vert \mathbf{D}\vert \vert\mathbf{F}\vert\vert \mathbf{p}\vert^{2}\cos\alpha\cos\beta\right)^{-1/2} \notag \\
&\phantom{{}={}}\times \Big(\vert \mathbf{p}\vert ^{2} - (2\vartheta_{1}/\vartheta) \vert\mathbf{D}\vert \vert \mathbf{F}\vert \vert \mathbf{p}\vert ^{2}\cos\alpha\cos\beta
\notag \\
&\phantom{{}={}}\hspace{0.5cm}+\vartheta_{1}^{2}/(4\vartheta^{2})\vert \mathbf{D}\vert^{2}\vert \mathbf{F}\vert ^{2}\vert \mathbf{p}\vert^{2}[\cos^{2}\beta+\cos^{2}\alpha \notag \\
&\phantom{{}={}}\hspace{2.4cm}+2 \cos\alpha\cos\beta \cos\Psi]\Big)^{1/2}\,,
\end{align}
as a function of the angles $\alpha$, $\beta$, and $\Psi$. The behavior of the norm of the group velocity is illustrated in Fig.~\ref{fig:group-velocity-spacelike-2}.
%%
According to the latter graph, we see that for different choices of $\alpha$ and $\beta$ the group velocity can exceed 1 above certain momenta, which implies superluminal signal propagation. The group velocity enters this regime at even lower momenta when the vectors $\mathbf{D}$ and $\mathbf{F}$ are antiparallel. Thus, we conclude that the generic dispersion relation of Eq.~\eqref{eq:dispersion-relation-CSDJ} corresponds to a propagating mode, whose dynamics comes from the presence of the higher-derivative terms. The specific cases of Eqs.~\eqref{RDCSD5A1}, \eqref{RDCSD5A2}, and \eqref{RDCSD5A3} exhibit radicands that are not necessarily positive definite. Therefore, they may be plagued by instabilities for certain values of the parameters involved. In this scenario, it is worthwhile to emphasize again that the pure CS Lagrangian, i.e., the regime $k\neq0,$ $\vartheta=0=\vartheta_{1}$ does not exhibit propagating modes.

Finally, the saturation of Eq.~\eqref{eq:saturated-propagator-estended-CSDJ} takes the rather simple form:
\begin{equation}
\mathit{SP}=-\frac{\mathrm{i}\Omega}{\vartheta (p_0^2-\mathbf{p}^2)(p_0^2-\omega^2)}\,,
	\label{SPMCSDJA}
\end{equation}
and the residues evaluated at the positive-energy dispersion relations are readily obtained as
\begin{subequations}
\begin{align}
\mathrm{Res}(\mathit{SP})|_{p_0=|\mathbf{p}|}&=\frac{\mathrm{i}\Omega|_{p_0=|\mathbf{p}|}}{2|\mathbf{p}|(k-\vartheta_1K_{pp})}\,, \\[2ex]
\mathrm{Res}(\mathit{SP})|_{p_0=\omega}&=-\frac{\mathrm{i}\Omega|_{p_0=\omega}}{2\omega(k-\vartheta_1K_{pp})}\,.
\end{align}
\end{subequations}
Here, the situation is comparable to that for the isotropic, timelike case, cf.~Eq.~\eqref{eq:residues-isotropic-timelike-configuration}. Taking $\omega>0$ and $|\mathbf{p}|>0$, independently of the sign of the common denominator and that of $\Omega$, one imaginary part becomes negative. This behavior indicates issues with unitarity at tree-level for one of the two poles.

\subsubsection{Classical field equations and solutions}

\begin{figure}
\includegraphics[scale=0.5]{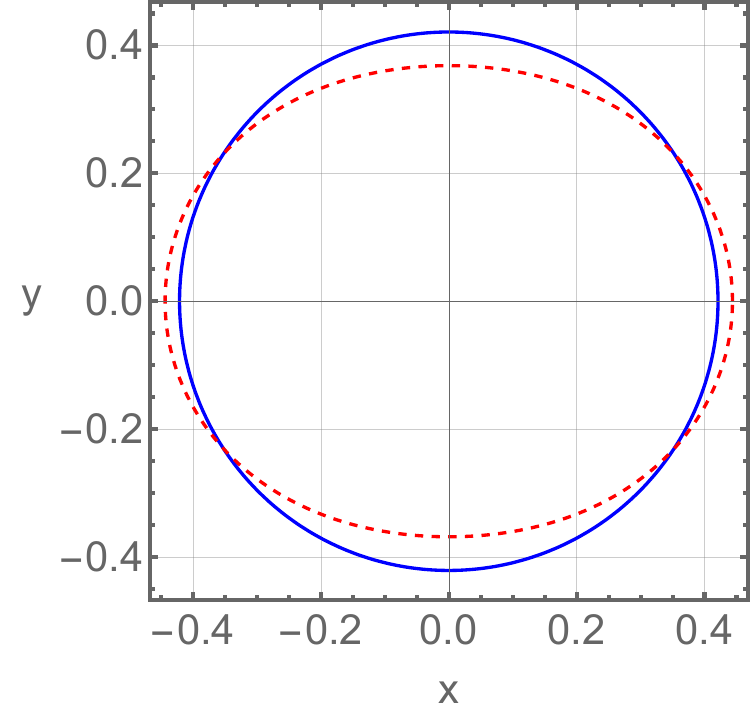}
\caption{Polar plot of Eq.~\eqref{eq:magnetic-field-extended-CSDJ} as a function of the angle $\alpha$. The value $\eta_2=1/2$ is used for the red (dashed) curve where the blue (plain) circle shows the behavior for $\eta_2=0$, for comparison.}
\label{fig:magnetic-field-polarplot}
\end{figure}

The current section is devoted to an extension of the analysis performed in Sec.~\ref{sec:field-equations-solutions-CSDJ} for CSDJ theory. To examine the impact of higher-derivative anisotropic terms on the solutions of the classical field equations explored in the latter section, we start from the extended CSDJ Lagrange density of Eq.~\eqref{LMCSD5A}, which is coupled to the external, conserved current $J^{\mu}$:
\begin{equation}
	\mathcal{L}=\mathcal{L}_{\mathrm{CSDJ}}'-J_{\mu}A^{\mu}\,.
\end{equation}
The Euler-Lagrange equations stated in Eq.~\eqref{E-L} lead to the following field equations:
\begin{align}
	J_{\mu}&=k\varepsilon _{\mu \nu \rho }\partial ^{\nu }A{}^{\rho }+\vartheta
	\varepsilon _{\mu \nu \rho }\partial ^{\nu }\square A{}^{\rho } \notag \\
&\phantom{{}={}}+\vartheta_{1}\varepsilon _{\mu \nu \rho }(K^{\lambda \beta }\partial _{\lambda}\partial _{\beta }) \partial ^{\nu }A{}^{\rho }\,,
\end{align}
which for $\mu=0$ yields
\begin{equation}
	kB+\vartheta \square B+\vartheta _{1}(K^{\lambda \beta }\partial_{\lambda }\partial _{\beta }) B=\rho\,,
\end{equation}
with the planar magnetic field $B=\varepsilon^{ij}\partial ^{i}A{}^{j}$. We already know that the DJ term is able to turn the solutions of the field equations of CS theory into dynamical modes. Now, we examine the effect of anisotropies on the solution for the magnetic field, which is modified because of the higher-derivative term. To do so, we consider the wave equation,
\begin{align}
	\left[ \square +\frac{k}{\vartheta }+\frac{\vartheta _{1}}{\vartheta }(K^{\lambda \beta }\partial _{\lambda }\partial _{\beta }) \right] B& =\frac{\rho }{\vartheta }\,,
\end{align}
which in the static limit reads
\begin{equation}
\left[ \nabla ^{2}-\frac{k}{\vartheta }-\frac{\vartheta _{1}}{\vartheta }(K^{ij}\partial _{i}\partial _{j}) \right] B =-\frac{\rho }{\vartheta }\,.
\end{equation}
The Green's function must fulfill
\begin{equation}
	\left[ \nabla ^{2}-\frac{k}{\vartheta }-\frac{\vartheta _{1}}{\vartheta }(K^{ij}\partial _{i}\partial _{j})\right]G(\mathbf{R})=\delta^{(2)}(\mathbf{R})\,,
\end{equation}
which, based on Eq.~\eqref{GreenF1a}, implies
\begin{equation}
 G(\mathbf{p})=-\frac{1}{\mathbf{p}^{2}+m^{2}+(\vartheta _{1}/\vartheta)(K^{ij}p^{i}p^{j})}\,.
\end{equation}
Computing the Fourier transformation of the latter is challenging for a generic $K^{ij}$. Hence, we take the choice of a $K^{ij}$ expressed in terms of a single two-component vector $\mathbf{T}$ as follows:
\begin{equation}
\label{eq:decomposition-background-T-vector}
	K^{ij}=T^{i}T^{j}\,,
\end{equation}
such that
\begin{equation}
	G(\mathbf{p})=-\frac{1}{\mathbf{p}^{2}+m^{2}+(\vartheta_{1}/\vartheta)(\mathbf{T}\cdot \mathbf{p}) ^{2}}\,.
\end{equation}
Further, we take into account that
\begin{equation}
	\mathbf{T}\cdot \mathbf{p}=\vert \mathbf{T}\vert\vert\mathbf{p}\vert \cos( \alpha -\phi)\,,
\end{equation}
\begin{figure}[t]
	\centering
	\includegraphics{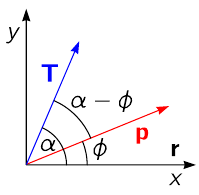}
	\caption{Vectors $\mathbf{T}$ of Eq.~\eqref{eq:decomposition-background-T-vector}, the momentum $\mathbf{p}$, and the position $\mathbf{r}$ in the plane.}%
	\label{vetores}
\end{figure}%%
where $\alpha$ is the angle between $\mathbf{T}$ and the position $\mathbf{r}$, see Fig.~\ref{vetores}. Then, the Green's function in momentum space reads
\begin{subequations}
\begin{equation}
\label{eq:green-function-spacelike-momentum-space}
	G(\mathbf{p}) =-\frac{1}{\mathbf{p}^{2}+m^{2}+\eta_{2}^{2}\mathbf{p}^{2}\cos ^{2}(\alpha -\phi)}\,,
\end{equation}
with the dimensionless parameter
\begin{equation}
	\eta_{2}^{2}=\frac{\vartheta _{1}}{\vartheta }\mathbf{T}^{2}\,.
\end{equation}
\end{subequations}
The Fourier transform of the Green's function in Eq.~\eqref{eq:green-function-spacelike-momentum-space} is challenging to evaluate analytically in its full generality. To be able to understand the deviation from the regime with $\eta_2=0$, we assume that $\eta_{2}^{2}\ll 1$. Then, the following expansion is justified:
\begin{equation}
\label{eq:green-function-spacelike-momentum-space-expanded}
	G(\mathbf{p})\simeq -\frac{1}{\mathbf{p}^{2}+m^{2}}+\frac{\eta_{2}^{2}\mathbf{p}^{2}}{(\mathbf{p}^{2}+m^{2})^{2}}\cos^{2}(\alpha-\phi)\,.
\end{equation}
Even the computation of the Fourier transform of the latter expanded Green's function is lengthy and App.~\ref{eq:green-function-spacelike-configuration-space-details} provides some details. Any reader who may not be interested in the technicalities can skip that part and jump to the final result, which is
\begin{align}
	G(\mathbf{R})&=-\frac{1}{2\pi} K_{0}(mR)\left(1-\frac{\eta_{2}^{2}}{2}\right) \notag\\
&\phantom{{}={}}-\frac{\eta_{2}^{2}}{4\pi}(mR)K_{1}(mR)\cos^2\alpha \,,
	\label{eq:green-function-spacelike-configuration-space}
\end{align}
where $K_1(x)$ is the first-order modified Bessel function of the second kind.
We consider a pointlike charge distribution at the origin: $\rho (\mathbf{r}^{\prime})=q\delta^{(2)}(\mathbf{r}^{\prime})$. Inserting Eq.~\eqref{eq:green-function-spacelike-configuration-space} as well as the latter into
\begin{equation}
	B(\mathbf{r})=-\frac{1}{\vartheta}\int \mathrm{d}^{2}r^{\prime }\,G(\mathbf{r}-\mathbf{r}^{\prime })\rho (\mathbf{r}^{\prime})\,,
\end{equation}
leads to the magnetic field that the charge generates:
\begin{align}
	B(\mathbf{r}) &=\frac{q}{2\pi\vartheta}\left[\left(1-\frac{\eta_{2}^{2}}{2}\right) K_{0}(mr)\right. \notag \\
	&\phantom{{}={}}\hspace{0.8cm}\left.{} +\frac{\eta_{2}^{2}}{2}mr K_{1}(mr)\cos^2\alpha \right]\,.
\label{eq:magnetic-field-extended-CSDJ}
\end{align}
The dependence of the magnetic field on the angle $\alpha$ is a clear manifestation of the anisotropy induced by the Lorentz-violating higher-derivative term at the end of Eq.~\eqref{LMCSD5A}. A polar plot of Eq.~\eqref{eq:magnetic-field-extended-CSDJ} is presented in Fig.~\ref{fig:magnetic-field-polarplot} for $\eta_2=1/2$. The latter value is already large considering that Eq.~\eqref{eq:magnetic-field-extended-CSDJ} is a perturbative result that holds for $\eta_2\ll 1$. However, this value is chosen for illustrative purpose, as otherwise the deviation from the result for $\eta_2=0$ would be difficult to perceive.

Finally, as we did in Eq.~\eqref{eq:integration-magnetic-field-CS}, we integrate the magnetic field of Eq.~\eqref{eq:magnetic-field-extended-CSDJ} over the entire plane. Note that the angle $\alpha$ is kept fixed and does not correspond to the angle in polar coordinates integrated over. By using Eq.~\eqref{eq:auxiliary-integral} and Eq.~(6.521.11) of Ref.~\cite{Gradshteyn:2007} we readily obtain
\begin{equation}
\int\mathrm{d}^2r\,B(\mathbf{r})=\frac{q}{k}\left[1+\frac{\eta_2^2}{2}\cos(2\alpha)\right]\,,
\end{equation}
which is an anisotropic perturbation of the result in Eq.~\eqref{eq:integration-magnetic-field-CS} in the regime $\eta_2\ll 1$.
%%

\section{Duality}
\label{sec:duality}

In Sec.~\ref{sec:MCS-electrodynamics-higher-derivatives} we have already emphasized the presence of a possible duality between MCSDJ theory of Eq.~\eqref{MCSDJ1} and MCSH theory stated in Eq.~\eqref{sec:MCSH-theory}. Dualities are interesting, as they always unravel fundamental relationships between apparently very different physics. In the following, another duality will be pointed out, which is based on some of the findings in Ref.~\cite{Deser:1984kw}. Actually, it is possible to recast the MCSDJ Lagrange density of Eq.~\eqref{LMCSD5B} into a different, but equivalent form:
\begin{subequations}
\begin{align}
S''_{\mathrm{MCSDJ}}&=\int\mathrm{d}^3x\,\mathcal{L}''_{\mathrm{MCSDJ}}\,, \displaybreak[0]\\[2ex]
\label{eq:lagrangian-MCSDJ-alternative}
\mathcal{L}''_{\mathrm{MCSDJ}}&=-\frac{1}{2}\tilde{F}^{\mu}\tilde{F}_{\mu}+\frac{k}{2}\tilde{F}^{\mu}A_{\mu} \notag \\
&\phantom{{}={}}+\frac{\vartheta}{2}\tilde{F}^{\mu}\square A_{\mu}+\frac{\vartheta_1}{2}\tilde{F}^{\mu}\hat{\mathcal{K}}^{(4)}A_{\mu}\,,
\end{align}
expressed in terms of the vector-valued quantity
\begin{equation}
\tilde{F}^{\mu}\equiv \frac{1}{2}\varepsilon^{\mu\alpha\beta}F_{\alpha\beta}=\varepsilon^{\mu\alpha\beta}\partial_{\alpha}A_{\beta}\,,
\end{equation}
\end{subequations}
with $[\tilde{F}^{\mu}]=1/2$. In this context it is worthwhile to note that
\begin{equation}
\tilde{F}^{\mu}\tilde{F}_{\mu}=\frac{1}{2}F_{\mu\nu}F^{\mu\nu}\,.
\end{equation}
To obtain the field equations associated with Eq.~\eqref{eq:lagrangian-MCSDJ-alternative}, we can either resort to the Euler-Lagrange equations of Eq.~\eqref{E-L} or, alternatively, compute the variation of the action for $A_{\varrho}$ directly. Let us pursue the second possibility:
\begin{align}
\frac{\delta S''_{\mathrm{MCSDJ}}}{\delta A_{\varrho}}&=-\tilde{F}_{\kappa}\frac{\delta \tilde{F}^{\kappa}}{\delta A^{\varrho}}+\frac{k}{2}\tilde{F}^{\varrho}-\frac{k}{2}\partial_{\alpha}A_{\mu}\varepsilon^{\mu\alpha\varrho} \notag \\
&\phantom{{}={}}+\frac{\vartheta}{2}\square \tilde{F}^{\varrho}-\frac{\vartheta}{2}\partial_{\alpha}\square A_{\mu}\varepsilon^{\mu\alpha\varrho} \notag \\
&\phantom{{}={}}+\frac{\vartheta_1}{2}\hat{\mathcal{K}}^{(4)} \tilde{F}^{\varrho}-\frac{\vartheta_1}{2}\partial_{\alpha}\hat{\mathcal{K}}^{(4)} A_{\mu}\varepsilon^{\mu\alpha\varrho} \notag \\
&=\partial_{\alpha}\tilde{F}_{\kappa}\varepsilon^{\kappa\alpha\varrho}+k\tilde{F}^{\varrho}+\vartheta\square \tilde{F}^{\varrho}+\vartheta_1\hat{\mathcal{K}}^{(4)}\tilde{F}^{\varrho}\,,
\end{align}
which implies
\begin{equation}
\label{eq:field-equations-MCSDJ-theory-alternative}
0=-\varepsilon^{\varrho\alpha\kappa}\partial_{\alpha}\tilde{F}_{\kappa}+k\tilde{F}^{\varrho}+\vartheta\square \tilde{F}^{\varrho}+\vartheta_1\hat{\mathcal{K}}^{(4)}\tilde{F}^{\varrho}\,.
\end{equation}
We also propose a self-dual Lagrange density, which is an extension of that given in Ref.~\cite{Deser:1984kw}:
\begin{subequations}
\begin{align}
S_{\mathrm{SD}}&=\int\mathrm{d}^3x\,\mathcal{L}_{\mathrm{SD}}\,, \displaybreak[0]\\[2ex]
\mathcal{L}_{\mathrm{SD}}&=\frac{1}{2}f_{\mu}f^{\mu}+\frac{\vartheta}{2k}f_{\mu}\square f^{\mu} \notag \\
&\phantom{{}={}}+\frac{\vartheta_1}{2k}f_{\mu}\hat{\mathcal{K}}^{(4)}f^{\mu}-\frac{1}{2k}\varepsilon^{\alpha\beta\gamma}f_{\alpha}\partial_{\beta}f_{\gamma}\,,
\end{align}
\end{subequations}
where $f_{\mu}$ is a vector field with $[f_{\mu}]=3/2$. The variation of the self-dual theory is readily obtained as
\begin{align}
\frac{\delta S_{\mathrm{SD}}}{\delta f_{\varrho}}&=f^{\varrho}+\frac{\vartheta}{k}\square f^{\varrho}+\frac{\vartheta_1}{k}\hat{\mathcal{K}}^{(4)}f^{\varrho} \notag \\
&\phantom{{}={}}-\frac{1}{2k}\varepsilon^{\varrho\beta\gamma}\partial_{\beta}f_{\gamma}+\frac{1}{2k}\varepsilon^{\alpha\beta\varrho}\partial_{\beta}f_{\alpha}\,,
\end{align}
leading to the field equations
\begin{equation}
\label{eq:field-equations-selfdual-theory}
0=f^{\varrho}+\frac{\vartheta}{k}\square f^{\varrho}+\frac{\vartheta_1}{k}\hat{\mathcal{K}}^{(4)}f^{\varrho}-\frac{1}{k}\varepsilon^{\varrho\beta\gamma}\partial_{\beta}f_{\gamma}\,.
\end{equation}
By comparing Eq.~\eqref{eq:field-equations-MCSDJ-theory-alternative} to Eq.~\eqref{eq:field-equations-selfdual-theory}, a duality between MCSDJ theory and the self-dual theory is evident. Note that a reformulation of the self-dual theory is possible as follows:
\begin{align}
\mathcal{L}_{\mathrm{SD}}&=-\frac{1}{2k}\varepsilon^{\mu\nu\varrho}f_{\mu}\partial_{\nu}f_{\varrho}+\frac{1}{2}f_{\mu}f^{\mu}+\frac{\vartheta}{2k}f_{\mu}\square f^{\mu} \notag \\
&\phantom{{}={}}+\frac{\vartheta_1}{2k}f_{\mu}\hat{\mathcal{K}}^{(4)}f^{\mu} \notag \displaybreak[0]\\
&=\left(\frac{\mathrm{i}}{\sqrt{2k}}f_{\mu}\right)\mathrm{i}(\mathrm{i}\varepsilon^{\mu\varrho\nu})\partial_{\nu}\left(\frac{\mathrm{i}}{\sqrt{2k}}f_{\varrho}\right) \notag \displaybreak[0]\\
&\phantom{{}={}}-k\left(\frac{\mathrm{i}}{\sqrt{2k}}f_{\mu}\right)\eta^{\mu\varrho}\left(\frac{\mathrm{i}}{\sqrt{2k}}f_{\varrho}\right) \notag \displaybreak[0]\\
&\phantom{{}={}}-\vartheta\left(\frac{\mathrm{i}}{\sqrt{2k}}f_{\mu}\right)\eta^{\mu\varrho}\square\left(\frac{\mathrm{i}}{\sqrt{2k}}f_{\varrho}\right) \notag \displaybreak[0]\\
&\phantom{{}={}}-\vartheta_1\left(\frac{\mathrm{i}}{\sqrt{2k}}f_{\mu}\right)\eta^{\mu\varrho}\hat{\mathcal{K}}^{(4)}\left(\frac{\mathrm{i}}{\sqrt{2k}}f_{\varrho}\right)\,.
\end{align}
The latter bears a major resemblance to a modified Dirac theory in $(2+1)$ spacetime dimensions of the form
\begin{equation}
\mathcal{L}_D'=\overline{\psi}\gamma^{\nu}\mathrm{i}\partial_{\nu}\psi-m_{\psi}\overline{\psi}\psi-\frac{1}{M}\overline{\psi}\square\psi-\overline{\psi}\hat{m}^{(4)}\psi\,,
\label{eq:modified-Dirac-theory}
\end{equation}
with a two-component Dirac spinor field $\psi$, its Dirac conjugate $\overline{\psi}=\psi^{\dagger}\gamma^0$, and the Dirac matrices $\gamma^{\mu}$ taken as the three Pauli matrices $\sigma^i$: $\gamma^0=\sigma^1$, $\gamma^1=\sigma^2$, and $\gamma^2=\sigma^3$. Note that $[\psi]=[\overline{\psi}]=1$ in $(2+1)$ spacetime dimensions. Moreover, $m_{\psi}$ is the fermion mass, $M$ another mass scale, and $\hat{m}^{(4)}=m^{(4)\mu\nu}(\mathrm{i}\partial_{\mu})(\mathrm{i}\partial_{\nu})$ the $(2+1)$-dimensional analog of a scalar higher-derivative operator of the Dirac fermion sector in the nonminimal SME~\cite{Kostelecky:2013rta}. Now, the following identifications can be made:
\begin{subequations}
\begin{align}
\overline{\psi}&\leftrightarrow \frac{\mathrm{i}}{\sqrt{2k}}f_{\mu}\,,\quad \psi\leftrightarrow \frac{\mathrm{i}}{\sqrt{2k}}f_{\mu}\,, \\[2ex]
\gamma^{\nu}&\leftrightarrow (\mathrm{i}\varepsilon^{\mu\varrho})^{\nu}\,,\quad \mathds{1}_2\leftrightarrow (\eta^{\mu\varrho})\,,\quad m_{\psi}\leftrightarrow k\,, \\[2ex] \frac{1}{M}&\leftrightarrow \vartheta\,,\quad \hat{m}^{(4)}\leftrightarrow \vartheta_1\hat{\mathcal{K}}^{(4)}\,.
\end{align}
\end{subequations}
Explicitly, the Dirac matrices are identified with
\begin{subequations}
\begin{align}
\gamma^0&\leftrightarrow\begin{pmatrix}
0 & 0 & 0 \\
0 & 0 & \mathrm{i} \\
0 & -\mathrm{i} & 0 \\
\end{pmatrix}\,,\quad \gamma^1\leftrightarrow\begin{pmatrix}
0 & 0 & -\mathrm{i} \\
0 & 0 & 0 \\
\mathrm{i} & 0 & 0 \\
\end{pmatrix}\,, \\[1ex]
\gamma^2&\leftrightarrow\begin{pmatrix}
0 & \mathrm{i} & 0 \\
-\mathrm{i} & 0 & 0 \\
0 & 0 & 0 \\
\end{pmatrix}\,.
\end{align}
\end{subequations}
The $(3\times 3)$ matrices on the right-hand sides form the adjoint (spin-1) representation of the $\mathfrak{su}(2)$ algebra. This makes sense, as Eq.~\eqref{eq:field-equations-MCSDJ-theory-alternative} is a spin-1 field theory. So there is a duality between the operator $\hat{\mathcal{K}}^{(4)}$ of Eq.~\eqref{kappa2}, which naturally occurs in extensions of CS theory in $(2+1)$ spacetime dimensions, and the $(2+1)$-dimensional equivalent of $\hat{m}^{(5)}$ being part of the nonminimal SME fermion sector. To the best of our knowledge, such a relationship has not been pointed out in the literature before.

\section{Application: Quantum Hall effect}
\label{sec:application-quantum-Hall-effect}

Chern-Simons electromagnetism in $(2+1)$ spacetime dimensions inspired us to introduce a number of models of an extended electrodynamics that incorporates violations of $\mathit{SO}(2,1)$ invariance. To get an overview of possible effects we investigated the modified mode structures where particular emphasis was put on identifying subluminal and superluminal regimes in the parameter spaces of the models. After gaining some theoretical understanding of the models and their properties, the next sensible step is to find applications. Note that the parameterization of Lorentz violation via the SME has already found its way into the description of certain condensed-matter systems and effects in material media, see, e.g., Refs.~\cite{Ajaib:2012wq,Silva:2020dli,Pedro2021}.

Interestingly, the concept of $\mathit{SO}(2,1)$ symmetry breaking potentially opens pathways beyond those already charted. A $(2+1)$-dimensional electrodynamics is obviously expected to be a theoretical playground for planar condensed-matter systems. In particular, CS theory is known to play a significant role in an effective description of both the integer and the fractional quantum Hall effect (QHE) \cite{Tong:2016kpv} where our focus will be on the first. Since we have been analyzing the physics of a Lorentz-violating planar electrodynamics, our initial interest is to find the potential impact of Lorentz violation on this celebrated phenomenon.

Von Klitzing discovered the integer QHE in 1980 \cite{vonKlitzing:1980pdk} and its theoretical description involves both quantum mechanical and profound geometrical concepts \cite{Laughlin:1981jd,Chruscinski:2004}. The theoretical foundation of this intriguing phenomenon is a two-dimensional electron gas and the associated nonrelativistic dispersion relation, i.e., the electron Hamiltonian is usually taken as isotropic. We now would like to incorporate anisotropies and explore what is their impact on the QHE. We propose
\begin{equation}
H=\frac{1}{2m_e}(\alpha p_x^2+\beta p_y^2)\,,
\end{equation}
with the two-dimensional momentum $\mathbf{p}=(p_x,p_y)$, the electron mass $m_e$, and generic modifications $\alpha$ and $\beta$ that can be parameterized by particular SME coefficients once a specific sector is chosen. One of the parameters $\alpha,\beta$ can be absorbed into the electron mass to give rise to an effective electron mass $m=m_e/\alpha$. The Hamiltonian is then recast into
\begin{equation}
\label{eq:hamiltonian-anisotropic}
H=\frac{1}{2m}(p_x^2+\Upsilon p_y^2)\,,
\end{equation}
with $\Upsilon\equiv\beta/\alpha$, introduced for brevity. Thus, without a restriction of generality it is sufficient to modify the dispersion relation either along the $x$ or the $y$ direction where the latter is chosen in this particular case.

\subsection{Density of states}

To evaluate the Hall resistivity, the density of states $D(E)$ of a two-dimensional particle system with the anisotropic Hamiltonian stated in Eq.~\eqref{eq:hamiltonian-anisotropic} is indispensable. According to the definition,
\begin{subequations}
\begin{align}
D(E)&=\int\frac{\mathrm{d}^2p}{(2\pi)^2}\delta(E-E_p)\,, \\[2ex]
\label{eq:free-dispersion-relation}
E_p&=\frac{\hbar^2}{2m}(p_x^2+\Upsilon p_y^2)\,,
\end{align}
\end{subequations}
with the dispersion relation $E_p$, which follows directly from Eq.~\eqref{eq:hamiltonian-anisotropic}. To evaluate $D(E)$, Cartesian coordinates are employed:
\begin{equation}
D(E)=\int_{-\infty}^{\infty} \frac{\mathrm{d}p_x}{2\pi} \int_{-\infty}^{\infty}\frac{\mathrm{d}p_y}{2\pi}\,\delta\left[E-\frac{\hbar^2}{2m}(p_x^2+\Upsilon p_y^2)\right]\,.
\end{equation}
Solving the argument of the $\delta$ function for $p_y$ leads to
\begin{equation}
p_y=\pm\frac{1}{\sqrt{\Upsilon}}\frac{\sqrt{2mE-\hbar^2p_x^2}}{\hbar}\,.
\end{equation}
Requiring that $p_y\in\mathbb{R}$, the component $p_x$ is restricted to the interval $[-P,P]$ with $P=\sqrt{2mE}/\hbar$. Then, the remaining integral over $p_x$ provides the final expression for the density of states:
\begin{align}
\label{eq:density-states}
D(E)&=\frac{1}{\sqrt{\Upsilon}}\frac{m}{2\pi^2\hbar}\int_{-P}^P \mathrm{d}p_x\,\frac{1}{\sqrt{2mE-\hbar^2p_x^2}} \notag \\
&=\left.\frac{1}{\sqrt{\Upsilon}}\frac{m}{2\pi^2\hbar^2}\arctan\left(\frac{\hbar p_x}{\sqrt{2mE-\hbar^2p_x^2}}\right)\right|^P_{-P} \notag \\
&=\frac{1}{\sqrt{\Upsilon}}\frac{m}{2\pi\hbar^2}\,.
\end{align}
Therefore, in comparison to the result for an isotropic electron dispersion, $D(E)$ gets modified by the global factor $1/\sqrt{\Upsilon}$, which is a measure for the anisotropy.

\subsection{Example for anisotropic modification}

Since the current paper is on nonminimal Lorentz-violating modifications in electromagnetism, we would also like to consult nonminimal operators for electrons. An intriguing possibility is to consider the $(2+1)$-dimensional analog of one of the scalar operators in the Dirac fermion sector of the nonminimal SME~\cite{Kostelecky:2013rta}, which does not have any counterpart in the minimal SME:%%
\begin{subequations}
\begin{align}
0&=\cancel{p}-(m_{\psi}+\hat{m})\mathds{1}_4\,, \\[2ex]
\hat{m}&=\sum_{\substack{d \text{ even} \\ d\geq 4}} m^{(d)\alpha_1\dots\alpha_{d-2}}p_{\alpha_1}\dots p_{\alpha_{d-2}}\,,
\end{align}
\end{subequations}
where we will truncate this series immediately after $d=4$. Note that the resulting operator corresponds to the one of Eq.~\eqref{eq:modified-Dirac-theory} that we found to be in a dual relationship with $\hat{\mathcal{K}}^{(4)}$ of Eq.~\eqref{kappa2}. So the latter finding should provide an excellent motivation for considering a fermionic modification governed by $\hat{m}^{(4)}$. Now, the relativistic particle Hamiltonian \cite{Kostelecky:2013rta} reads
\begin{align}
h&=E_0+\frac{\hat{m}m_{\psi}}{E_0}=\sqrt{p^2+m_{\psi}^2}+\frac{\hat{m}m_{\psi}}{\sqrt{p^2+m_{\psi}^2}} \notag \\
&=m_{\psi}\sqrt{1+\frac{p^2}{m_{\psi}^2}}+\frac{\hat{m}}{\sqrt{1+\frac{p^2}{m_{\psi}^2}}}\,,
\end{align}
whose form does not depend on the number of spacetime dimensions considered.
Expanding for $m_{\psi}^2\gg p^2$ provides the nonrelativistic limit, which is the relevant one for the system to be studied, cf.~Eq.~\eqref{eq:hamiltonian-anisotropic}. Hence,
\begin{subequations}
\begin{align}
h_{\mathrm{nonrel}}&=m_{\psi}\left(1+\frac{p^2}{2m_{\psi}^2}+\dots\right)+\hat{m}\left(1-\frac{p^2}{2m_{\psi}^2}+\dots\right) \notag \displaybreak[0]\\
&=m_{\psi}+\hat{m}+\frac{p^2}{2m_{\psi}}+\dots \notag \displaybreak[0]\\
&=m_{\psi}+\frac{p^2}{2m_{\psi}}+m^{(4)xx}p_x^2+m^{(4)yy}p_y^2 \notag \displaybreak[0]\\
&=m_{\psi}+\frac{1}{2m_{\psi}}\Big[(1+2m_{\psi}m^{(4)xx})p_x^2 \notag \displaybreak[0]\\
&\phantom{{}={}}\hspace{1.9cm}+(1+2m_{\psi}m^{(4)yy})p_y^2\Big]\,,
\end{align}
where we invoked the special choice
\begin{equation}
\hat{m}=m^{(4)xx}\frac{\partial^2}{\partial x^2}+m^{(4)yy}\frac{\partial^2}{\partial y^2}\,.
\end{equation}
\end{subequations}
Now, if the electron mass is redefined as before,
\begin{equation}
m=\frac{m_e}{1+2m_{\psi}m^{(4)xx}}\,,
\end{equation}
the quantity $\Upsilon$ takes the form
\begin{equation}
\Upsilon=1+2m(m^{(4)yy}-m^{(4)xx})\,.
\end{equation}
Thus, $\Upsilon\neq 1$ arises for $m^{(4)yy}\neq m^{(4)xx}$, which emphasizes the anisotropic properties of this particular model.

\subsection{Modified Landau problem}

From a quantum mechanical perspective, the Landau problem is at the heart of the QHE. This problem treats the quantum motion of an electron in a magnetic field, i.e., the energy eigenvalues and eigenfunctions are to be determined. To do so, the free Hamiltonian of Eq.~\eqref{eq:hamiltonian-anisotropic} is minimally coupled to a vector potential $\mathbf{A}=(A_x,A_y)$ to describe the two-dimensional electron gas in the presence of a magnetic field $\mathbf{B}$ of field strength $B=|\mathbf{B}|$ being perpendicular to the sample. This gives rise to
\begin{equation}
H_A=\frac{1}{2m}\left[(p_x+eA_x)^2+\Upsilon(p_y+eA_y)^2\right]\,.
\end{equation}
Now we employ the Landau gauge $A_x=0$ and $A_y=Bx$. Then, the modified Landau Hamiltonian takes the form
\begin{align}
\label{eq:landau-hamiltonian}
H_L&=H_A|_{\substack{\text{Landau} \\ \text{gauge}}} \notag \\
&=\frac{1}{2m}\left[-\hbar^2\frac{\partial^2}{\partial x^2}+\Upsilon\left(\frac{\hbar}{\mathrm{i}}\frac{\partial}{\partial y}+eBx\right)^2\right]\,.
\end{align}
Due to $[H_L,p_y]=0$, the momentum along the $y$ direction is conserved and it makes sense to label energy eigenfunctions by the eigenvalues of $p_y$ that we denote by $k_y$. Therefore, the following separation \textit{ansatz} for the total wavefunction is reasonable:
\begin{equation}
\label{eq:landau-hamiltonian-wavefunction-ansatz}
\psi(x,y)=\phi_{k_y}(x)\exp(\mathrm{i}k_yy)\,.
\end{equation}
Applying Eq.~\eqref{eq:landau-hamiltonian} to the latter wavefunction leads to
\begin{subequations}
\begin{align}
H_L\psi(x,y)=\left[-\frac{\hbar^2}{2m}\frac{\mathrm{d}^2}{\mathrm{d}x^2}+\frac{m}{2}\Upsilon\omega_c^2(x-x_k)^2\right]\psi(x,y)\,,
\end{align}
with the cyclotron frequency $\omega_c$, the orbit center coordinate $x_k$, and the magnetic length scale $l_B$, which are explicitly given as follows:
\begin{equation}
\omega_c=\frac{eB}{m}\,,\quad x_k=l_B^2k_y\,,\quad l_B=\sqrt{\frac{\hbar}{eB}}\,.
\end{equation}
\end{subequations}
Now, the Schr\"{o}dinger equation
\begin{equation}
H_L\psi(x,y)=E\psi(x,y)\,,
\end{equation}
can be interpreted as that of a harmonic oscillator. So the electron energy is quantized and expressed in the usual way in terms of a modified cyclotron frequency:
\begin{subequations}
\begin{align}
E_n&=\hbar\omega_c'\left(n+\frac{1}{2}\right)\,, \\[2ex]
\omega_c'&=\sqrt{\Upsilon}\omega_c\,,
\end{align}
\end{subequations}
with the quantum number $n\in\mathbb{N}_0$.

\subsection{Quantized Hall resistivity}

The curves of constant electron energy in momentum space are ellipses,
\begin{equation}
E=\frac{\hbar^2}{2m}(k_x^2+\Upsilon k_y^2)=\hbar\omega_c'\left(n+\frac{1}{2}\right)\,,
\end{equation}
i.e., they are reformulated as follows:
\begin{subequations}
\begin{align}
1&=\frac{k_x^2}{A^2}+\frac{k_y^2}{B^2}\,, \displaybreak[0]\\[1ex]
A&=\sqrt{\frac{2m}{\hbar}\Upsilon\omega_c\left(n+\frac{1}{2}\right)}\,,\quad B=\sqrt{\frac{2m}{\hbar}\omega_c\left(n+\frac{1}{2}\right)}\,.
\end{align}
\end{subequations}
The area of such an ellipse associated with momentum $\mathbf{k}$ amounts to
\begin{equation}
S_{k,n}=\pi AB=\frac{2\pi\sqrt{\Upsilon}eB}{\hbar}\left(n+\frac{1}{2}\right)\,,
\end{equation}
such that the difference between the areas $S_{k,n}$ and $S_{k,n-1}$ of neighbouring ellipses is
\begin{equation}
\label{eq:difference-areas-momentum-space}
\Delta S_k\equiv S_{k,n}-S_{k,n-1}=\frac{2\pi\sqrt{\Upsilon} eB}{\hbar}\,.
\end{equation}
In configuration space, a modified particle Lagrangian is needed to describe the semiclassical motion of the particle. It is proposed that
\begin{equation}
L=-m\sqrt{1-v_x^2-\frac{v_y^2}{\Upsilon}}\,,
\end{equation}
with the particle velocity $\mathbf{v}=(v_x,v_y)$. In principle, the latter follows from the Hamiltonian of Eq.~\eqref{eq:hamiltonian-anisotropic} and the associated dispersion relation via the procedure developed in Ref.~\cite{Kostelecky:2010hs}. The equations of motion are then solved by elliptical trajectories satisfying
\begin{equation}
\label{eq:energy-trajectories-configuration-space}
E=\frac{m}{2}\Upsilon\omega_c^2\left(x^2+\frac{y^2}{\Upsilon}\right)=\hbar\sqrt{\Upsilon}\omega_c\left(n+\frac{1}{2}\right)\,,
\end{equation}
which are alternatively cast into
\begin{subequations}
\begin{align}
1&=\frac{x^2}{\tilde{A}^2}+\frac{y^2}{\tilde{B}^2}\,, \\[1ex]
\tilde{A}&=\sqrt{\frac{2\hbar}{m\omega_c}\frac{1}{\sqrt{\Upsilon}}\left(n+\frac{1}{2}\right)}\,,\quad \tilde{B}=\sqrt{\frac{2\hbar}{m\omega_c}\sqrt{\Upsilon}\left(n+\frac{1}{2}\right)}\,.
\end{align}
\end{subequations}
So the area enclosed by such an ellipse in configuration space takes the form
\begin{equation}
S_{r,n}=\pi\tilde{A}\tilde{B}=\frac{2\pi\hbar}{m\omega_c}\left(n+\frac{1}{2}\right)\,.
\end{equation}
Then, the difference between the areas $S_{r,n}$ and $S_{r,n-1}$ of neighbouring ellipses reads
\begin{equation}
\label{eq:difference-areas-configuration-space}
\Delta S_r=\frac{2\pi\hbar}{eB}\,.
\end{equation}
The product of Eqs.~(\ref{eq:difference-areas-momentum-space}) and (\ref{eq:difference-areas-configuration-space}) provides a constant: $\Delta S_r\Delta S_k=4\pi\sqrt{\Upsilon}$. Equation~\eqref{eq:difference-areas-configuration-space} allows for defining the magnetic-flux quantum $\Phi_0$:
\begin{equation}
\label{eq:flux-quantum}
\Delta S_rB=\frac{h}{e}\equiv \Phi_0\,.
\end{equation}
To compute the number of states in a single Landau level, we need the density of states $D(E)$ of Eq.~\eqref{eq:density-states}, which is based on the anisotropic dispersion relation of free electrons stated in Eq.~\eqref{eq:free-dispersion-relation}. Note that the external magnetic field is not capable of changing the number of states. The energy window of a single Landau level is $\Delta E\equiv E_n-E_{n-1}=\hbar\omega_c'$. Then, the number of states in each Landau level follows from
\begin{align}
N_s&=(2s+1)D(E)\Delta E S=\frac{2s+1}{\sqrt{\Upsilon}}\frac{m}{2\pi\hbar^2}\hbar\omega_c'S \notag \\
&=(2s+1)\frac{e}{h}BS=(2s+1)\frac{\Phi}{\Phi_0}\,,
\end{align}
where $\Phi=BS$ is the magnetic flux permeating a sample of area $S$ and $\Phi_0$ is the flux quantum of Eq.~\eqref{eq:flux-quantum}. Furthermore, we can also take into account the electron spin degeneracy $2s+1=2$ with the spin quantum number $s=1/2$. However, when the magnetic field is strong enough, the energy levels associated with different spin projections become nondegenerate, whereupon the energy levels separate. In this case the electrons are usually considered as spinless, i.e., we employ $s=0$ in $N_s$. The corresponding number density of states per area then amounts to
\begin{equation}
\label{eq:number-density}
n_s=\frac{N_s}{S}=\frac{e}{h}B\,.
\end{equation}
Finally, the filling factor $\nu$, i.e., the number of occupied Landau levels corresponds to the ratio between the electron density $n_e$ and $n_s$ of Eq.~\eqref{eq:number-density}:
\begin{equation}
\label{eq:filling-factor}
\nu\equiv\frac{n_e}{n_s}=\frac{n_e}{B}\frac{h}{e}\,.
\end{equation}
The occurrence of Landau levels that are occupied successively implies a quantized Hall resistivity in units of~$h/e^2$:
\begin{equation}
\label{eq:Hall-resistivity-quantized}
R_{xy}=\frac{B}{n_ee}=\frac{h}{\nu e^2}\,,
\end{equation}
for $\nu\in\mathbb{N}$. So an anisotropy in the electron dispersion of the form of Eq.~\eqref{eq:hamiltonian-anisotropic} does not modify the very essence of the QHE. The outcome makes perfect sense, as this phenomenon is of topological nature, i.e., local properties of the electron such as perturbations of its dispersion relation are not expected to lead to a different behavior. This will become clearer in the forthcoming section.

\subsection{CS theory}

It is possible to describe the quantization of the Hall resistivity of Eq.~\eqref{eq:Hall-resistivity-quantized} by means of an effective theory that is of CS form. Let $S[\psi,\varphi,A]$ be the action of the field theory taking into account all degrees of freedom of the sample, i.e., fermionic ones for the electron incorporated into $\psi$, scalar ones for impurities contained in $\varphi$, and electromagnetic degrees of freedom described by the vector field $A_{\mu}$. Integrating out the fermionic and scalar degrees of freedom in the path integral implies a CS theory based on the action $S_{\mathrm{CS}}[A]$ as follows:
\begin{subequations}
\begin{align}
\int\mathcal{D}\psi\mathcal{D}\varphi\mathcal{D}A\,&\exp\left(\frac{\mathrm{i}}{\hbar}S[\psi,\varphi,A]\right) \notag \\
&=\int\mathcal{D}A\exp\left(\frac{\mathrm{i}}{\hbar}S_{\mathrm{CS}}[A]\right)\,,
\end{align}
with
\begin{equation}
\label{eq:CS-theory-quantum-Hall-effect}
S_{\mathrm{CS}}[A]=\int\mathrm{d}^3x\,\left(\frac{\zeta e^2}{4\pi\hbar}\varepsilon^{\mu\nu\lambda}A_{\mu}\partial_{\nu}A_{\lambda}+\frac{j_{\mu}}{c}A^{\mu}\right)\,.
\end{equation}
\end{subequations}
Here we reinstated the constants of nature, which renders the parameter $\zeta$ dimensionless. Integrating out $\psi$ and $\varphi$ of the generic action $S$ is challenging. Instead, it is more straightforward to derive Ohm's law from the field equations of $S_{\mathrm{CS}}[A]$, which we will be doing as follows. The variation of the action is:
\begin{align}
\frac{\delta S_{\mathrm{CS}}[A]}{\delta A_{\varrho}}&=\frac{\zeta e^2}{4\pi\hbar}(\varepsilon^{\varrho\nu\lambda}\partial_{\nu}A_{\lambda}-\varepsilon^{\mu\nu\varrho}\partial_{\nu}A_{\mu})+\frac{j^{\varrho}}{c} \notag \\
&=\frac{\zeta e^2}{2\pi\hbar}\varepsilon^{\varrho\nu\lambda}\partial_{\nu}A_{\lambda}+\frac{j^{\varrho}}{c}=\frac{\zeta e^2}{4\pi\hbar}\varepsilon^{\varrho\nu\lambda}F_{\nu\lambda}+\frac{j^{\varrho}}{c}\,.
\end{align}
The principle of stationary action leads to the field equations
\begin{equation}
\frac{\zeta e^2}{4\pi\hbar}\varepsilon^{\varrho\nu\lambda}F_{\nu\lambda}=-\frac{j^{\varrho}}{c}\,.
\end{equation}
Let us consider the spatial components of the current density:
\begin{equation}
j^k=-\frac{\zeta e^2}{4\pi\hbar}\varepsilon^{kl0}cF_{l0}=\frac{\zeta e^2}{2\pi\hbar}\varepsilon^{kl}E^l\,,
\end{equation}
which allows us to read off the Hall conductivity $\sigma_{xy}$:
\begin{equation}
j^x=\sigma_{xy}E^y\,,\quad \sigma_{xy}=\frac{\zeta e^2}{2\pi\hbar}=\frac{\zeta e^2}{h}\,.
\end{equation}
The latter is the inverse of the Hall resistivity:
\begin{equation}
R_{xy}=\frac{1}{\sigma_{xy}}=\frac{h}{\zeta e^2}\,,
\end{equation}
where we can simply choose $\zeta=\nu\in\mathbb{N}$ with the filling factor of Eq.~\eqref{eq:filling-factor} to identify the latter $R_{xy}$ with that of Eq.~\eqref{eq:Hall-resistivity-quantized}. Based on this finding, the topological nature of the QHE is understood in terms of a topological invariant known as the first Chern number. This is the reason for the QHE being independent of the microscopic details of the sample, as, in principle, measurements of the Hall resistivity correspond to measuring a topological invariant.

Note that Eq.~\eqref{eq:free-dispersion-relation} can be understood via an effective spatial metric $\tilde{\eta}_{ij}$:
\begin{equation}
\label{eq:free-dispersion-relation-effective-metric}
E=\frac{\hbar^2}{2m}k^i\tilde{\eta}_{ij}k^j\,,\quad (\tilde{\eta}_{ij})=\begin{pmatrix}
1 & 0 \\
0 & \Upsilon \\
\end{pmatrix}\,.
\end{equation}
It is possible to interpret the form of the elliptic trajectories in configuration space, see Eq.~\eqref{eq:energy-trajectories-configuration-space}, as
\begin{equation}
E=\frac{\hbar^2}{2m}r^i\tilde{\eta}^{-1}_{ij}r^j\,,\quad (\tilde{\eta}^{-1}_{ij})=\begin{pmatrix}
1 & 0 \\
0 & 1/\Upsilon \\
\end{pmatrix}\,,
\end{equation}
with $\tilde{\eta}^{-1}_{ij}$ being the inverse of the effective metric. However, Eq.~\eqref{eq:CS-theory-quantum-Hall-effect} does not involve any metric tensor, as it is a topological field theory. Hence, there is no possibility of the effective metric occurring in the Hall resistivity. This is an alternative way to explain why the Hall resistivity of Eq.~\eqref{eq:Hall-resistivity-quantized} cannot involve $\Upsilon$, which describes a geometric property of the electron dispersion relation.

\subsection{(Extended) CSDJ theory}

Now we extend the previous analysis by a nonminimal contribution to the CS term; cf.~the extended CSDJ Lagrange density of Eq.~\eqref{LMCSD5A}:
\begin{align}
S[A]&=\int\mathrm{d}^3x\,\left[\frac{\zeta e^2}{4\pi\hbar}\varepsilon^{\mu\nu\lambda}A_{\mu}\partial_{\nu}(1+K^{\alpha\beta}\partial_{\alpha}\partial_{\beta})A_{\lambda}\right. \notag \\
&\phantom{{}={}}\hspace{1.1cm}\left.{}+\frac{j_{\mu}}{c}A^{\mu}\right]\,.
\end{align}
The variation of the action provides
\begin{align}
\frac{\delta S[A]}{\delta A_{\varrho}}&=\frac{\zeta e^2}{4\pi\hbar}\Big[\varepsilon^{\varrho\nu\lambda}\partial_{\nu}(1+K^{\alpha\beta}\partial_{\alpha}\partial_{\beta})A_{\lambda} \notag \\
&\phantom{{}={}}\hspace{0.8cm}-\varepsilon^{\mu\nu\varrho}\partial_{\nu}(1+K^{\alpha\beta}\partial_{\alpha}\partial_{\beta})A_{\mu}\Big]+\frac{j^{\varrho}}{c} \notag \\
&=\frac{\zeta e^2}{2\pi\hbar}\varepsilon^{\varrho\nu\lambda}\partial_{\nu}(1+K^{\alpha\beta}\partial_{\alpha}\partial_{\beta})A_{\lambda}+\frac{j^{\varrho}}{c} \notag \\
&=\frac{\zeta e^2}{4\pi\hbar}\varepsilon^{\varrho\nu\lambda}(1+K^{\alpha\beta}\partial_{\alpha}\partial_{\beta})F_{\nu\lambda}+\frac{j^{\varrho}}{c}\,.
\end{align}%%
Although additional time derivatives are likely to cause problems with unitarity, as we remarked under Eq.~\eqref{eq:saturation-residue-MCSDJ-theory}, the coefficients $K^{00}$ and $K^{0i}=K^{i0}$ contracted with time derivatives can parameterize interesting physics. So we keep them in the setting of the effective field theory, which leads to
\begin{equation}
j^k=\frac{\zeta e^2}{2\pi\hbar}\varepsilon^{kl}(1-K^{\alpha\beta}p_{\alpha}p_{\beta})E^l\,,
\end{equation}
in momentum space. Thus, the modified Hall resistivity reads
\begin{equation}
\label{eq:Hall-conductivity-nonminimal-operator}
R_{xy}=\frac{h}{(1-K^{\alpha\beta}p_{\alpha}p_{\beta})\zeta e^2}\,,
\end{equation}
which is now both energy-, momentum-, and direction-dependent via the presence of $K^{\alpha\beta}$. As the previous considerations show, anisotropies at the level of Eqs.~\eqref{eq:free-dispersion-relation}, \eqref{eq:free-dispersion-relation-effective-metric} do not have any impact on the QHE, which is topological. However, anisotropies can be incorporated into the Hall resistivity at the effective-field theory level via the presence of nonminimal SME field operators. These are then automatically accompanied by a momentum dependence of the electromagnetic field.

The reason for this being possible is that nonminimal extensions of CS theory such as the CSDJ model of Eq.~\eqref{LMCSD5A} naturally involve the spacetime metric. So the topological nature of the QHE, which applies to the infrared regime, is now superimposed by phenomena depending on local properties of the sample. Hence, based on Eq.~\eqref{eq:Hall-conductivity-nonminimal-operator}, our prediction is that materials should exist whose quantized Hall conductivity changes in the ultraviolet regime. A possible behavior is shown in Eq.~\eqref{eq:Hall-conductivity-nonminimal-operator}. Note, however, that the latter is still an effective description of potential ultraviolet phenomena.

Modifications of the Hall conductivity that depend on the energy-momentum $(E,\mathbf{p})$ of the electromagnetic field, but are isotropic, have already been proposed in Ref.~\cite{VanMechelen:2019ebr}. In the latter paper four regimes of the Hall conductivity $\sigma_{xy}=\sigma_{xy}(E,|\mathbf{p}|)$ have been pointed out. First, there is the static regime $\sigma_{xy}(0,0)$, which is the standard case investigated since von Klitzing's experimental discovery of the QHE \cite{vonKlitzing:1980pdk}. In the context of extended CSDJ theory this regime is characterized by $K^{\mu\nu}=0$.

The regime with an energy-dependent Hall conductivity, $\sigma_{xy}(E,0)$, is called dynamical and is described by $K^{00}\neq 0$ and $K^{0i}=K^{ij}=0$. The authors of the aforementioned reference denote the momentum-dependent regime $\sigma_{xy}(0,p)$ as viscous and it is of their principal interest. Referring to extended CSDJ theory, this case involves $K^{ij}=\delta^{ij}$ and $K^{00}=K^{0i}=0$. The latter two choices for $K^{\mu\nu}$ violate $\mathit{SO}(2,1)$ invariance, but they are still isotropic, i.e., $\mathit{SO}(2)$ symmetry is maintained.

Last but not least, the dynamical and viscous settings can be joined giving rise to $\sigma_{xy}(E,p)$, which is the most involved case to be studied. It should be effectively described via $K^{\mu\nu}=\eta^{\mu\nu}$, which implies the $\mathit{SO}(2,1)$-invariant CSDJ model. The Hall conductivities $\sigma_{xy}(0,p)$ and $\sigma_{xy}(E,p)$, respectively, give rise to a nonvanishing photonic Chern number, i.e., topological phases of the electromagnetic field now become essential. Our $\sigma_{xy}=R_{xy}^{-1}$ with $R_{xy}$ of Eq.~\eqref{eq:Hall-conductivity-nonminimal-operator} and $K^{\mu\nu}\neq \eta^{\mu\nu}$ is a possible extension of the proposal in Ref.~\cite{VanMechelen:2019ebr}, which incorporates anisotropies in momentum space. In principle, a possible generalization of Eq.~\eqref{eq:Hall-conductivity-nonminimal-operator} including higher orders of the momentum is
\begin{equation}
R_{xy}=\frac{h}{(1+\hat{\mathcal{K}}(p))\zeta e^2}\,.
\end{equation}
with $\hat{\mathcal{K}}$ of Eq.~\eqref{KHO1} transformed to momentum space.

\section{Conclusions and final remarks}
\label{sec:conclusions}

In this paper we proposed a series of modified electrodynamics in $(2+1)$ spacetime dimensions. Each was based on CS theory and three different types of extensions: 1) a Maxwell term, 2) higher-derivative Lorentz-invariant terms, and 3) higher-derivative Lorentz-violating contributions. The specific terms incorporated into these theories are briefly summarized in Tab.~\ref{tab:planar-electrodynamics-models}.

We determined the propagators of each model in Tab.~\ref{tab:planar-electrodynamics-models}, which served as a base for deriving their physical dispersion relations. Table~\ref{tab:number-modes} provides the number of massless and massive modes that we found for each model. If the corresponding extended theories do not involve higher-order time derivatives, the modes are affected by the presence of Lorentz violation, but the number of modes remains the same. If higher-order time derivatives occur, though, additional modes show up that are nonperturbative in the controlling coefficients, i.e., they do not approach one of the Lorentz-invariant modes for vanishing coefficients.

Note that the propagator suitably contracted with conserved currents, known as the saturation, poses a means to gain a first understanding of possible unitarity issues at the quantum level. In contrast to what we found in previous works, e.g., Refs.~\cite{Leticia2,Leticia1}, it is not just higher-order time derivatives that are able to spoil unitarity. In fact, CS-type terms in $(2+1)$ spacetime dimensions give rise to contributions of topological nature, which were clearly not present in the $(3+1)$-dimensional theories studied in the latter articles. These topological terms shroud the behavior of the saturation and are a hurdle to making clear statements on the fate of tree-level unitarity. However, the criterion used indicates unitarity violations for at least certain regions in the parameter spaces of the models. Our analysis can be refined by checking the validity of the optical theorem for tree-level processes, cf.~Refs.~\cite{Marat1,Marat2,Schreck,Avila,Lopez,OpticalTheorem}, amongst others, which is beyond the scope of the current work.

We also looked at the classical propagation properties of the modes, i.e., we computed their group and front velocities. Having these results at our disposal, our intention was to identify sub- and superluminal regimes in the parameter spaces. Signals can, in fact, propagate faster than the speed of light when parameters and controlling coefficients are chosen suitably. However, this finding does not pose a problem with classical causality, as we interpreted the $(2+1)$-dimensional electrodynamics in the context of planer condensed-matter systems. Therefore, the symmetry group $\mathit{SO}(2,1)$ is not governed by a velocity characterizing the propagation of information at a fundamental level.
\begin{table}
\subfloat[]{\label{tab:planar-electrodynamics-models}\begin{tabular}{ccccc}
\toprule
 & CS & Maxwell & HD LI & HD LV \\
\colrule
(extended) MCS & \checkmark & \checkmark & & $(\checkmark)$ \\
(extended) MCSDJ & \checkmark & \checkmark & \checkmark & $(\checkmark)$ \\
(extended) CSDJ & \checkmark & & \checkmark & $(\checkmark)$ \\
\botrule
\end{tabular}} \\
\subfloat[]{\label{tab:number-modes}\begin{tabular}{cccc}
\toprule
      & Massless modes & Massive modes \\
\colrule
MCSDJ & 1              & 2 \\
MCS   & 1              & 1 \\
CSDJ  & 1              & 1 \\
\botrule
\end{tabular}}
\caption{\protect\subref{tab:planar-electrodynamics-models} Models introduced and studied with abbreviations HD (higher-derivative), LI (Lorentz-invariant), and LV (Lorentz-violating). \protect\subref{tab:number-modes} Number of physical modes identified in each of the models}
\end{table}%%

Another interest of ours were dualities between apparently very different models. We found one duality between an extended MCSDJ theory and a modified Dirac theory in $(2+1)$ spacetime dimensions that involves a Lorentz-invariant higher-derivative term and a $(2+1)$-dimensional version of the dimension-5 $\hat{m}$ coefficients of the SME. This finding is complementary to the results of Refs.~\cite{Deser:1984kw,VanMechelen:2019ebr} and reveals a kind of supersymmetry between spin-1 and spin-1/2 excitations in the plane, which are subject to $\mathit{SO}(2,1)$ invariance violation.

Our final objective was to apply a subset of the models introduced to a real planar condensed-matter system. It has been well-known for some time that CS theory can be employed as an effective description of the quantum Hall effect, i.e., it is possible to describe the quantization of the Hall resistivity by means of a CS theory. An anisotropic modification of the electron dispersion relation was shown to not have any impact on the Hall resistivity, which was expected, as the quantum Hall effect is topological and local properties of the sample do not feed into it --- at least not at leading order in the momentum of the electron or the electromagnetic fields.

In spite of that, the effective description of the quantum Hall effect via CS theory allows for proposing possible modifications of the latter that, in principle, could be tested experimentally. CSDJ theory leads to a correction of the usual quantized Hall resistivity that involves the energy-momentum of the electromagnetic field. Besides, extended CSDJ theory would even imply an anisotropic Hall resistivity. The models introduced in this paper and the results obtained have the potential of finding additional application in other planar condensed-matter systems.

\bigskip

\begin{acknowledgments}

It is a pleasure to thank Z.~Jacob for valuable discussions and for pointing out to us some useful references in the context of dualities.
M.S. is indebted to FAPEMA Universal 00830/19 and CNPq/Produtividade 310076/2021-8. M.M.F. is supported by CNPq/Produtividade 311220/2019-3, CNPq/Universal 422527/2021-1, and FAPEMA/POS-GRAD-02575/21. L. Lisboa-Santos acknowledges support by FAPEMA BPD-11962/22. Furthermore, we are indebted to CAPES/Finance Code 001.

\end{acknowledgments}

\appendix

\section{Fourier transform of anisotropic Green's function}
\label{eq:green-function-spacelike-configuration-space-details}

Here we intend to Fourier-transform the expanded Green's function of Eq.~\eqref{eq:green-function-spacelike-momentum-space-expanded} from momentum to configuration space, i.e.,
\begin{subequations}
\begin{align}
G(\mathbf{R})&=\int\frac{\mathrm{d}^2p}{(2\pi)^2}\,\exp(-\mathrm{i}Rp\cos\phi)G(p)\,, \displaybreak[0]\\[2ex]
G(p)&\simeq -\frac{1}{p^{2}+m^{2}}+\frac{\eta_{2}^{2}p^{2}}{(p^{2}+m^{2})^{2}}\cos^{2}(\alpha-\phi)\,,
\end{align}
\end{subequations}
whose first part is already known from Eq.~\eqref{eq:green-function-configuration-space-first-model} such that%%
\begin{subequations}
\label{eq:green-function-configuration-space}
\begin{align}
G(\mathbf{R})&\simeq -\frac{1}{2\pi}K_0(mR)+\frac{\eta_2^2}{(2\pi)^2}I\,, \displaybreak[0]\\[2ex]
I&=\int\mathrm{d}^2p\,\frac{p^2\exp(-\mathrm{i}Rp\cos \phi)}{(p^{2}+m^{2})^{2}}\cos^{2}(\alpha-\phi)\,.
\end{align}
\end{subequations}
The second part is to be evaluated as follows. By parametric differentiation for the mass,
\begin{subequations}
\begin{align}
\label{eq:integral-I}
I&=-\frac{1}{2m}\frac{\partial\tilde{I}}{\partial m}\,, \\[2ex]
\tilde{I}&=\int\mathrm{d}^2p\,\left(\frac{p^{2}\exp(-\mathrm{i}Rp\cos \phi)}{p^{2}+m^{2}}\right)\cos^{2}(\alpha-\phi)\,.
\end{align}
\end{subequations}
As this point it makes sense to rewrite the trigonometric function as
\begin{align}
\cos^2(\alpha-\phi)&=\sin^2\alpha+\cos(2\alpha)\cos^2\phi \notag \\
&\phantom{{}={}}+\frac{1}{2}\sin(2\alpha)\sin(2\phi)\,.
\end{align}
The angular integral over the third of these contributions vanishes,
\begin{equation}
\int_0^{2\pi} \exp(-\mathrm{i}Rp\cos\phi)\sin(2\phi)=0\,,
\end{equation}
as the integrand is antisymmetric with respect to the middle point of the interval, i.e., $\phi=\pi$. Thus,
\begin{subequations}
\begin{align}
\tilde{I}&=\sin^2\alpha \tilde{I}_1+\cos(2\alpha)\tilde{I}_2\,, \displaybreak[0]\\[2ex]
\tilde{I}_1&=\int\mathrm{d}^2p\,\frac{p^2\exp(-\mathrm{i}Rp\cos\phi)}{p^{2}+m^2}\,, \displaybreak[0]\\[2ex]
\tilde{I}_2&=\int\mathrm{d}^2p\,\frac{p^2\exp(-\mathrm{i}Rp\cos\phi)}{p^{2}+m^2}\cos^2\phi\,.
\end{align}
\end{subequations}
The first of the latter integrals gives
\begin{align}
\tilde{I}_1&=\int_0^{\infty}\mathrm{d}p\,\frac{p^3}{p^{2}+m^2}\int_0^{2\pi} \mathrm{d}\phi\,\exp(-\mathrm{i}Rp\cos\phi) \notag \\
&=2\pi\int_0^{\infty} \mathrm{d}p\,\frac{p^3J_0(Rp)}{p^{2}+m^2}=-2\pi m^2K_0(mR)\,.
\end{align}
The second is more involved and requires another parameter differentiation:
\begin{align}
\label{eq:integral-I2-tilde}
\tilde{I}_2&=-\frac{\partial^2}{\partial R^2}\int\mathrm{d}^2p\,\frac{\exp(-\mathrm{i}Rp\cos\phi)}{p^{2}+m^2} \notag \\
&=-2\pi\frac{\partial^2}{\partial R^2}K_0(mR)\,,
\end{align}
i.e., it is traced back to the second derivative of a modified Bessel function. Note that first-order derivatives of modified Bessel functions can again be expressed in terms of modified Bessel functions of different orders. In particular,
\begin{equation}
\label{eq:relation-bessel-functions-1}
\frac{\mathrm{d}}{\mathrm{d}x}K_n(x)=-\frac{1}{2}[K_{n-1}(x)+K_{n+1}(x)]\,.
\end{equation}
Further valuable relationships are
\begin{equation}
\label{eq:relation-bessel-functions-2}
-\frac{2n}{x}K_n(x)=K_{n-1}(x)-K_{n+1}(x)\,,
\end{equation}
as well as $K_{-n}(x)=K_n(x)$, see Eqs.~(8.486.11), (8.486.10), and (8.486.16) in Ref.~\cite{Gradshteyn:2007}. Applying these allows us to express Eq.~\eqref{eq:integral-I2-tilde} as follows:
\begin{equation}
\tilde{I}_2=-2\pi m^2\left[K_0(mR)+\frac{1}{mR}K_1(mR)\right]\,.
\end{equation}
Finally, for Eq.~\eqref{eq:integral-I} we must evaluate the parameter derivative for $m$. Here, the previous relationships of Eqs.~\eqref{eq:relation-bessel-functions-1}, \eqref{eq:relation-bessel-functions-2} are again useful and we obtain
\begin{equation}
I=\pi\left[K_0(mR)-mRK_1(mR)\cos^2\alpha \right]\,.
\end{equation}
After the dust settles, the Green's function in configuration space stated in Eq.~\eqref{eq:green-function-configuration-space} can be cast into its final form:
\begin{align}
G(\mathbf{R})&=-\frac{1}{2\pi}K_0(mR) \notag \\
&\phantom{{}={}}+\frac{\eta_2^2}{4\pi}\left[K_0(mR)-mRK_1(mR)\cos^2\alpha\right]\,,
\end{align}
which is reprinted in a slightly different shape in Eq.~\eqref{eq:green-function-spacelike-configuration-space}.

\end{document}